\documentclass[prx, twocolumn,superscriptaddress]{revtex4-1}

\usepackage{amsmath,amssymb, amsfonts}
\usepackage{bm}
\usepackage{graphicx}
\usepackage{braket}
\usepackage{ascmac}
\usepackage{color}
\usepackage{multirow}

\usepackage[T1]{fontenc}

\usepackage{here}

\newcommand{\pmat}[1]{\begin{pmatrix} #1 \end{pmatrix}}%
\newcommand{\tr}{\mathrm{tr}}
\newcommand{\Pf}{\mathrm{Pf}}
\newcommand{\sgn}{\mathrm{sgn}}

\begin{document}
\title{Extrinsic topology of Floquet anomalous boundary states in quantum walks}

\author{Takumi Bessho}
\email{takumi.bessho@yukawa.kyoto-u.ac.jp}
\affiliation{Yukawa Institute for Theoretical Physics, Kyoto University, Kyoto 606-8502, Japan}

\author{Ken Mochizuki}
\email{ken_mochizuki@eng.hokudai.ac.jp}
\affiliation{Advanced Institute for Materials Research (WPI-AIMR), Tohoku University, Sendai 980-8577, Japan}

\author{Hideaki Obuse}
\email{hideaki.obuse@eng.hokudai.ac.jp}
\affiliation{Department of Applied Physics, Hokkaido University, Sapporo 060-8628, Japan}
\affiliation{Institute of Industrial Science, The University of Tokyo, 5-1-5 Kashiwanoha, Kashiwa,Chiba 277-8574, Japan}

\author{Masatoshi Sato}
\email{msato@yukawa.kyoto-u.ac.jp}
\affiliation{Yukawa Institute for Theoretical Physics, Kyoto University, Kyoto 606-8502, Japan}

\date{\today}

\begin{abstract}
Bulk-boundary correspondence is a fundamental principle for topological phases where bulk topology determines gapless boundary states. On the other hand, it has been known that corner or hinge modes in higher order topological insulators may appear due to "extrinsic" topology of the boundaries even when the bulk topological numbers are trivial.  In this paper, we find that Floquet anomalous boundary states in quantum walks have similar extrinsic topological natures.  In contrast to higher order topological insulators, the extrinsic topology in quantum walks is manifest even for first-order topological phases. We present the topological table for extrinsic topology in quantum walks and illustrate extrinsic natures of Floquet anomalous boundary states in concrete examples.
\end{abstract}

\maketitle
\section{introduction}
\label{sec:introduction}
Recently, it has become clear that a class of gapless boundary states may appear without bulk topological numbers because of topological properties of boundaries. 
Such boundary states are called "extrinsic" \cite{Geier18}.
In particular, extrinsic topological phases are realized in higher-order topological insulators with gapless corner and hinge modes \cite{Benalcazar17,Benalcazar17_Science,Benalcazar19,Song17,Langbehn17,Schindler18,Ezawa18,Khalaf18,Matsugatani18}.
For example, attaching a one-dimensional Su-Schrieffer-Heeger chain \cite{SSH79} onto an edge of a two-dimensional chiral-symmetric topologically trivial insulator, we obtain extrinsic zero-energy corner states. The zero-energy gapless modes are robust against continuous deformation of the system unless the bulk energy gap closes.
The topological invariant of the edge Su-Schrieffer-Heeger chain determines the types and the numbers of these corner modes.

Floquet systems, of which Hamiltonians are periodic in time \cite{Kitagawa10_Floquet,Bukov15,Eckardt17,Oka19,Jiang11,Carpentier15,Nathan15,Nathan18}, and quantum walks, where periodically multiplying unitary operators describe their dynamics  \cite{Aharonov93,Kempe03,Kitagawa10,Obuse11,Kitagawa12,Gross12,Asboth12,Asboth13,Tarasinski14,Obuse15,Asboth15,Cedzich16,Cardano16,Barkhofen17,Cedzich18,Chen18,Cedzich18-1,Cedzich21,Matsuzawa20,Suzuki19}, have attracted uprising interests due to topological phenomena intrinsic to non-equilibrium systems.
Both Floquet systems and quantum walks have the $2\pi/T$ energy periodicity because of the Bloch-Floquet theorem for time translation symmetry with time period $T$ \cite{Sambe73}.
In addition to a conventional gap, usually at $\epsilon=0$,
the $2\pi/T$ periodicity 
allows another gap at the energy zone boundary $\epsilon=\pi/T$.
If the bulk energy gaps at $\epsilon=0$ and/or $\epsilon=\pi/T$ is open,
gapless boundary states at $\epsilon=0$ and/or $\epsilon=\pi/T$ can appear both in Floquet systems \cite{Rudner13,Kitagawa10_Floquet,Carpentier15,Jiang11} and quantum walks \cite{Kitagawa10,Kitagawa12,Asboth13}.
Due to the shared properties above, quantum walks are often regarded as a kind of Floquet systems in many previous literatures. 

In this paper, however, we clarify an intrinsic difference between Floquet systems and quantum walks in the bulk-boundary correspondence. The Floquet systems are governed by the continuous-time evolution of time-periodic Hamiltonians. In contrast, the quantum walks are governed by the discrete-time evolution of unitary operators. This difference leads to the absence of the bulk-boundary correspondence in quantum walks. In quantum walks, bulk topological invariants are insufficient to determine the number of the gapless boundary states, and the gapless boundary states can depend on the boundary topology. 
We also reveal that the extrinsic boundary states have the same form as Floquet anomalous edge states. Contrary to higher-order topological insulators in equilibrium, quantum walks may host the extrinsic boundary states even in first-order topological phases. 

This paper is organized as follows.
In Sec.~\ref{sec:simple}, we give an illustrative example showing the difference between Floquet systems and quantum walks, and explain the extrinsic topological nature of quantum walks. 
In Sec.~\ref{sec:general}, we classify extrinsic topological phases in quantum walks.
In Sec.~\ref{sec:example}, we examine extrinsic topological phases in one-dimensional quantum walks in detail.
In Sec.~\ref{sec:Floquet_vs_qw}, the relation between the  topological classification of Floquet systems and that of quantum walks is discussed.
In Sec.~\ref{sec:CSQW}, we argue the bulk-boundary correspondence for chiral-symmetric quantum walks in one dimension (1D).
For chiral symmetric quantum walks in 1D, it has been shown that the bulk topological numbers fully determines the number of boundary zero modes \cite{Asboth13}, 
{\it i.e.} no extrinsic topological phase appears. On the other hand, our classification indicates the presence of an extrinsic topological phase in this case.
We clarify that this difference comes from the difference in the definition of chiral symmetry. Indeed, Ref.~\cite{Asboth13} introduces chiral symmetry in a specific manner, and 
we demonstrate that the extrinsic topological phase becomes trivial under the special realization of chiral symmetry. 
We also find that class CII quantum walks in 1D has a similar property: Using an appropriate realization of symmetries, 
the bulk topological numbers fully determine the number of boundary zero modes. 
In Sec.~\ref{sec:implementation}, we present several physical implementations of extrinsic topological phases, and examine their properties. Finally, we give the conclusion in Sec.~\ref{sec:conclusion}

\section{Illustrative example}
\label{sec:simple}

\begin{figure}[t]
\centering
\includegraphics[width=55mm]{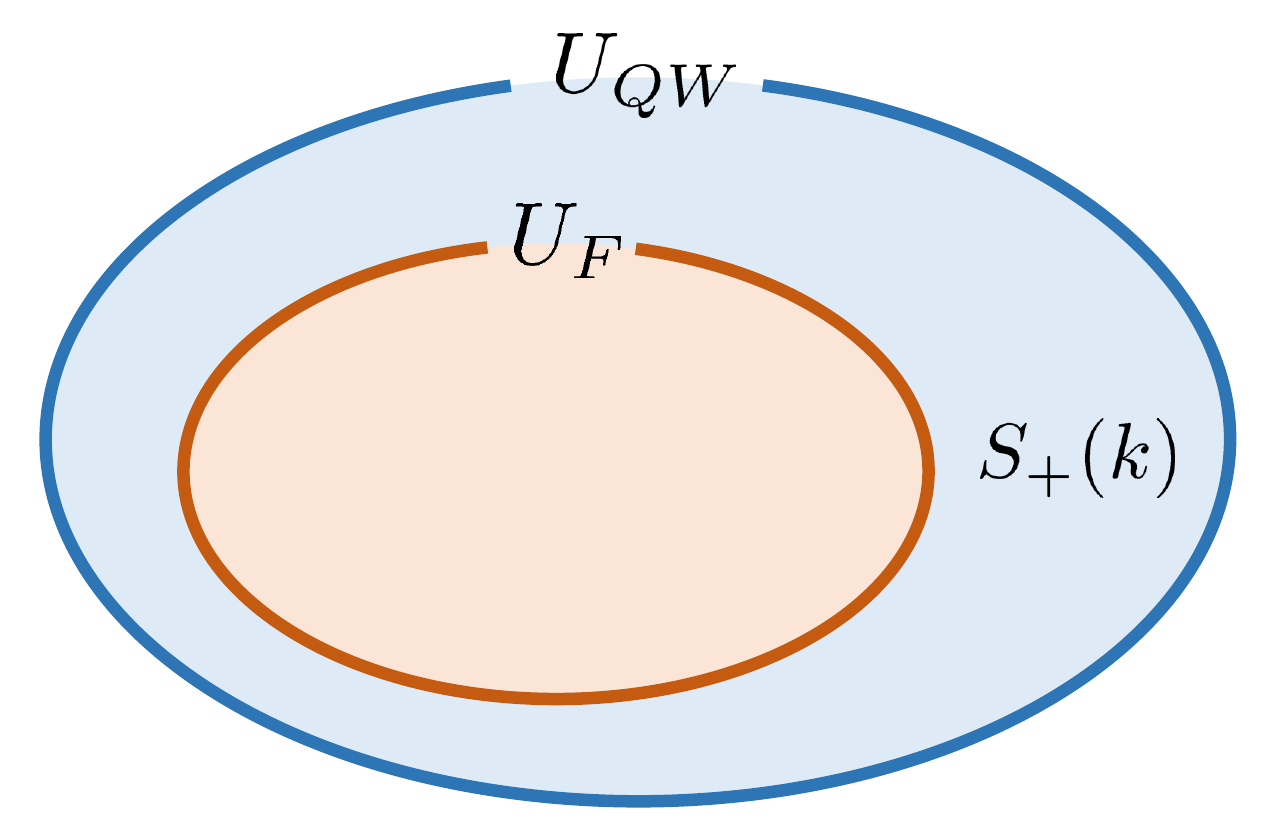} 
\caption{
Relation between quantum walks and Floquet systems. Any Floquet operator $U_F$ can be regarded as a quantum walk $U_{\text{QW}}$.
On the other hand, some quantum walks $U_{\text{QW}}$ such as $S_+(k)$ in Eq.~(\ref{eq:QW_simple}) cannot be realized as a Floquet operator $U_F$.
}
	\label{fig:Venn}
\end{figure}

In this section, we give an illustrative example to clarify an essential difference between Floquet systems and quantum walks.
We first compare the definition of quantum walks and Floquet systems.
In conventional Floquet systems, the time-evolution operator $U(\bm{k},t_1 \to t_2)$ in the momentum representation is given by a time-dependent microscopic Hamiltonian $H(\bm{k},t)$,
\begin{align}
    U(\bm{k},t_1 \to t_2) = \mathcal{T} \exp \left[ -i \int_{t_1}^
    {t_2} dt H(\bm{k},t)
    \right],
    \label{eq:Floquet_continuous}
\end{align}
where $\mathcal{T}$ is the time ordering operator. The one-cycle time-evolution operator $U_F(\bm{k}):=U(\bm{k},0\to T)$ is called the Floquet operator, where $T$ is the time period of the microscopic Hamiltonian $H(\bm{k},t+T)=H(\bm{k},t)$.
In quantum walks, however, the one-cycle time evolution is given directly by a series of unitary operators $U_j (\bm{k})$,
\begin{align}
    U_{\text{QW}}(\bm{k})=\prod_j U_j (\bm{k}).
    \label{eq:QW}
\end{align}
For both Floquet systems and quantum walks, we can define the effective Hamiltonians $H_F$ and $H_{\text{QW}}$ as 
\begin{align}
    U_F=e^{-iH_F T}, \quad U_{\text{QW}}=e^{-iH_{\text{QW}}},
    \label{eq:effH}
\end{align}
which describe stroboscopic dynamics of these systems.
The quasi-energies of these systems are defined as the eigenvalues of the effective Hamiltonians and the periodicity $2\pi/T$ of the energy can be understood as the periodicity in the phases of $U_F$ and $U_{\rm QW}$, where $T=1$ for quantum walks.
When $U_j(\bm{k})$'s are written by a microscopic Hamiltonian: 
$U_j(\bm{k}) = \mathcal{T} \exp \left[ -i \int_{t_{j}}^{t_{j+1}} d t H(\bm{k},t)\right]$, the quantum walk reduces to a Floquet system. 
However, this is not always possible for general quantum walks [Fig.~\ref{fig:Venn}]. 
An illustrative example is a single step quantum walk $U_{\text{QW}}(k)=S_+(k)$, where the operation $S_+(k)$ shifts the walker to the right if its spin is up \cite{Kempe03,Kitagawa10},
\begin{align}
    S_+(k)=\pmat{e^{-ik} & 0 \\ 0 & 1}.
    \label{eq:QW_simple}
\end{align}

As one can check immediately, this model has a non-trivial winding number $w_1[U_{\rm QW}(k)]=1$ with $w_1[U(k)]$ defined by
\begin{align}
    w_1[U(k)]= \int_0^{2\pi} \frac{dk}{2\pi} 
    \tr[U(k)^{-1} i\partial_k U(k)].
    \label{eq:w1}
\end{align}
On the other hand, for any Floquet continuous time evolution $U(k,t_1 \to t_2)$, the winding number becomes zero 
\footnote{We note that 
the Thouless pumping model in the adiabatic limit shows a nonzero winding number for the low-energy bands and the opposite winding number for the high-energy bands
\cite{Privitera18}, and thus the total winding number is zero.
In general, if a one-cycle time evolution operator has a block-diagonal structure:
\[
    U_F(k) = \pmat{U_1(k) & \\ & U_2(k)},
\]
$w_1 [U_1 (k)]$ can take nonzero value while we have $w_1 [U_F (k)]=w_1 [U_1 (k)]+w_1 [U_2 (k)]=0$ \cite{Nakagawa20}.
}:
\begin{align}
    w_1[U(k,t_1 \to t_2)] = w_1[U(k,0 \to 0)]=0,
\end{align}
as $U(k,t_1 \to t_2)$ is continuously deformed into $U(k,0 \to 0)=1$.
Therefore, the quantum walk $U_{\text{QW}}(k)=S_+(k)$ cannot be written by a Floquet continuous time evolution.

\begin{figure}[t]
\centering
\includegraphics[width=86mm]{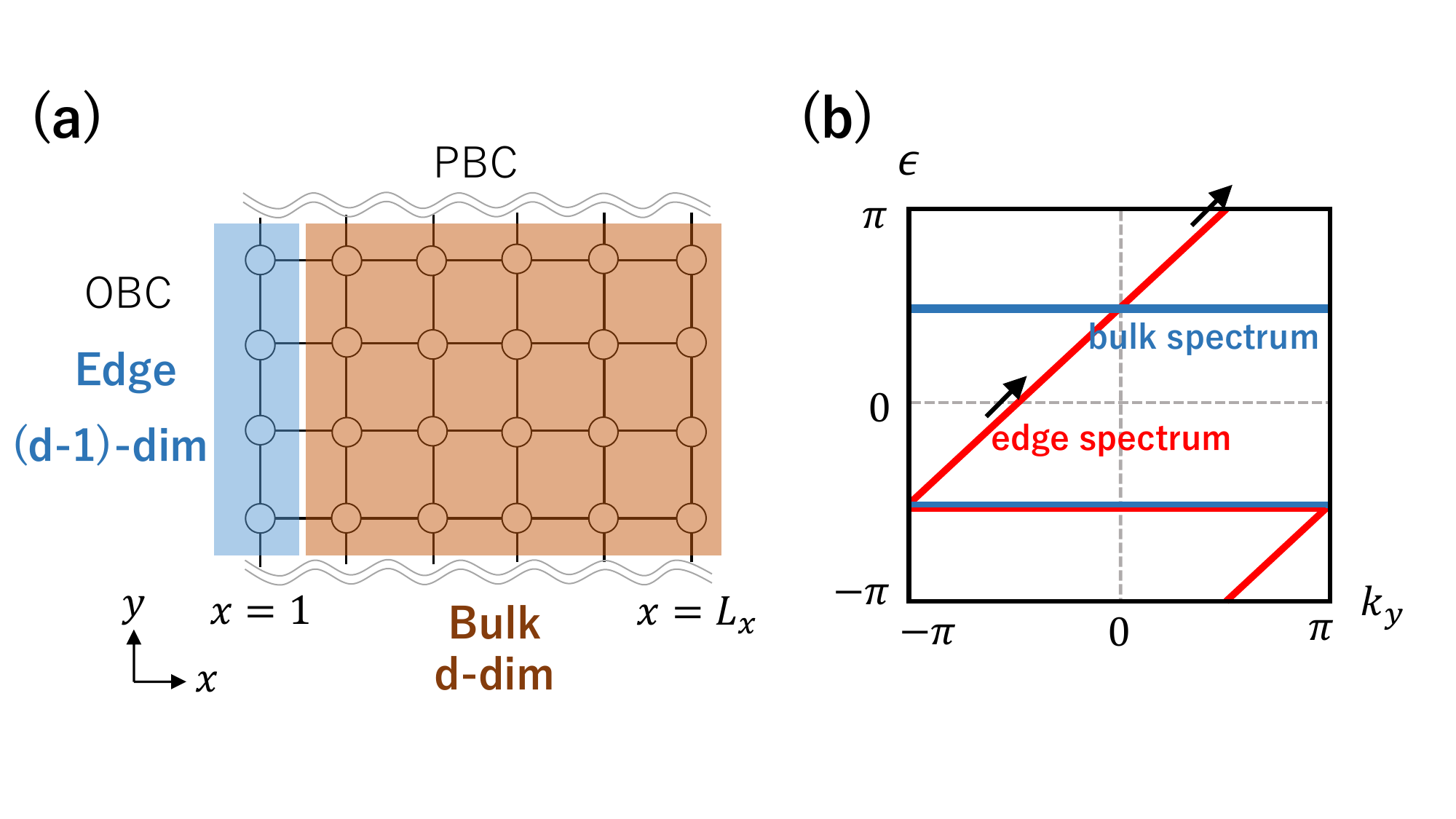} 
\caption{
Lattice structure and energy spectrum of the simple extrinsic topological model in Eq.~(\ref{eq:simple_extrinsic}).
(a) We attach the ($d-1$)-dimensional boundary onto the $d$-dimensional bulk. 
In the simple mode, we consider $d=2$. We impose the periodic boundary conditions (PBC) in the $y$ direction and the open boundary conditions (OBC) in the $x$ direction. 
(b) The energy spectrum of Eq.~(\ref{eq:simple_extrinsic}). The blue (red) line indicates the bulk (edge) energy spectrum. 
}
	\label{fig:simple_extrinsic}
\end{figure}

In this paper, we argue that this difference enables extrinsic topological phases in quantum walks.
A simple example of the extrinsic topological phase in a quantum walk is given as follows [Fig. \ref{fig:simple_extrinsic}].
First let us prepare a topologically trivial bulk model in two dimensions (2D) with energy gaps at $\epsilon=0$ and $\epsilon=\pi/T$ by using the coin operator $C(\theta)$. In the momentum space representation, the bulk model is
\begin{align}
U_{\text{bulk}}(\bm{k}) = C(\frac{\pi}{2}), \ C(\theta):=e^{-i\theta \sigma_z}.
\end{align}
Here, we used the Pauli matrix $\sigma_z$.
The real space representation of the bulk model is
\begin{align}
U_{\text{bulk}}
=\sum_{x=2}^{L_x} \sum_{y=1}^{L_y} -i\ket{\bm{r},+}\bra{\bm{r},+} +i\ket{\bm{r},-}\bra{\bm{r},-},
\label{eq:rbulk}
\end{align}
where $\bm{r}$ represents the $(x,y)$ position, and $\pm$ indicate two orthogonal internal states of the walker.
The effective Hamiltonian $H_{\rm QW}$ of this model is
\begin{align}
H_{\rm QW}(\bm{k})=\frac{\pi}{2}\sigma_z,    
\end{align}
which has the eigenvalues $\epsilon (\bm{k})= \pm \pi/2$. Thus, the model has energy gaps both at $\epsilon = 0,\pi$.
Furthermore, as the Hamiltonian is a constant, the bulk topological number is zero, so no gapless boundary mode is expected.

However, we can obtain a gapless chiral edge mode by decorating the boundary at $x=1$ of this model with a local unitary operator.
For instance, we can consider the following edge unitary operator 
\begin{align}
U_{\text{edge}}(k_y)  = S_+(k_y) C(\frac{\pi}{2}),
\label{eq:example_edge}
\end{align}
where $k_y$ is the momentum along the edge.
The effective Hamiltonian $H_{\rm QW}^{\rm edge}$ of the edge unitary operator is
\begin{align}
H_{\rm QW}^{\rm edge}(k_y)=
\begin{pmatrix}
k_y+\pi/2 & 0\\
0 & -\pi/2
\end{pmatrix},
\end{align}
and thus the edge unitary operator supports a chiral gapless mode with the linear dispersion $\epsilon (k_y)=k_y+\pi/2$.
The real space representation of the edge unitary operator is
\begin{align}
U_{\text{edge}} = \sum_{y=1}^{L_y} -i \ket{1,y+1,+}\bra{1,y,+} +i \ket{1,y,-}\bra{1,y,-},
\end{align}
where the periodic boundary condition is imposed in the $y$ direction. By attaching this to the real space representation of the bulk model $U_{\rm bulk}$ in Eq.~(\ref{eq:rbulk}) at $x=1$, we have a decorated bulk model,
\begin{align}
U_{\text{bulk}} \oplus U_{\text{edge}}.
\label{eq:simple_extrinsic}
\end{align}
In spite of the trivial bulk topological number,
the decorated model has a chiral edge mode, 
as shown in Fig.~\ref{fig:simple_extrinsic}~(b).

The extrinsic gapless chiral mode originates from the winding number of the edge unitary operator.
As explained above, in contrast to conventional Floquet systems, a unitary operator in a quantum walk may have a nonzero winding number, and the edge unitary operator $U_{\rm edge}(k_y)$ in Eq.~(\ref{eq:example_edge}) has $w_1[U_{\rm edge}(k_y)]=1$. 
Following the theorem proved in Ref.~\cite{BS20}, 
this implies the existence of gapless chiral modes both at $\epsilon=0,\pi$ at the same time:
The theorem in 1D indicates the relation between 
the winding number of $U_{\rm edge}(k_y)$ and the gapless modes, 
\begin{align}
\label{eq:NN1d}
    \sum_{\epsilon_p(k_{p\alpha})=0} \sgn ~v_{p\alpha} = 
    \sum_{\epsilon_p(k_{p\alpha})=\pi} \sgn ~v_{p\alpha} = w_1 [U_{\text{edge}}(k_y)],
\end{align}
where $k_{p\alpha}$ is the $\alpha$-th gapless point of band $p$ of the edge unitary operator defined by $\epsilon (k_{p\alpha}) = 0$ or $\pi$, and
$v_{p\alpha}=\left(\partial \epsilon_p / \partial k_y \right)_{k_y=k_{p\alpha}}$ is the group velocity of the gapless mode at $k_{p\alpha}$. Since ${\rm sgn}\ v_{p\alpha}$ measures the chirality of the gapless mode, a nonzero winding number implies the existence of chiral gapless modes both at $\epsilon=0, \pi$.

On the basis of the theorem in Ref.~\cite{BS20}, we classify  extrinsic topological phases in quantum walks in arbitrary dimensions.
Superficially, the classification coincides with that for gapless modes in  ordinary topological insulators and superconductors [Table \ref{tb:extrinsic}].
This is because the gapless boundary states in quantum walks have the same topological charges as those in usual topological insulators and superconductors.
However, contrary to the conventional gapless modes, gapless modes in the extrinsic topological phases appear in the absence of bulk topological numbers because of the non-trivial topology of boundary unitary operators.
Furthermore, the extrinsic gapless modes
always appear in a pair at $\epsilon=0,\pi/T$.
In even (odd) dimensions, the net topological charge of extrinsic gapless modes at $\epsilon=0$ is the same as (opposite to) the net topological charge of those at $\epsilon=\pi/T$.

The extrinsic topological phase in quantum walks is closely related to the so-called Floquet anomalous topological phase \cite{Roy17}.
In Floquet systems, there are two types of topological phases: one is the conventional topological phase defined by the effective Hamiltonian in Eq.~(\ref{eq:effH}), and the other is the anomalous one determined from a refined dynamics of the microscopic Hamiltonian in Eq.~(\ref{eq:Floquet_continuous}).
These two types of topological phases are needed to fully determine the boundary states both at $\epsilon=0$ and $\epsilon=\pi/T$.
In quantum walks, however, the microscopic Hamiltonian does not always exist, and thus anomalous topological phase is not possible in general. 
As a result, the bulk topology is insufficient to determine the boundary gapless states. 
Instead, we find that boundary operators can be topological, which enables us to fully control gapless states on the  boundary.

We also discuss possible physical implementations of such extrinsic topological phases.
We numerically and analytically study the robustness of extrinsic chiral gapless modes against disorders in the  Anderson localization Hamiltonian \cite{Anderson58}.
We also show that suitable modulations of boundaries can eliminate gapless boundary states in the split step quantum walk in one dimension \cite{Kitagawa10,Kitagawa12,Obuse15} and the five step model in two dimensions \cite{Rudner13}, which clearly illustrates the extrinsic nature of gapless modes in the quantum walks.

\section{Classification of extrinsic topology in quantum walks}\label{sec:general}
\begin{figure}[t]
\centering
\includegraphics[width=56mm]{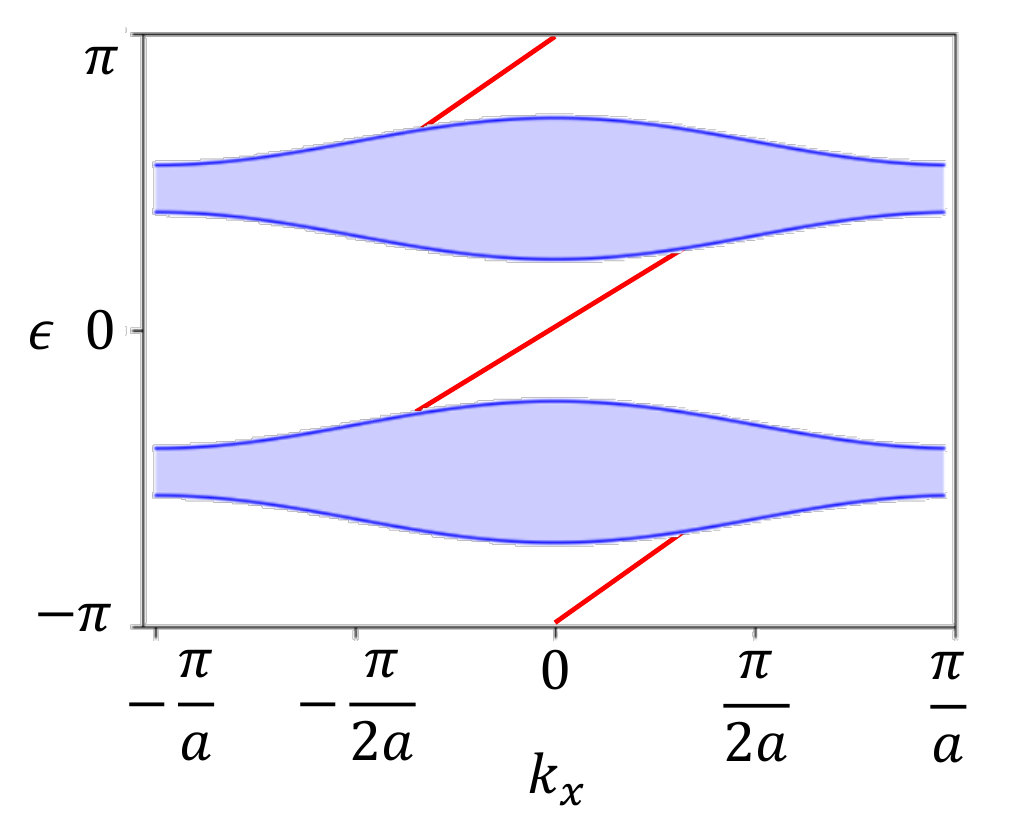} 
\caption{
A typical energy spectrum of a quantum walk: 
The energy has $2\pi$ periodicity, so there exist two high-symmetric bulk energy gaps at $\epsilon=0$ and $\pi$.
Gapless boundary states may appear at both bulk gaps. Here $a$ is a lattice constant.
}
	\label{fig:gap}
\end{figure}

Let us consider a general one-cycle time-evolution unitary operator $U_{\rm QW}({\bm k})$ and the effective Hamiltonian $H_{\rm QW}({\bm k})$ defined by $U_{\text{QW}}(\bm{k})=e^{-iH_{\text{QW}}({\bm k})}$ in $d$-dimensions with $\bm{k}=(k_1,k_2,\ldots,k_d)$. We first introduce the Altland-Zirnbauer (AZ) symmetry classes \cite{Altland97}.
$H_{\text{QW}}(\bm{k})$ possibly satisfies time-reversal symmetry (TRS), particle-hole symmetry (PHS) and/or chiral symmetry (CS):
\begin{align}
T H_{\text{QW}}(\bm{k}) T^{-1} &= H_{\text{QW}}(-\bm{k}), \\
C H_{\text{QW}}(\bm{k}) C^{-1} &= -H_{\text{QW}}(-\bm{k}), \\
\Gamma H_{\text{QW}}(\bm{k}) \Gamma^{-1} &= -H_{\text{QW}}(\bm{k}).
\end{align}
Here, $T$ and $C$ are anti-unitary operators with $T^2=\pm 1$ and $C^2=\pm 1$, and $\Gamma$ is a unitary operator with $\Gamma^2=1$.
The AZ symmetry classes are defined by the presence or absence of TRS, PHS and/or CS [Table \ref{tb:twogaps}].
In quantum walks, it is beneficial to rewrite the symmetries as those for $U_{\text{QW}}(\bm{k})$:
\begin{align}
\label{eq:TRSQW}
T U_{\text{QW}}(\bm{k}) T^{-1} &= U_{\text{QW}}(-\bm{k})^\dag, \\
\label{eq:PHSQW}
C U_{\text{QW}}(\bm{k}) C^{-1} &= U_{\text{QW}}(-\bm{k}), \\
\label{eq:CSQW}
\Gamma U_{\text{QW}}(\bm{k}) \Gamma^{-1} &= U_{\text{QW}}(\bm{k})^\dag.
\end{align}
When there are particle-hole and/or chiral symmetries, we have a symmetry constraint $\epsilon_n = -\epsilon_m \ (\text{mod} \ 2\pi)$ for the energy bands $n,m$ of $H_{\rm QW}$, and obtain high-symmetric energy gaps at $\epsilon=0,\pi$.
We assume that gaps are open at these levels for bulk bands in the following arguments [Fig. \ref{fig:gap}].

Gapless boundary states appear at $\epsilon=0$ and $\epsilon=\pi$ individually. They are described by a gapless Dirac Hamiltonian:
\begin{align}
    H_{\text{QW}}(\bm{k}_{\parallel})= \sum_{j=1}^{d-1} k_j \gamma_j + \epsilon_{\text{Gap}}\hat{1},
    \label{eq:gaplessDirac}
\end{align}
up to continuous deformations.
Here, we take the open boundary condition in the $x_d$ direction, $\bm{k}_{\parallel}=(k_1,k_2,\ldots,k_{d-1})$ is the momentum along the boundary, and
$\epsilon_{\text{Gap}}=0,\pi$ indicates the energy gaps we consider.
The gamma matrix $\gamma_j$ is Hermitian and obeys $\{\gamma_i,\gamma_j\}=2\delta_{ij}$. This boundary Hamiltonian is taken to be compatible with AZ symmetries.
The gapless Dirac Hamiltonian in Eq.~(\ref{eq:gaplessDirac}) has the same form as that for boundary states in conventional topological insulators and superconductors in equilibrium \cite{Kobayashi14,Matsuura13,Chiu14,Chiu16}.
In the latter systems, from the bulk-boundary correspondence, the topological classification of gapless boundary states in $(d-1)$-dimensions is the same as that of insulators and superconductors in $d$-dimensions.
Therefore, the topological classification of $(d-1)$-dimensional boundary states in quantum walks coincides with that of ordinary insulators and superconductors in $d$-dimensions [Table. \ref{tb:twogaps}].
Note that the topological numbers in Table \ref{tb:twogaps}
are doubled because each gap at $\epsilon=0,\pi$ may host gapless states.

\begin{table*}[t]
\centering
\caption{Periodic table for $(d-1)$-dimensional gapless boundary states of $d$-dimensional quantum walks in AZ symmetry class.
We assume bulk gaps at $\epsilon=0$ and $\pi$.
The table has the periodicity $d=d+2$ for class A and AII, and has the periodicity $d=d+8$ for other symmetry classes.}
\begin{tabular}{cccccccccccc}
  \hline\hline
  ~AZ class~~~~ & ~~$T$~~ & ~~$C$~~ & ~~$\Gamma$~~~ & ~$d=1$~ & ~~~2~~~ & ~~~3~~~ & ~~~4~~~ & ~~~5~~~ & ~~~6~~~ & ~~~7~~~ & ~~~8~~~ \\
  \hline
  A & 0 & 0 & 0 & 0 & $\mathbb{Z}^2$ & 0 & $\mathbb{Z}^2$ & 0 & $\mathbb{Z}^2$ & 0 & $\mathbb{Z}^2$ \\
  AIII & 0 & 0 & 1 & $\mathbb{Z}^2$ & 0 & $\mathbb{Z}^2$ & 0 & $\mathbb{Z}^2$ & 0 & $\mathbb{Z}^2$ & 0 \\
  \hline
  AI & $+1$ & 0 & 0 & 0 & 0 & 0 & $(2\mathbb{Z})^2$ & 0 & $\mathbb{Z}_2^2$ & $\mathbb{Z}_2^2$ & $\mathbb{Z}^2$
  \\
  BDI & $+1$ & $+1$ & 1 & $\mathbb{Z}^2$ & 0 & 0 & 0 & $(2\mathbb{Z})^2$ & 0 & $\mathbb{Z}_2^2$ & $\mathbb{Z}_2^2$ 
  \\
  D & 0 & $+1$ & 0 & $\mathbb{Z}_2^2$ & $\mathbb{Z}^2$ & 0 & 0 & 0 & $(2\mathbb{Z})^2$ & 0 & $\mathbb{Z}_2^2$
  \\
  DIII & $-1$ & $+1$ & 1 & $\mathbb{Z}_2^2$ & $\mathbb{Z}_2^2$ & $\mathbb{Z}^2$ & 0 & 0 & 0 & $(2\mathbb{Z})^2$ & 0 
  \\
  AII & $-1$ & 0 & 0 & 0 & $\mathbb{Z}_2^2$ & $\mathbb{Z}_2^2$ & $\mathbb{Z}^2$ & 0 & 0 & 0 & $(2\mathbb{Z})^2$ 
  \\
  CII & $-1$ & $-1$ & 1 & $(2\mathbb{Z})^2$ & 0 & $\mathbb{Z}_2^2$ & $\mathbb{Z}_2^2$ & $\mathbb{Z}^2$ & 0 & 0 & 0
  \\
  C & 0 & $-1$ & 0 & 0 & $(2\mathbb{Z})^2$ & 0 & $\mathbb{Z}_2^2$ & $\mathbb{Z}_2^2$ & $\mathbb{Z}^2$ & 0 & 0 
  \\
  CI & $+1$ & $-1$ & 1 & 0 & 0 & $(2\mathbb{Z})^2$ & 0 & $\mathbb{Z}_2^2$ & $\mathbb{Z}_2^2$ & $\mathbb{Z}^2$ & 0 
  \\
  \hline\hline
\end{tabular}
\label{tb:twogaps}
\end{table*}

The gapless boundary states in quantum walks have two different topological origins. 
The first one is the bulk topology of  the effective Hamiltonian $H_{\text{QW}}(\bm{k})$.
The two gaps $\epsilon=0$ and $\pi$ separate bulk bands of 
$H_{\text{QW}}(\bm{k})$ into two, from which one can define "occupied" and "empty" bands like conventional insulators. 
Therefore, in a manner similar to usual topological insulators, $H_{\rm QW}({\bm k})$
can be topological, which gives gapless boundary states at $\epsilon=0$.
One of the doubled topological numbers $\mathbb{Z}^2$, $\mathbb{Z}_2^2$ and $(2\mathbb{Z})^2$ in Table \ref{tb:twogaps} corresponds to the boundary states determined from the bulk topological invariant of $H_{\text{QW}}(\bm{k})$.

Another origin is the main subject of this paper, {\it i.e.}, the extrinsic topology of boundary operators.
Below, we discuss possible lattice terminations of $d$-dimensional quantum walks, which is given by $(d-1)$-dimensional unitary operators $U_{\text{BDQW}}(\bm{k}_\parallel)$. 

For this purpose, we employ a theory on gapless states of unitary operators \cite{BS20}.
For a boundary unitary operator $U_{\text{BDQW}}(\bm{k}_\parallel)$ in a symmetry class specified by the presence or absence of AZ symmetries,
we have gapless states when the unitary operator has a non-trivial topological number.
The extended Nielsen-Ninomiya theorem \cite{BS20} summarizes
the explicit  relation between gapless states and the topological number for the $(d-1)$-dimensional unitary operator $U_{\text{BDQW}}(\bm{k}_\parallel)$:
\begin{align}
\label{eq:NN}
\sum_{\epsilon_\alpha=0}\nu_{\alpha}^0 =(-1)^{d} \sum_{\epsilon_\alpha=\pi}\nu^\pi_{\alpha}=n,
\end{align}
where $\nu_\alpha^{0,\pi}$ is the $\alpha$-th topological charge of gapless states at $\epsilon=0,\pi$, and
$n$ is the topological number of $U_{\text{BDQW}}(\bm{k}_\parallel)$. 
Equation (\ref{eq:NN1d}) is an example of the extended Nielsen-Ninomiya theorem for class A with $d=2$, and 
we will give 
the explicit forms of $\nu_\alpha^{0,\pi}$ and $n$ for $d=1$ in Sec. \ref{sec:example}.
In general, we can define $n$ as follows:
We introduce the doubled Hamiltonian ${\cal H}_U(\bm{k}_\parallel)$ \cite{Higashikawa19}:
\begin{align}
    {\cal H}_U(\bm{k}_\parallel)=\pmat{ 0 & U_{\text{BDQW}}(\bm{k}_\parallel) \\ U_{\text{BDQW}}^\dag(\bm{k}_\parallel) & 0},
\end{align}
which is Hermitian, and has eigenvalues $\pm 1$ due to ${\cal H}_U(\bm{k}_\parallel)^2=\hat{1}$. It also obeys CS,
\begin{align}
    \label{eq:properCS_HU}
    \Sigma_z {\cal H}_U(\bm{k}_\parallel) \Sigma_z = - {\cal H}_U(\bm{k}_\parallel), \ \Sigma_z = \pmat{\hat{1} & 0 \\ 0 & -\hat{1}}.
\end{align}
The AZ symmetries of $U_{\text{BDQW}}(\bm{k}_\parallel)$, which should be compatible with those for the bulk operator $U_{\rm QW}({\bm k})$ in Eqs.~(\ref{eq:TRSQW})-(\ref{eq:CSQW}), lead to
\begin{align}
\label{eq:TRS_HU}
    \tilde{T} {\cal H}_U(\bm{k}_\parallel) \tilde{T}^{-1} &= {\cal H}_U(-\bm{k}_\parallel), \ \tilde{T}=\pmat{0 & T \\ T & 0},
    \\
\label{eq:PHS_HU}
    \tilde{C} {\cal H}_U(\bm{k}_\parallel) \tilde{C}^{-1} &= {\cal H}_U(-\bm{k}_\parallel), \ \tilde{C}=\pmat{C & 0 \\ 0 & C},
    \\
\label{eq:CS_HU}
    \tilde{\Gamma} {\cal H}_U(\bm{k}_\parallel) \tilde{\Gamma}^{-1} &=  {\cal H}_U(\bm{k}_\parallel), \ \tilde{\Gamma}=\pmat{0 & \Gamma \\ \Gamma & 0}.
\end{align}
Therefore, ${\cal H}_U(\bm{k}_\parallel)$ can be regarded as a topological insulator or superconductor with symmetries in Eq.~(\ref{eq:properCS_HU}) and Eqs.~(\ref{eq:TRS_HU})-(\ref{eq:CS_HU}).
As shown in Appendix \ref{sec:classifyU},
by using the standard Clifford algebra extension method,  \cite{Kitaev09,Morimoto13}, we can perform the topological classification of ${\cal H}_U(\bm{k}_\parallel)$, which also provides a topological classification of the boundary operator $U({\bm k}_\parallel)$.
The resultant classification of $U_{\text{BDQW}}(\bm{k}_\parallel)$ in $(d-1)$-dimensions coincides with the classification of conventional topological insulators and superconductors in $d$-dimensions [Table \ref{tb:extrinsic}].
The topological number of ${\cal H}_U({\bm k}_\parallel)$ gives 
the topological number $n$ in Eq.~(\ref{eq:NN}).

\begin{table*}[t]
\centering
\caption{Periodic table of extrinsic topological phases in $(d-1)$-dimensional boundaries of $d$-dimensional quantum walks. 
This table shows the presence or absence of the $\mathbb{Z}$ or $\mathbb{Z}_2$ topological invariant $n$ for $(d-1)$-dimensional boundary unitary operators.
The table has the periodicity $d=d+2$ for class A and AII, and has the periodicity $d=d+8$ for other symmetry classes.
}
\begin{tabular}{cccccccccccc}
  \hline\hline
  ~AZ class~~~~ & ~~$T$~~ & ~~$C$~~ & ~~$\Gamma$~~~ & ~$d=1$~ & ~~~2~~~ & ~~~3~~~ & ~~~4~~~ & ~~~5~~~ & ~~~6~~~ & ~~~7~~~ & ~~~8~~~ \\
  \hline
  A & 0 & 0 & 0 & 0 & $\mathbb{Z}$ & 0 & $\mathbb{Z}$ & 0 & $\mathbb{Z}$ & 0 & $\mathbb{Z}$ \\
  AIII & 0 & 0 & 1 & $\mathbb{Z}$ & 0 & $\mathbb{Z}$ & 0 & $\mathbb{Z}$ & 0 & $\mathbb{Z}$ & 0 \\
  \hline
  AI & $+1$ & 0 & 0 & 0 & 0 & 0 & $2\mathbb{Z}$ & 0 & $\mathbb{Z}_2$ & $\mathbb{Z}_2$ & $\mathbb{Z}$
  \\
  BDI & $+1$ & $+1$ & 1 & $\mathbb{Z}$ & 0 & 0 & 0 & $2\mathbb{Z}$ & 0 & $\mathbb{Z}_2$ & $\mathbb{Z}_2$ 
  \\
  D & 0 & $+1$ & 0 & $\mathbb{Z}_2$ & $\mathbb{Z}$ & 0 & 0 & 0 & $2\mathbb{Z}$ & 0 & $\mathbb{Z}_2$
  \\
  DIII & $-1$ & $+1$ & 1 & $\mathbb{Z}_2$ & $\mathbb{Z}_2$ & $\mathbb{Z}$ & 0 & 0 & 0 & $2\mathbb{Z}$ & 0 
  \\
  AII & $-1$ & 0 & 0 & 0 & $\mathbb{Z}_2$ & $\mathbb{Z}_2$ & $\mathbb{Z}$ & 0 & 0 & 0 & $2\mathbb{Z}$ 
  \\
  CII & $-1$ & $-1$ & 1 & $2\mathbb{Z}$ & 0 & $\mathbb{Z}_2$ & $\mathbb{Z}_2$ & $\mathbb{Z}$ & 0 & 0 & 0
  \\
  C & 0 & $-1$ & 0 & 0 & $2\mathbb{Z}$ & 0 & $\mathbb{Z}_2$ & $\mathbb{Z}_2$ & $\mathbb{Z}$ & 0 & 0 
  \\
  CI & $+1$ & $-1$ & 1 & 0 & 0 & $2\mathbb{Z}$ & 0 & $\mathbb{Z}_2$ & $\mathbb{Z}_2$ & $\mathbb{Z}$ & 0 
  \\
  \hline\hline
\end{tabular}
\label{tb:extrinsic}
\end{table*}

Attaching the boundary unitary operator with nonzero $n$ to a boundary of quantum walks, we can change the number of gapless modes there, in accordance with the extended Nielsen-Ninomiya theorem in Eq.~(\ref{eq:NN}):
The number of boundary states changes as
\begin{align}
    \sum_{\epsilon_\alpha=0}\nu_{\alpha}^0 
    \to & \sum_{\epsilon_\alpha=0}\nu_{\alpha}^0 + n,
    \\
    \sum_{\epsilon_\alpha=\pi}\nu_{\alpha}^\pi 
    \to & \sum_{\epsilon_\alpha=\pi}\nu_{\alpha}^\pi + (-1)^{d} n.
    \label{eq:QW_bulk_boundary}
\end{align}
On the other hand, for even (odd) $d$, the difference (summation) of the boundary states between $\epsilon=0$ and $\epsilon=\pi$ does not change, so it is intrinsically determined by the bulk topological number $n_{\text{bulk}}$ of $H_{\rm QW}({\bm k})$.
\begin{align}
    \sum_{\epsilon_\alpha=0}\nu_{\alpha}^0 - (-1)^d \sum_{\epsilon_\alpha=\pi}\nu_{\alpha}^\pi = n_{\text{bulk}}.
    \label{eq:nbulk}
\end{align}
Thus, the bulk-boundary correspondence partially holds in quantum walks.

For classes A, AI and AII, we can choose the energy gaps arbitrarily, because there is no symmetry constraint in the quasi-energy.
If there are $l$ energy gaps $\epsilon=\mu_i \ (i=1,\ldots,l)$, the topological classification of the boundary states in Table \ref{tb:twogaps} changes as $\mathbb{Z}^2\to\mathbb{Z}^l$, $\mathbb{Z}_2^2\to\mathbb{Z}_2^l$,
$(2\mathbb{Z})^2\to (2\mathbb{Z})^l$.
In these cases, the extended Nielsen-Ninomiya theorem in Eq.~(\ref{eq:NN}) takes the form of 
\begin{align}
    \sum_{\epsilon_\alpha=\mu} \nu^\mu_\alpha = n,
\end{align}
where $\nu^\mu_\alpha$ is the topological charge of
$\alpha$-th gapless state at $\epsilon=\mu$ \cite{BS20}.
We note that this formula is compatible with Eq.~(\ref{eq:NN}) with odd $d$ since in class A, AI and AII, the topological numbers $n$ of $U_{\text{BDQW}}(\bm{k}_\parallel)$ is always $0$ or $\mathbb{Z}_2$ for odd $d$.

\section{Extrinsic boundary states of quantum walks in 1D}
\label{sec:example}
Topological phases of quantum walks have been studied mainly in 1D \cite{Kitagawa10,Obuse11,Kitagawa12,Gross12,Asboth12,Asboth13,Tarasinski14,Barkhofen17,Cedzich16,Cardano16,Cedzich18}.
In this section, we study extrinsic topological phases of quantum walks in 1D, {\it i.e.} zero-dimensional extrinsic boundary states.
According to the periodic table in Table \ref{tb:extrinsic}, nontrivial extrinsic topological phases appear for classes AIII, BDI, D, DIII and CII in 1D.
We identify the topological numbers $n$ and $\nu^{0,\pi}$ for these classes, and show that they obey the extended Nielsen-Ninomiya theorem in Eq.~(\ref{eq:NN}). 

We have also studied 2D and 3D cases in Appendix.~\ref{sec:example_higher}.

\subsection{class AIII}
\label{sec:AIII}
A boundary quantum walk operator $U_{\rm BDQW}$ in a class AIII quantum walk in 1D
is a unitary matrix obeying
\begin{align}
\Gamma U_{\text{BDQW}}^{\dagger}\Gamma^{-1}=U_{\text{BDQW}},\quad
\Gamma^2=1, 
\label{eq:AIII0d}
\end{align}
with a unitary matrix $\Gamma$.
This relation implies that $U_{\text{BDQW}}\Gamma$ is Hermitian, and 
it has nonzero real eigenvalues because of ${\rm det}[U_{\rm BDQW}\Gamma]\neq 0$.
Therefore, we can define the $\mathbb{Z}$ topological number $n$ in Eq.~(\ref{eq:NN}) as follows,
\begin{align}
n=\frac{1}{2}\left[N_+(U_{\text{BDQW}}\Gamma)-N_-(U_{\text{BDQW}}\Gamma)\right], 
\label{eq:nFAIII0d}
\end{align}
where $N_{+}(U_{\text{BDQW}}\Gamma)$ ($N_{-}(U_{\text{BDQW}}\Gamma)$) is the number of positive (negative) eigenvalues of $U_{\text{BDQW}}\Gamma$.
We can also introduce the topological charge $\nu^{0,\pi}$ of gapless modes at $\epsilon=0,\pi$
in the following manner.
For a gapless mode $|u_{0}\rangle$ at $\epsilon=0$, 
we have 
\begin{align}
U_{\rm BDQW}|u_{0}\rangle=|u_{0}\rangle,
\quad
U_{\rm BDQW}^\dagger|u_{0}\rangle=|u_{0}\rangle,
\end{align}
and CS implies
\begin{align}
U_{\rm BDQW}\Gamma|u_{0}\rangle=\Gamma|u_{0}\rangle. \end{align}
Thus, by taking a linear combination of $|u_0\rangle$ and $\Gamma|u_0\rangle$,  the gapless mode can be an eigenstate of $\Gamma$,
\begin{align}
\Gamma \ket{u_{0}}=\pm \ket{u_{0}}. 
\end{align}
Then, the eigenvalue of $\Gamma$ defines the topological number $\nu^0$ of $|u_0\rangle$. In a similar manner, we can also define $\nu^\pi$ for a gapless mode $|u_\pi\rangle$ at $\epsilon=\pi$. In summary, the topological numbers $\nu^{0,\pi}$ are written as 
\begin{align}
    \nu^{0,\pi} = \bra{u_{0,\pi}}\Gamma \ket{u_{0,\pi}},
    \label{eq:chargeCS1d}
\end{align}
with the normalization $\langle u_{0,\pi}|u_{0,\pi}\rangle=1$. 

When $n$ is non-zero, we have gapless modes according to the extended Nielsen-Ninomiya theorem in Eq.~(\ref{eq:NN}).
To check the theorem, we consider a general $2\times 2$ unitary matrix in class AIII 
\footnote{
We note that it is possible to construct other forms of $U_{\text{BDQW}}$ for larger matrix sizes.
},
\begin{align}
U_{\text{BDQW}}=a_0\sigma_0+i a_1\sigma_1+i a_2\sigma_2+a_3\sigma_3, \quad \Gamma=\sigma_3, 
\label{eq:AIII2by2}
\end{align}
where $a_\mu$ are real parameters. 
From the unitarity condition $U_{\text{BDQW}} U_{\text{BDQW}}^\dag = 1$, we obtain three possible cases:
\begin{align}
    \left\{\begin{array}{l}
    (\text{i}) \ a_3=0, \ a_0^2+a_1^2+a_2^2=1, \\
    (\text{ii}) \ a_3= 1, \\
    (\text{iii}) \ a_3=-1.
    \end{array}\right.
\end{align}
The eigenvalues of $U_{\text{BDQW}}$ in each case are given by
\begin{align}
\left\{\begin{array}{l}
(\text{i}) \ \lambda_{\pm} = a_0 \pm i \sqrt{a_1^2+a_2^2},
\\
(\text{ii}),(\text{iii}) \ \lambda_\pm = \pm 1.
\end{array}\right.
\end{align}
The corresponding eigenstates are
\begin{align}
\left\{\begin{array}{l}
(\text{i}) \ \ket{u_\pm}= \frac{1}{\sqrt{2(a_1^2+a_2^2)}}\pmat{
\pm i \sqrt{a_1^2+a_2^2} \\
ia_1-a_2
},
\\
(\text{ii}) \ \ket{u_+}=(1,0)^\text{T}, \ 
\ket{u_-}=(0,1)^\text{T},
\\
(\text{iii}) \ \ket{u_+}=(0,1)^\text{T}, \ 
\ket{u_-}=(1,0)^\text{T}.
\end{array}\right.
\end{align}
which satisfy
\begin{align}
    \nu_\pm = \bra{u_\pm}\Gamma \ket{u_\pm} = \left\{\begin{array}{ll}
    0 & \text{for} \ (\text{i}), \\
    \pm 1 & \text{for} \ (\text{ii}), \\
    \mp 1 & \text{for} \ (\text{iii}),
    \end{array}\right.
\end{align}
Therefore, we obtain
\begin{align}
\sum \nu^0 =-\sum \nu^\pi=
\left\{\begin{array}{ll}
0 & \text{for} \ (\text{i}), \\
+1 & \text{for} \ (\text{ii}), \\
-1 & \text{for} \ (\text{iii}).
\end{array}\right.
\label{eq:AIIInu}
\end{align}

Now we compare this result with the topological number $n$. The Hermitian matrix $U_{\text{BDQW}} \Gamma$ is given by
\begin{align}
    U_{\text{BDQW}} \Gamma = 
    \left\{\begin{array}{ll}
    a_0 \sigma_3 + a_1 \sigma_2 - a_2 \sigma_1 & \text{for} \ (\text{i}), \\
    +\sigma_0 & \text{for} \ (\text{ii}), \\
    -\sigma_0 & \text{for} \ (\text{iii}), \\
    \end{array}\right.
\end{align}
of which eigenvalues are
\begin{align}
    \mathcal{E} =
    \left\{\begin{array}{ll}
    \pm 1 & \text{for} \ (\text{i}), \\
    +1 & \text{for} \ (\text{ii}), \\
    -1 & \text{for} \ (\text{iii}).
    \end{array}\right.
\end{align}
Thus, $n$ in Eq.~(\ref{eq:nFAIII0d}) is evaluated as
\begin{align}
    n = \left\{\begin{array}{ll}
    0 & \text{for} \ (\text{i}), \\
    +1 & \text{for} \ (\text{ii}), \\
    -1 & \text{for} \ (\text{iii}).
    \end{array}\right.
    \label{eq:AIIInF}
\end{align}
Equations.~(\ref{eq:AIIInu}) and (\ref{eq:AIIInF}) reproduce the extended Nielsen-Ninomiya theorem in Eq.~(\ref{eq:NN}).

\subsection{class BDI}
In class BDI, a boundary unitary operator  obeys TRS and PHS,
\begin{align}
T U_{\text{BDQW}} T^{-1}=U_{\text{BDQW}}^\dag,
\nonumber\\
C U_{\text{BDQW}} C^{-1}=U_{\text{BDQW}}, 
\end{align}
where $T$ and $C$ are anti-unitary operators with $CT=TC$ and $T^2=C^2=1$. 
By combining  TRS with PHS,  the boundary operator also has CS,
\begin{align}
\Gamma U_{\text{BDQW}}^{\dagger}\Gamma^{-1}=U_{\text{BDQW}}, \quad \Gamma=TC. 
\end{align}  
Using CS, 
the topological number $n$ of $U_{\rm BDQW}$ and the topological charge $\nu^{0,\pi}$ of gapless modes at $\epsilon=0,\pi$ are defined in the same manner as in class AIII.

The theorem in Eq. (\ref{eq:NN}) can be checked in a manner similar to class AIII. A general $2\times 2$ unitary matrix in class BDI is given by
\begin{align}
&U_{\rm BDQW}=a_0\sigma_0+i a_1 \sigma_1+ a_3 \sigma_3, 
\nonumber\\
&T=K, \quad C=\sigma_3 K,  
\label{eq:BDI2by2}
\end{align}
where $a_\mu$ are real parameters, and $K$ is the complex conjugation operator. From $U_{\text{BDQW}} U_{\text{BDQW}}^\dag=1$, we obtain three possible cases:
\begin{align}
\left\{\begin{array}{l}
(\text{i}) \ a_3=0, \ a_0^2+a_1^2=1, \\
(\text{ii}) \ a_3=1, \\
(\text{iii}) \ a_3=-1.
\end{array}\right.
\end{align}
Then, $U_{\text{BDQW}}$ in Eq.~(\ref{eq:BDI2by2}) obeys 
the theorem since 
it is a special case of Eq.~(\ref{eq:AIII2by2}) with $a_2=0$.

\subsection{class D}

For class D, a boundary operator $U_{\text{BDQW}}$ satisfies
\begin{align}
C U_{\text{BDQW}} C^{-1}=U_{\text{BDQW}}, \quad C^2=1, 
\label{seq:0dD}
\end{align}
with an anti-unitary operator $C$.
This relation implies that ${\rm det}(U_{\text{BDQW}})$ is real, and thus it takes $\pm 1$. 
Therefore, we can define the $\mathbb{Z}_2$ topological invariant $n$ of $U_{\rm BDQW}$ by
\begin{align}
(-1)^{n}={\rm det}(U_{\text{BDQW}}).
\label{eq:nFD0d}
\end{align}
On the other hand, the presence or absence of a gapless state at $\epsilon=0,\pi$ defines the
$\mathbb{Z}_2$ invariant $\nu^{0,\pi}$ for the gapless state. Note that an even number of gapless states trivializes $\nu^{0,\pi}$.

A general $2\times 2$ unitary matrix in class D is given by
\begin{align}
U_{\text{BDQW}}=a_0\sigma_0+a_1\sigma_1+ia_2\sigma_2+a_3\sigma_3, \quad C=K,
\label{eq:D2by2}
\end{align}
Then, the unitarity of $U_{\text{BDQW}}$ leads to two possible cases:
\begin{align}
    \left\{\begin{array}{l}
    (\text{i}) \ a_0=a_2=0, \ a_1^2+a_3^2=1, \\
    (\text{ii}) \ a_1=a_3=0, \ a_0^2+a_2^2=1,
    \end{array}\right.
\end{align}
where the eigenvalues of $U_{\text{BDQW}}$ are
\begin{align}
    \left\{\begin{array}{l}
    (\text{i}) \ \lambda_{\pm}=\pm 1, \\
    (\text{ii}) \ \lambda_{\pm}=a_0 \pm i a_2.
    \end{array}\right.
\end{align}
Thus, only for (i), a single gapless mode exists at $\epsilon=0,\pi$. 
From this, we obtain
\begin{align}
\sum \nu^0 =-\sum \nu^\pi=
\left\{\begin{array}{ll}
1 & \text{for} \ (\text{i}), \\
0 & \text{for} \ (\text{ii}).
\end{array}\right.
\label{eq:Dnu}
\end{align}

On the other hand, a direct calculation shows
\begin{align}
    \det (U_{\text{BDQW}}) = a_0^2+a_2^2-a_1^2-a_3^2,
\end{align}
and thus, the $\mathbb{Z}_2$ invariant $n$ in Eq.~(\ref{eq:nFD0d}) is
\begin{align}
     n = \left\{\begin{array}{ll}
1 & \text{for} \ (\text{i}), \\
0 & \text{for} \ (\text{ii}).
\end{array}\right.
    \label{eq:DnF}
\end{align}
The topological numbers in Eqs.~(\ref{eq:Dnu}) and (\ref{eq:DnF}) satisfy 
the extended Nielsen-Ninomiya theorem in Eq.~(\ref{eq:NN}).

\subsection{class DIII}

For a boundary unitary operator $U_{\text{BDQW}}$ in class DIII, 
we have
\begin{align}
T U_{\text{BDQW}} T^{-1}=U_{\text{BDQW}}^\dag, 
\nonumber\\
C U_{\text{BDQW}} C^{-1}=U_{\text{BDQW}},
\end{align}
where $T$ and $C$ are anti-unitary operators with $T^2=-1$, $C^2=1$ and $CT=TC$. 
For convenience, we decompose $T$ and $C$ into the  unitary parts ${\cal T}$ and ${\cal C}$ and the complex conjugation operator $K$:
\begin{align}
    T=\mathcal{T}K,\quad C=\mathcal{C}K.
\end{align}
Then, the matrix $U_{\text{BDQW}}\mathcal{T}$ is found to be antisymmetric, so we can introduce the Pfaffian ${\rm Pf}(U_{\text{BDQW}}\mathcal{T})$.
We can also show
\begin{align}
[{\rm Pf}(U_{\text{BDQW}}\mathcal{T})]^*={\rm det}(\mathcal{C}^*){\rm Pf}(U_{\text{BDQW}}\mathcal{T}),
\end{align} 
and thus, we have ${\rm Pf}(U_{\text{BDQW}}\mathcal{T})=\pm 1$ in the basis with ${\rm det}(\mathcal{C}^*)=1$.
Then, the sign of the Pfaffian defines the $\mathbb{Z}_2$ invariant $n$ for $U_{\rm BDQW}$:
\begin{align}
(-1)^{n}=-{\rm Pf}(U_{\text{BDQW}}\mathcal{T}).
\label{eq:nFDIII0d}
\end{align}
On the other hand, for gapless states at $\epsilon=0,\pi$,
the presence or absence of a Kramers pair of gapless states defines the $\mathbb{Z}_2$ invariant $\nu^{0,\pi}$. 
Note that any eigenstate of $U_{\rm BDQW}$ doubly degenerates due to the Kramers theorem for TRS.

To obtain a non-trivial example, we need at least a $4\times 4$ unitary matrix in this class. 
Let us consider a general $4\times 4$ unitary matrix in class DIII,
\begin{align}
U_{\text{BDQW}}=a_{00} \tau_0\sigma_0+a_{10}\tau_1\sigma_0+a_{30}\tau_3\sigma_0
\nonumber \\
+ia_{21}\tau_2\sigma_1 
+ia_{22}\tau_2\sigma_2+a_{23}\tau_2\sigma_3.
\label{eq:DIII4by4}
\end{align}
Here, $\tau_i\sigma_j$ is the tensor product of the Pauli matrices $\tau_i$ and $\sigma_j$. 
TRS and PHS are given by  $T=\tau_0\sigma_2 K$ and $C=\tau_0\sigma_1 K$.
From $U_{\text{BDQW}} U_{\text{BDQW}}^\dag = 1$, we obtain two possibilities,
\begin{align}
    \left\{\begin{array}{l}
    (\text{i}) \ a_{00}=a_{21}=a_{22}=0, \ a_{30}^2+a_{10}^2+a_{23}^2=1, \\
    (\text{ii}) \ a_{30}=a_{10}=a_{23}=0, \ a_{00}^2+a_{21}^2+a_{22}^2=1.
    \end{array}\right.
\end{align}
The eigenvalues of $U_{\rm BDQW}$ with Kramers degeneracy are given by
\begin{align}
    \left\{\begin{array}{l}
    (\text{i}) \ \lambda_{\pm}=\pm 1, \\
    (\text{ii}) \ \lambda_{\pm}=a_{00} \pm i \sqrt{a_{21}^2+a_{22}^2},
    \end{array}\right.
\end{align}
and thus, the system supports a single Kramers pair of gapless states at $\epsilon=0,\pi$ only for (i). Therefore, we obtain
\begin{align}
\sum \nu^0 =-\sum \nu^\pi=
\left\{\begin{array}{ll}
1 & \text{for} \ (\text{i}), \\
0 & \text{for} \ (\text{ii}).
\end{array}\right.
\label{eq:DIIInu}
\end{align}
On the other hand, 
$\Pf(U_{\text{QW}}\mathcal{T})$ becomes
\begin{align}
    \Pf (U_{\text{BDQW}}\mathcal{T}) = a_{30}^2+a_{10}^2+a_{23}^2 - a_{00}^2-a_{21}^2-a_{22}^2,
\end{align}
so $n$ in Eq.~(\ref{eq:nFDIII0d}) is evaluated as
\begin{align}
     n = \left\{\begin{array}{ll}
1 & \text{for} \ (\text{i}), \\
0 & \text{for} \ (\text{ii}),
\end{array}\right.
    \label{eq:DIIInF}
\end{align}
which obeys the theorem in Eq.~(\ref{eq:NN}).

\subsection{class CII}

Finally, we examine boundary unitary operators for class CII quantum walks in 1D.
The boundary unitary operator obeys 
\begin{align}
T U_{\text{BDQW}} T^{-1} = U_{\text{BDQW}}^\dag, 
\nonumber\\
C U_{\text{BDQW}} C^{-1}=U_{\text{BDQW}}, 
\label{eq:CII0d}
\end{align}
where $T$ and $C$ are anti-unitary operators with $T^2=-1$, $C^2=-1$ and $CT=TC$.
Combining TRS with PHS, we also have CS,
\begin{align}
\Gamma U_{\text{BDQW}}^{\dagger} \Gamma^{-1}=U_{\text{BDQW}}, \quad \Gamma=TC. 
\end{align}
Using CS, we can introduce the topological numbers $n$ and $\nu^{0,\pi}$ in Eqs.~(\ref{eq:nFAIII0d}) and (\ref{eq:chargeCS1d}) in the same manner as those in class AIII.
However, in contrast to class AIII, because of additional CS and TRS in Eq.~(\ref{eq:CII0d}), 
these topological numbers only take even integers.
First, the Hermitian matrix $U_{\text{BDQW}}\Gamma$ has its own TRS defined by $C$
\begin{align}
C[U_{\text{BDQW}}\Gamma]C^{-1}=U_{\text{BDQW}}\Gamma,
\end{align}
which results in the Kramers degeneracy for eigenstates of $U_{\text{BDQW}}\Gamma$.
Therefore, 
\begin{align}
n=\frac{1}{2}\left[N_+(U_{\text{BDQW}}\Gamma)-N_-(U_{\text{BDQW}}\Gamma)\right], \label{eq:nF_0dCII}
\end{align}
in Eq.~(\ref{eq:nFAIII0d}) becomes a $2\mathbb{Z}$ topological invariant.
Furthermore,  gapless modes at $\epsilon=0,\pi$ also form  Kramers pairs due to the original TRS by $T$. The Kramers pair has a common eigenvalue of $\Gamma$ since $T$ commutes with $\Gamma$, so
the total charges $\sum \nu^{0,\pi}$ of gapless modes with $\nu^{0,\pi}$ defined by Eq.~(\ref{eq:chargeCS1d}) also become even integers.

A general $4\times 4$ unitary matrix in class CII is given by
\begin{align}
U_{\text{BDQW}}&=a_{00}\tau_0\sigma_0+ia_{10}\tau_1\sigma_0+a_{30}\tau_3\sigma_0
\nonumber \\
&+ia_{21}\tau_2\sigma_1+ia_{22}\tau_2\sigma_2+ia_{23}\tau_2\sigma_3, 
\label{eq:CII4by4}
\end{align}
with $T=\tau_0\sigma_2 K$ and $C=\tau_3\sigma_2 K$.
The unitarity condition $U_{\text{BDQW}} U_{\text{BDQW}}^\dag = 1$ leads to the following three possibilities:
\begin{align}
    \left\{\begin{array}{l}
    (\text{i}) \ a_{30}=0, \ a_{00}^2+a_{10}^2+a_{21}^2+a_{22}^2+a_{23}^2=1,\\
    (\text{ii}) \ a_{30}= 1, \ a_{00}=a_{10}=a_{21}=a_{22}=a_{23}=0, \\
    (\text{iii}) \ a_{30}=- 1, \ a_{00}=a_{10}=a_{21}=a_{22}=a_{23}=0.
    \end{array}\right.
\end{align}
The eigenvalues of $U_{\rm BDQW}$ with Kramers degeneracy are
\begin{align}
    \left\{\begin{array}{l}
    (\text{i}) \ \lambda_{\pm}=a_{00} \pm i \sqrt{a_{10}^2+a_{21}^2+a_{22}^2+a_{23}^2}, \\
    (\text{ii}) \ \lambda_{\pm}=\pm 1, \\
    (\text{iii}) \ \lambda_{\pm}=\pm 1.
    \end{array}\right.
\end{align}
and thus the boundary operator supports gapless states at $\epsilon=0,\pi$ for (ii) and (iii).
We can show that each Kramers pair satisfies
\begin{align}
    \nu_\pm = \bra{u_\pm}\Gamma \ket{u_\pm} = \left\{\begin{array}{ll}
    0 & \text{for} \ (\text{i}), \\
    \mp 1 & \text{for} \ (\text{ii}), \\
    \pm 1 & \text{for} \ (\text{iii}),
    \end{array}\right.
\end{align}
and thus, we obtain
\begin{align}
\sum \nu^0 =-\sum \nu^\pi=
\left\{\begin{array}{ll}
0 & \text{for} \ (\text{i}), \\
-2 & \text{for} \ (\text{ii}), \\
+2 & \text{for} \ (\text{iii}).
\end{array}\right.
\label{eq:CIInu}
\end{align}
On the other hand, the Hermitian matrix $U_{\text{BDQW}} \Gamma$ is given by
\begin{align}
    U_{\text{BDQW}} \Gamma = &a_{00}\tau_3\sigma_0 +a_{10}\tau_2\sigma_0 +a_{30}\tau_0\sigma_0 
    \nonumber \\
    &- a_{21}\tau_1\sigma_1 -a_{22}\tau_1\sigma_2 -a_{23}\tau_1\sigma_3
\end{align}
of which eigenvalues are
\begin{align}
    \mathcal{E} =-a_{30} \pm \sqrt{a_{00}^2+a_{10}^2+a_{21}^2+a_{22}^2+a_{23}^2}
\end{align}
Thus, by taking into account the Kramers degeneracy, $n$ in Eq.~(\ref{eq:nFAIII0d}) is evaluated as
\begin{align}
    n = \left\{\begin{array}{ll}
    0 & \text{for} \ (\text{i}), \\
    -2 & \text{for} \ (\text{ii}), \\
    +2 & \text{for} \ (\text{iii}).
    \end{array}\right.
    \label{eq:CIInF}
\end{align}
The results in Eqs.~(\ref{eq:CIInu}) and (\ref{eq:CIInF}) satisfy the extended Nielsen-Ninomiya theorem in Eq.~(\ref{eq:NN}).

\section{Classification of Floquet systems v.s. quantum walks}
\label{sec:Floquet_vs_qw}
In this section, we compare the topological classification of quantum walks and that of Floquet systems.
Let us start with a brief review of the topological classification of Floquet topological insulators and superconductors \cite{Roy17}.
We consider a general time periodic Hamiltonian $H(\bm{k},t+T)=H(\bm{k},t)$
and the time-evolution operator
\begin{align}
    U(\bm{k},t) = \mathcal{T} \exp \left[ -i \int_{0}^
    {t} d t H(\bm{k},t)
    \right].
    \label{eq:time_evolution}
\end{align}
From the one-cycle time evolution $U_F(\bm{k})=U(\bm{k}, T)$, we define the effective Hamiltonian by $U_F(\bm{k})=e^{-iH_F(\bm{k})T}$.
The fundamental symmetries, TRS, PHS and CS in Floquet systems are defined in the microscopic Hamiltonian:
\begin{align}
T H(\bm{k},t) T^{-1} &= H(-\bm{k},-t), \\
C H(\bm{k},t) C^{-1} &= -H(-\bm{k},t), \\
\label{eq:CSFloquet}
\Gamma H(\bm{k},t) \Gamma^{-1} &= -H(\bm{k},-t).
\end{align}
Here, $T$ and $C$ are anti-unitary operators with $T^2=\pm 1$ and $C^2=\pm 1$, and $\Gamma$ is a unitary operator with $\Gamma^2=1$. 
These symmetries lead to TRS, PHS and CS for the effective Hamiltonian $H_F(\bm{k})$:
\begin{align}
T H_{F}(\bm{k}) T^{-1} &= H_{F}(-\bm{k}), \\
C H_{F}(\bm{k}) C^{-1} &= -H_{F}(-\bm{k}), \\
\Gamma H_{F}(\bm{k}) \Gamma^{-1} &= -H_{F}(\bm{k}).
\end{align}
For convenience, we also rewrite the symmetries as those for time-evolution operator $U(\bm{k},t)$:
\begin{align}
\label{eq:TRSconti}
T U(\bm{k},t) T^{-1} = U(-\bm{k},-t), \\
\label{eq:PHSconti}
C U(\bm{k},t) C^{-1} = U(-\bm{k},t), \\
\label{eq:CSconti}
\Gamma U(\bm{k},t) \Gamma^{-1} = U(\bm{k},-t).
\end{align}

Instead of the  microscopic Hamiltonian $H(\bm{k},t)$,  we classify the time-evolution operator $U(\bm{k},t)$ because it has the same information as $H({\bm k},t)$.
We decompose $U(\bm{k},t)$ into two parts:
\begin{align}
    C(\bm{k},t)=e^{-iH_F(\bm{k})t},\quad
    L(\bm{k},t)=U(\bm{k},t)C(\bm{k},t)^{-1},
    \label{eq:decomposition}
\end{align}
where $L(\bm{k},t)$ is periodic in $t$.
We call $C(\bm{k},t)$ and $L(\bm{k},t)$ as the constant time evolution and the loop unitary, respectively.
When $H_F(\bm{k})$ is gapped at $\epsilon=0,\ \pi/T$ and has the branch cut at $\epsilon=\pi/T$,
this decomposition is shown to be unique up to homotopy equivalence \cite{Roy17}.
Thus the topological classification of $U(\bm{k},t)$ reduces to those of the constant time evolution $C(\bm{k},t)$ and the loop unitary $L(\bm{k},t)$.

The topological classification of $C(\bm{k},t)$ is the same as that of $H_F(\bm{k})$, and thus the same as that of ordinary topological insulators and superconductors. This coincides with the classification of $H_{\text{QW}}(\bm{k})$ in the previous section.

On the other hand, the loop unitary $L({\bm k},t)$ realizes the Floquet anomalous topological phase intrinsic to dynamical systems \cite{Rudner13}.  
Interestingly, for a very different reason, the topological classification of $L(\bm{k},t)$ also coincides with that of ordinary topological insulators and superconductors as shown below:
One can classify $L({\bm k},t)$ by using the doubled Hamiltonian ${\cal H}_L(\bm{k},t)$,
\begin{align}
    {\cal H}_L(\bm{k},t)=\pmat{ 0 & L(\bm{k},t) \\ L^\dag(\bm{k},t) & 0},
\end{align}
which is Hermitian, gapped due to ${\cal H}_L(\bm{k},t)^2=\hat{1}$, periodic both in ${\bm k}$ and $t$, and obeys CS
\begin{align}
    \Sigma_z {\cal H}_L(\bm{k},t) \Sigma_z = - {\cal H}_L(\bm{k},t), \ \Sigma_z = \pmat{\hat{1} & 0 \\ 0 & -\hat{1}}.
    \label{eq:properCS_HL}
\end{align}
Since $L(\bm{k},t)$ obeys the same symmetries as $U(\bm{k},t)$ in Eqs.~(\ref{eq:TRSconti})-(\ref{eq:CSconti}), ${\cal H}_L(\bm{k},t)$ may have
\begin{align}
\label{eq:TRS_HL}
    \tilde{T} {\cal H}_L(\bm{k},t) \tilde{T}^{-1} &= {\cal H}_L(-\bm{k},-t), \ \tilde{T}=\pmat{T & 0 \\ 0 & T},
    \\
\label{eq:PHS_HL}
    \tilde{C} {\cal H}_L(\bm{k},t) \tilde{C}^{-1} &= {\cal H}_L(-\bm{k},t), \ \tilde{C}=\pmat{C & 0 \\ 0 & C},
    \\
\label{eq:CS_HL}
    \tilde{\Gamma} {\cal H}_L(\bm{k},t) \tilde{\Gamma}^{-1} &=  {\cal H}_L(\bm{k},-t), \ \tilde{\Gamma}=\pmat{\Gamma & 0 \\ 0 & \Gamma}.
\end{align}
Therefore, by regarding $t$ as a space direction and then ${\cal H}_L(\bm{k},t)$ as a $(d+1)$-dimensional topological insulator with proper symmetries defined above, we can classify $L({\bm k},t)$.
As shown in Appendix \ref{sec:classifyL},
one can perform the classification by using the 
Clifford algebra extension method \cite{Kitaev09,Morimoto13}, and 
find that the classification coincides with that of ordinary topological insulators and superconductors in $d$-dimensions.
Thus, the periodic table of $L(\bm{k},t)$ is the same as that of extrinsic topological phases in quantum walks [Table \ref{tb:extrinsic}].

Therefore, combining the classifications of $C({\bm k},t)$ and $L({\bm k},t)$, we find that the topological periodic table of $U(\bm{k},t)$ agrees with that for quantum walks in Table \ref{tb:twogaps}.
Furthermore, the bulk-boundary correspondence in Floquet systems is summarized as \cite{Roy17}
\begin{align}
    \sum_{\epsilon_\alpha=0}\nu_{\alpha}^0 = n_{\text{C}} + n_{\text{L}},
    \\
    \sum_{\epsilon_\alpha=\pi/T}\nu_{\alpha}^{\pi} = (-1)^{d} n_{\text{L}},
    \label{eq:FloquetBEC}
\end{align}
where $n_C$ is the topological invariant of $C(\bm{k},t)$ or equivalently $H_F(\bm{k})$, and $n_L$ is the topological invariant of $L(\bm{k},t)$, and
$\nu_\alpha^{0,\pi}$ is the topological charge of $\alpha$-th gapless states at $\epsilon=0$ or $\pi/T$.
These formulas correspond to Eq.~(\ref{eq:QW_bulk_boundary}) with Eq.~(\ref{eq:nbulk}) in quantum walks.

In summary, extrinsic boundary states determined by the boundary topology of quantum walks correspond to boundary states determined by $L(\bm{k},t)$ of Floquet systems. 
In other words, the extrinsic boundary states in quantum walks correspond to the Floquet anomalous boundary states \cite{Rudner13}.
Note that no well-defined loop unitary exists in quantum walks due to the absence of the microscopic Hamiltonian $H(\bm{k},t)$:
Whereas one can introduce $L({\bm k}, t)$ for a quantum walk in a specific manner \cite{Asboth15}, it is not unique. For instance, the same one-cycle time-evolution operator for the quantum walk can be obtained by the effective Hamiltonian $H({\bm k},t)=H_{\rm QW}(\bm{k})$, for which $L({\bm k},t)=\hat{1}$.
As a result, we can construct a microscopic Hamiltonian with $n_\text{L}=0$ and $n_\text{L}$ cannot be uniquely determined from time-evolution operators for any quantum walks.
This observation is also consistent with the extrinsic nature of Floquet anomalous boundary states in quantum walks.

\section{Bulk-Boundary correspondence in 1D chiral-symmetric quantum walks}
\label{sec:CSQW}
For 1D chiral symmetric quantum walks, it has been known that the bulk-boundary correspondence holds \cite{Asboth13} with a different definition of CS from ours.
In this section, we explain why the bulk topological numbers fully determine gapless boundary states in their definition of CS, and also discuss another possibility of a similar bulk-boundary correspondence in other symmetry classes.

We first review a specific realization of CS introduced in Ref.~\cite{Asboth13}.
To define CS, Asb\'{o}th and Obuse decomposed the time-evolution unitary operator of a quantum walk into two parts
\begin{align}
    U_{\text{QW}}=U_2 U_1,
\end{align}
where $U_1$ and $U_2$ may consist of multiple unitary operators.
Then, they introduced the decomposed CS as \cite{Asboth13},
\begin{align}
    \Gamma U_1 \Gamma^{-1} = U_2^\dag,
    \quad \Gamma^2=1,
    \label{eq:CSasboth}
\end{align}
with a unitary operator $\Gamma$, which leads to the original CS in Eq.~(\ref{eq:CSQW}).
Note that Floquet systems with CS in Eq.~(\ref{eq:CSFloquet}) 
naturally realize Eq.~(\ref{eq:CSasboth}) by regarding $U_1$ and $U_2$ as $U_1=U(0\to T/2)$ and $U_2=U(T/2 \to T)$.

Under this special realization of CS, we can show that the bulk-boundary correspondence holds \cite{Asboth13, Mochizuki20}. 
For this purpose, one can take the basis where $\Gamma$ and $U_1$ are given by
\begin{align}
    \Gamma=\pmat{\hat{1} & 0 \\ 0 & -\hat{1} }, \
    U_1=\pmat{a & b \\ c & d},
    \label{eq:basis}
\end{align}
and show that if the band gap at $\epsilon=0$ ($\epsilon=\pi$) is open, $b$ and $c$ ($a$ and $d$) have the well-defined 1D winding number $w_1[b]$ and $w_1[c]$ ($w_1[a]$ and $w_1[d]$) \cite{Mochizuki20}.
Then, we have the relation \cite{Asboth13},
\begin{align}
\left\{\begin{array}{l}
    \displaystyle\sum_{\epsilon_\alpha=0} \nu^0_\alpha = \frac{w_1[b]-w_1[c]}{2}, \\ 
    \displaystyle\sum_{\epsilon_\alpha=\pi} \nu^\pi_\alpha = \frac{w_1[a]-w_1[d]}{2},
    \label{eq:BECasboth}
\end{array}\right.
\end{align}
where $\nu_\alpha^{0,\pi}$ is the topological charge of boundary zero modes defined by Eq.~(\ref{eq:chargeCS1d}).
Therefore, the net topological charges of boundary modes at $\epsilon=0$ and $\pi$ are determined by the bulk topological numbers. In other words, no extrinsic boundary modes are possible in this case. 

We can easily show why extrinsic boundary modes are prohibited under CS in Eq.~(\ref{eq:CSasboth}).
From the decomposed CS, $U_{\rm QW}\Gamma$ is recast into
\begin{align}
    U_{\text{QW}}\Gamma=U_2 U_1\Gamma=U_2(\Gamma^{-1} U_2^\dag \Gamma) \Gamma = U_2\Gamma U_2^\dag,
    \label{eq:Argument_CS_trivial}
\end{align}
which takes the form of unitary transformation of $\Gamma$ by $U_2$. Therefore, any boundary operator $U_{\rm BDQW}$ decomposed into two parts $U'_2 U'_1$ by CS also has the same form,
\begin{align}
U_{\rm BDQW}\Gamma=U'_2\Gamma U'_2{}^{\dagger}. \label{eq:CS_bdry} 
\end{align}
From this, we can show that the zero-dimensional (0D) $\mathbb{Z}$ topological number defined by Eq.~(\ref{eq:nFAIII0d}) becomes zero: Because of the above relation, $U_{\rm BDQW}\Gamma$ has the same eigenvalues as $\Gamma$, so we have 
\begin{align}
    n= \frac{1}{2} [N_+(U_{\text{BDQW}}\Gamma) - N_-(U_{\text{BDQW}}\Gamma)]=0.
\end{align}
Therefore, the boundary unitary operator gives no additional gapless state in 0D boundaries. 
Thus, in 1D quantum walks with decomposed CS Eq.~(\ref{eq:CSasboth}),
the bulk topological numbers fully determine the numbers of  boundary modes at $\epsilon=0,\pi$.

Whereas the decomposed CS in Eq.~(\ref{eq:CSasboth}) prohibits the extrinsic zero modes at boundaries of 1D systems, it may allow extrinsic topological phases in other dimensions.
For instance, extrinsic zero modes of 2D boundaries are not prohibited by the decomposed CS in Eq.~(\ref{eq:CSasboth}).
The topological invariant for a 2D boundary operator $U_{\rm BDQW}(k_x,k_y)$ with CS is the Chern number of $U_{\rm BDQW}(k_x,k_y)\Gamma$.
Equation (\ref{eq:CS_bdry}) merely implies that $U'_2$ diagonalizes $U_{\rm BDQW}\Gamma$, so we can realize any  Chern number by choosing a proper $U'_2$. 
Consequently, we can add arbitrary numbers of 2D extrinsic boundary states to  three-dimensional quantum walks under the decomposed CS in Eq.~(\ref{eq:CSasboth}).

One may ask a question if there is any other symmetry class that can recover the bulk-boundary correspondence with an appropriate definition of symmetries.
The answer is yes.
We find that 1D quantum walks in class CII also have the same property.
To see this, we again decompose the time-evolution unitary operator of a quantum walk into two parts,
\begin{align}
    U_{\text{QW}}=U_2 U_1.
\end{align}
Then, we consider decomposed TRS and PHS, which give the original TRS and PHS for $U_{\rm QW}$ in Eq.~(\ref{eq:TRSQW}) and (\ref{eq:PHSQW}),
\begin{align}
T U_1(k) T^{-1} &= U_2^\dag(-k), 
\label{eq:TRS_CII}
\\ 
C U_1(k) C^{-1}&=U_1(-k), \ C U_2(k) C^{-1}=U_2(-k). 
\label{eq:PHS_CII}
\end{align}
Here $T$ and $C$ are anti-unitary operators with $CT=TC$ and $T^2=C^2=-1$.
Combining TRS with PHS, we also have decomposed CS,
\begin{align}
\Gamma U_1 \Gamma^{-1}=U_2^\dag, \quad \Gamma=TC.
\label{eq:CS_CII}
\end{align}
Using the decomposed CS, we can again take the basis in Eq.~(\ref{eq:basis}) and prove the bulk-boundary correspondence in Eq.~(\ref{eq:BECasboth}).
Note that the decomposed PHS in Eq.~(\ref{eq:PHS_CII}) leads to two-fold degeneracy similar to the Kramers doublet, 
and thus topological numbers on both sides of Eq.~(\ref{eq:BECasboth}) take only even integers.
(See Appendix \ref{sec:CII_2Zwinding}.)
In a manner similar to the above,
we can also show that 0D extrinsic boundary states are prohibited by Eq.~(\ref{eq:CS_CII}). The 0D $2\mathbb{Z}$ topological invariant in Eq.~(\ref{eq:nF_0dCII}) is always zero by the decomposed CS.

\section{Physical implementations}
\label{sec:implementation}
In this section, we present three possible physical implementations of the extrinsic topological phases of quantum walks.

\subsection{2D disordered systems with extrinsic edge states}
In this section, we examine robustness of extrinsic edge modes against disorders, which is an analog of robustness of quantum Hall edge states against impurity scatterings \cite{Halperin1982}.
We consider a single-band model with an extrinsic edge mode in quantum walks. The edge mode is robust against impurities, and shows a directed position displacement along the edge characterized by the topological winding number \cite{Thouless83,Kitagawa10_Floquet,Titum16,Nakagawa20,Cardano16,Fedorova20}.

Let us consider the following single-band tight-binding model with random onsite potentials in 2D, which is typically used for the study of the Anderson localization:
\begin{align}\label{eq:Anderson}
    & H_{A} = \sum_{x,y} J \ket{x+1,y} \bra{x,y} + J \ket{x,y+1} \bra{x,y}
    \nonumber \\
    &  + h.c. + V_{x,y} \ket{x,y} \bra{x,y},
\end{align}
where $J$ is the hopping amplitude and $V_{x,y}\in [-W,W]$ is a random potential uniformly distributed within $[-W,W]$. Here $J$ and $V_{x,y}$ are real, so the above Hamiltonian belongs to class AI and is topologically trivial. For class AI in 2D, it is known that all the eigenstates are localized for any nonzero $W$ if the system is large enough \cite{Anderson58}.
Below, we impose the periodic boundary condition in the $x$ direction, and the open boundary condition in the $y$ direction. 

\begin{figure}[t]
\centering
\includegraphics[width=85mm]{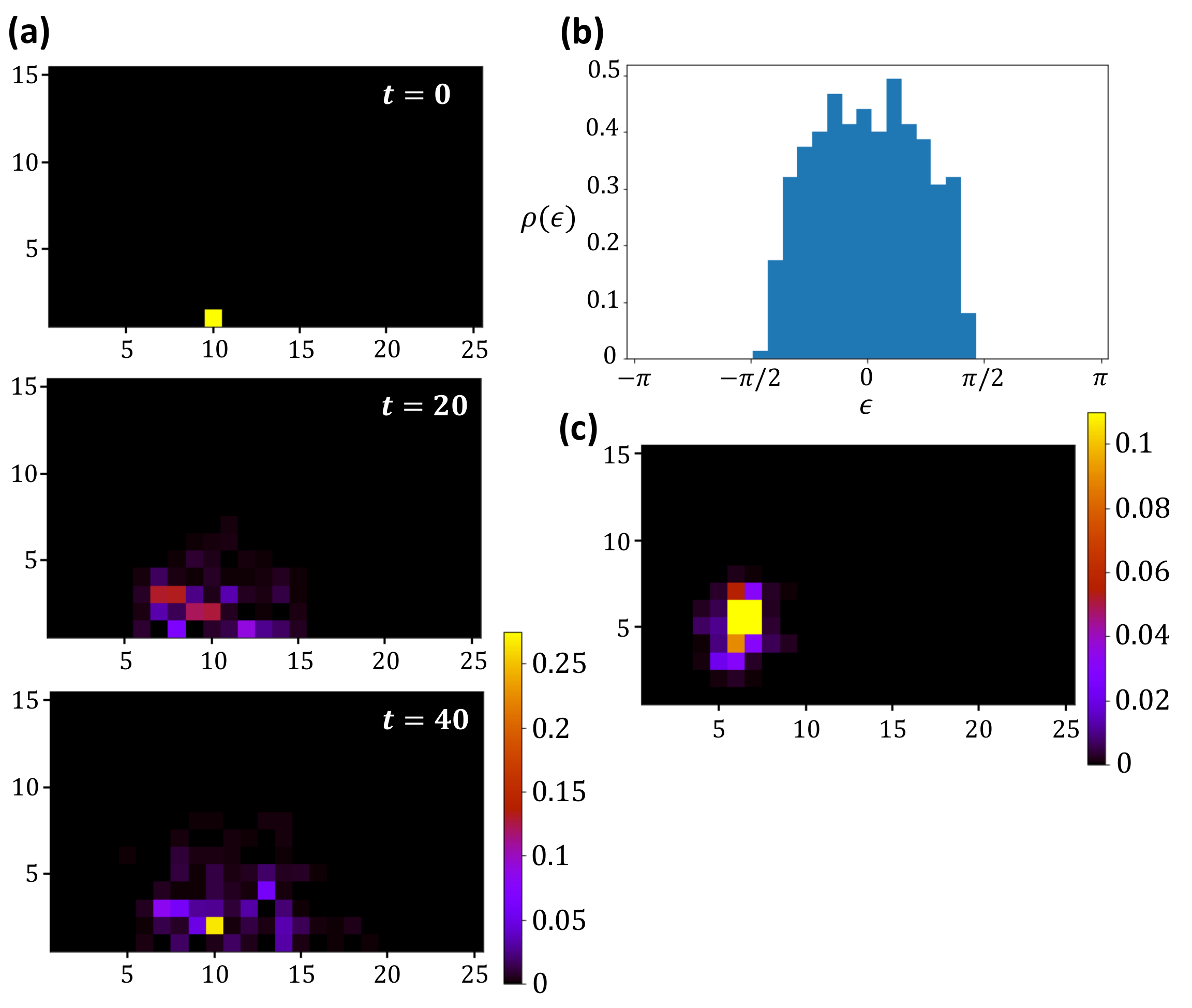} 
\caption{(a) Dynamics of a wave packet, (b) DOS and (c) a typical eigenstate of the Anderson model in Eq.~(\ref{eq:Anderson}). The parameters are $J=0.2, W=1, L_x=25$ and $L_y=15$. The initial state of the wave packet is $\ket{x=10,y=1}$. 
We consider the strongly localized regime $W \gg J$, and thus all the eigenstates are localized.
(a) We see that a wave packet starting at an edge does not diffuse into the bulk 
and shows a localization with almost the same radius as a bulk eigenstate.
(b) The width of the energy band is broader than $2J$ due to the random potential $V_{x,y}\in [-W,W]$.
(c) A typical eigenstate shows the Anderson localization.
}
	\label{fig:Anderson_model}
\end{figure}

The one-cycle time evolution by the Hamiltonian in Eq.~(\ref{eq:Anderson}) is described by 
\begin{align}
U_A=e^{-i H_A T}.
\end{align}
with $T=1$. The wavefunction at time step $t$ is derived by multiplying the state by $U_A^t$.
Figure~\ref{fig:Anderson_model} 
shows (a)
the wave packet dynamics at time step $t$ starting at an edge of the system, (b) the density of states (DOS) histogram, and (c) a typical eigenstate of $U_A$. 
The system exhibits localized behavior in the finite system, indicating the localization length is smaller than the system size.
After long time steps, the wave packet on the edge shows a localization with almost the same radius as the typical eigenstate, and does not diffuse into the bulk.

\begin{figure}[t]
\centering
\includegraphics[width=85mm]{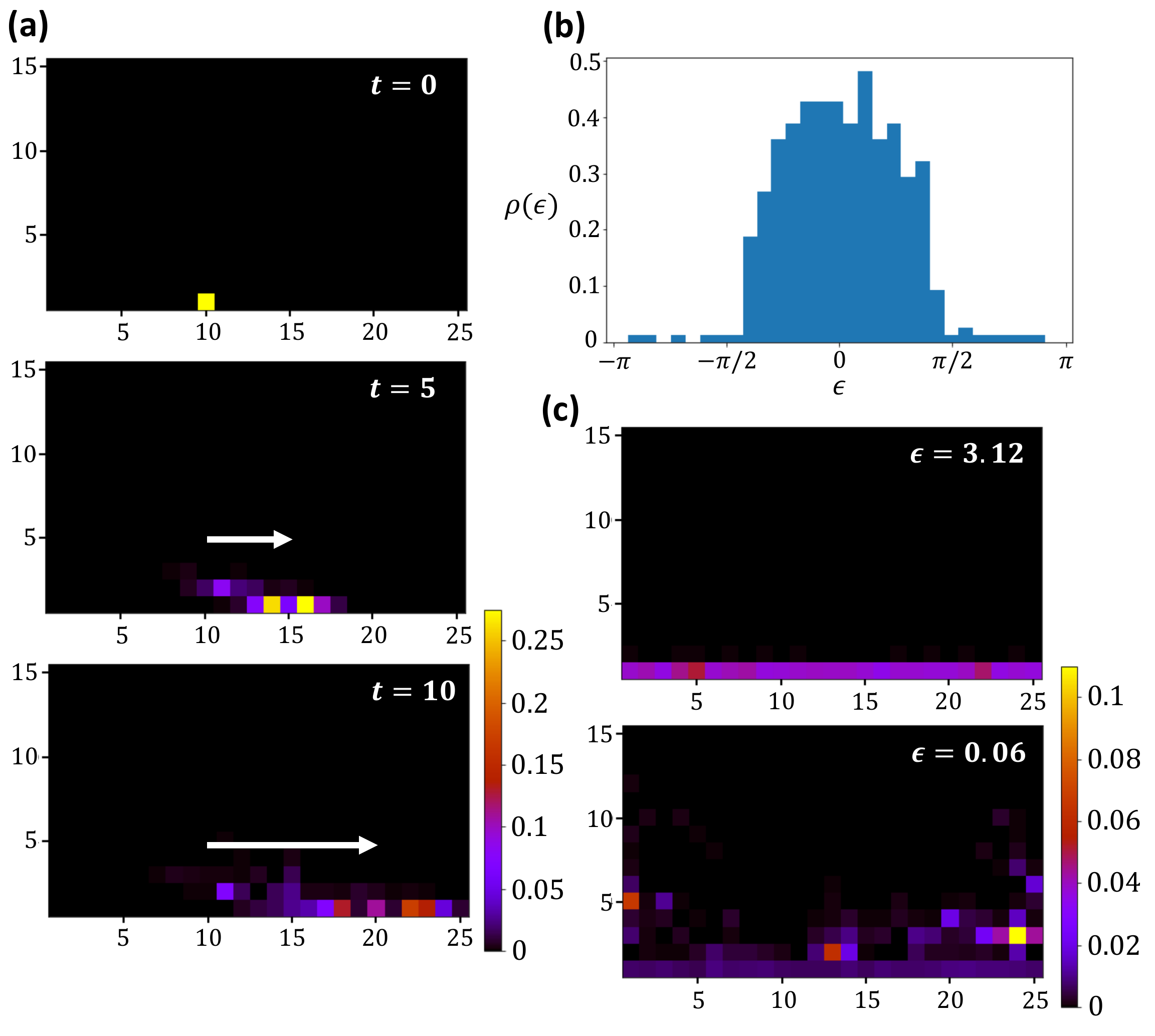} 
\caption{(a) Dynamics of a wave packet, (b) DOS and (c) delocalized eigenstates of the Anderson model with the extrinsic edge mode in Eq.~(\ref{eq:Anderson_edge}). The parameters are $J=0.2$, $W=1$, $L_x=25$, $L_y=15$ and $T=1$. 
The color scale is different between (a) and (c). The initial state of the wave packet is $\ket{x=10,y=1}$. 
Since $W \gg J$, all bulk eigenstates are localized.
(a) The wave packet propagates to the $+x$-direction.
(b) The DOS around $\epsilon=\pi$ indicates the presence of an anomalous chiral edge mode.
(c) Delocalized eigenstates with eigenenergies $\epsilon=3.12$ and $\epsilon=0.06$ correspond to the anomalous chiral edge mode. 
}
	\label{fig:Anderson_model_A}
\end{figure}

We next introduce an extrinsic chiral edge state onto the boundary at $y=1$. We multiply $U_A$ by the unitary operator $A$,
\begin{align}
    A(k_x) = U_{\text{edge}}(k_x)\otimes \ket{y=1}\bra{y=1} + \sum_{y=2}^{L_y} 1\otimes \ket{y}\bra{y}.
    \label{eq:A_Anderson}
\end{align}
where the edge unitary operator is $U_{\text{edge}}(k_x)=e^{-ik_x}$,
and consider the decorated time-evolution operator 
\begin{align}
    U_{A}'= A U_A.
\label{eq:Anderson_edge}
\end{align}
In the above sections, we have considered attaching completely decoupled boundary unitary operators. In this section, however, we consider attaching boundary unitary operators coupled with the bulk.
As discussed in Sec.~\ref{sec:introduction},  $U_{\text{edge}}(k_x)$ provides a chiral edge mode with $\epsilon=k_x$, which should be robust against disorders.

Figure \ref{fig:Anderson_model_A} shows (a)
the wave packet dynamics starting at the decorated edge with $y=1$, (b) the DOS histogram, and (c) a typical eigenstate profile of $U_A'$.
The wave packet propagates to the $+x$-direction, as we expected. 
Thus, the boundary operator $A$ induces a robust edge state analogous to a quantum Hall edge state. 
We also find that the extended state along the edge survives even when its energy is overlapped with that of the original bulk band.
This is because of the Anderson localization of the bulk states, which suppresses the mixing with the edge state.
In  Appendix \ref{sec:random_phase}, we also show that the extrinsic chiral edge state is robust even in the presence of random phases along the edge at $y=1$.

In general, we can characterize the directed wave packet movement due to extrinsic chiral modes by 
the winding number in Eq.~(\ref{eq:w1}).
To show this, we consider the time-evolution by $\hat{U}$ in 1D for a while, then apply the result to the extrinsic edge modes in 2D.

For a quantum walk in 1D, 
we introduce the polarization at $x$ as
\begin{align}
    P_x = \sum_{\alpha} \bra{x,\alpha} \hat{x} \ket{x,\alpha},
\end{align}
where $\ket{x,\alpha}$ represents a state at position $x$ with internal degrees of freedom $\alpha$ such as spin, orbital and so on.
After the one-cycle time evolution by $\hat{U}$,  
the polarization becomes
\begin{align}
    P_x (T) 
    = \sum_\alpha \bra{x,\alpha} \hat{U}^\dag \hat{x} \hat{U} \ket{x,\alpha}.
    \label{eq:polarizationT}
\end{align}
If $\hat{U}$ has translation symmetry, one can show that 
the position displacement after the one-cycle evolution 
is given by the winding number
\cite{Thouless83,Kitagawa10_Floquet,Titum16,Nakagawa20,Cardano16,Fedorova20}
\begin{align}
P_x (T) - P_x  = w_1[U(k)],
    \label{eq:dP=w}
\end{align}
where $U(k)$ is the momentum space representation of $\hat{U}$. 
The proof is as follows:
From the Fourier transformation, 
the polarization in Eq.~(\ref{eq:polarizationT}) is rewritten as 
\begin{align}
    P_x (T) &= \sum_\alpha \bra{x}\bra{\alpha} \hat{U}^\dag \hat{x} \hat{U} \ket{x}\ket{\alpha}
    \nonumber \\
    &= \sum_{\alpha} \left[ \frac{1}{\sqrt{L}} \sum_k e^{ikx} \bra{k} \right] \bra{\alpha} \hat{U}^\dag \left[ \sum_{x'} x' \ket{x'}\bra{x'} \right] 
    \nonumber \\
    & \quad \times \hat{U} \left[ \frac{1}{\sqrt{L}} \sum_{k'} e^{-ik'x} \ket{k'} \right] \ket{\alpha}
    \nonumber \\
    &= \frac{1}{L^2} \sum_{\alpha,k,k',x'} e^{ikx} \bra{\alpha} U^\dag(k) e^{-ikx'} [-i\partial_{k'} e^{ik'x'}] 
    \nonumber \\
    & \quad \times U(k') e^{-ik'x} \ket{\alpha},
\end{align}
where $U(k)$ acts on the space of internal degrees of freedom $\alpha$. Using the partial integration and 
the relation
\begin{align}
   \frac{1}{L}\sum_{x'} e^{i(k'-k)x'} = \delta_{k,k'}, 
\end{align}
we obtain 
\begin{align}
    P_x (T) &= \frac{1}{L} \sum_{\alpha,k} e^{ikx} \bra{\alpha} U^\dag(k) i\partial_{k}[ U(k) e^{-ikx}] \ket{\alpha}
    \nonumber \\
    &= \frac{1}{L} \sum_{\alpha,k} \bra{\alpha} U^\dag(k) [i\partial_{k} U(k)] \ket{\alpha}
    + \sum_{\alpha} x .
\end{align}
The first term in the right hand side in the above is the winding number in Eq.~(\ref{eq:w1})
\begin{align}
    &\frac{1}{L} \sum_{\alpha,k} \bra{\alpha} U^\dag(k) [i\partial_{k} U(k)] \ket{\alpha}
    \nonumber \\
    &= \int_{0}^{2\pi} \frac{dk}{2\pi} \tr[U^\dag (k) i\partial_k U(k)] = w_1[U(k)],
\end{align}
while the second term is the polarization $P_x$
\begin{align}
    \sum_{\alpha} x = \sum_{\alpha} \bra{x,\alpha} \hat{x} \ket{x,\alpha} = P_x.
\end{align}
Therefore, we obtain $P_x (T) = w_1[U(k)] + P_x$, and thus the formula in Eq.~(\ref{eq:dP=w}).

Let us check this formula in Eq.~(\ref{eq:dP=w}) for a simple example
$U_{\text{QW}} = S_+ R(\theta)$ with
\begin{align}
    S_+ &= \pmat{ e^{-ik} & 0 \\ 0 & 1} , \ R(\theta)= \pmat{ \cos \theta & - \sin \theta \\ \sin \theta & \cos \theta},
\end{align}
which has $w_1[U_{\text{QW}}(k)]=1$. 
Consider the initial states
\begin{align}
    \ket{x,\uparrow}=\pmat{\ket{x} \\ 0}, \ket{x,\downarrow}=\pmat{0 \\ \ket{x}},
\end{align}
with the polarization
\begin{align}
    P_x &= \bra{x,\uparrow} \hat{x} \ket{x,\uparrow} + \bra{x,\downarrow} \hat{x} \ket{x,\downarrow} 
    \nonumber \\
    &= 2x.
\end{align}
The states after one cycle are
\begin{align}
    U_{\text{QW}} \pmat{\ket{x} \\ 0} = \pmat{\cos\theta\ket{x+1} \\ \sin\theta \ket{x}}, \\
    U_{\text{QW}} \pmat{0 \\ \ket{x}} = \pmat{-\sin\theta\ket{x+1} \\ \cos\theta\ket{x}},
\end{align}
and thus, we have
\begin{align}
    P_x (T) &= \bra{x,\uparrow} U_{\text{QW}}^\dag \hat{x} U_{\text{QW}} \ket{x,\uparrow} 
    \nonumber \\
    & \quad + \bra{x,\downarrow} U_{\text{QW}}^\dag \hat{x} U_{\text{QW}} \ket{x,\downarrow} 
    \nonumber \\
    &= 2x + 1 = P_x + 1.
\end{align}
Therefore, we obtain $P_x(T) -P_x = w_1[U_{\text{QW}}]=1$, which coincides with the formula in Eq.~(\ref{eq:dP=w})

We can generalize the above relation between the polarization and the winding number even in the presence of disorders.
For this purpose, 
we consider the flux inserted unitary operator $\hat{U}(\Phi)$,
where the hopping terms in $\hat{U}$ are modified by the uniform gauge potential $A_x=\Phi/L$ as  $\ket{x+q}\bra{x} \to e^{-i(\Phi/L)q}\ket{x+q}\bra{x}$ \cite{Niu85}.
Then, we introduce its winding number \cite{Gong18}
\begin{align}
    w_1[\hat{U}(\Phi)]= \int_0^{2\pi} \frac{d\Phi}{2\pi} 
    \tr[\hat{U}^\dagger(\Phi) i\partial_{\Phi} \hat{U}(\Phi)].
    \label{eq:w1mag}
\end{align}
We note that the flux inserted unitary operator $\hat{U}(\Phi)$ is periodic in $\Phi$ with the period $2\pi$ up to the large gauge transformation $\hat{U}_G = e^{-\frac{2\pi i}{L}\hat{x}}$,
$
    \hat{U}(\Phi+2\pi)=\hat{U}_G \hat{U}(\Phi) \hat{U}_G^\dag.
$
Thus, the winding number takes an integer.
In a manner similar to Eq.~(\ref{eq:dP=w}), we can prove
\begin{align}
P(T) -P= w_1[\hat{U}(\Phi)],
    \label{eq:dP=wPhi}
\end{align}
where $P(T)-P$ is an averaged version of the position displacement $P_x(T)-P_x$ after one-cycle evolution. 
See Appendix \ref{sec:dP=wPhi} for details.

Now, we extend the above results 
to extrinsic edge modes in 2D.
Since an extrinsic edge mode is localized at an edge of the system, say at $y=1$, we consider the polarization at $y=1$,
\begin{align}
    P_x|_{y=1} 
    = \sum_{\alpha} \bra{x,y=1,\alpha} \hat{x} \ket{x,y=1,\alpha}.
\end{align}
Then, the polarization after the one cycle time evolution is
\begin{align}
    P_x (T)|_{y=1} 
    = \sum_{\alpha} \bra{x,y=1,\alpha} \hat{U}^\dag \hat{x} \hat{U} \ket{x,y=1,\alpha},
\end{align}
where $\hat{U}$ is the time-evolution operator in 2D. 
As shown in Appendix \ref{sec:dP=wP}, if the system has translation symmetry, the formula in Eq.~(\ref{eq:dP=w}) can be generalized as
\begin{align}
    P_x (T)|_{y=1} - P_x|_{y=1}
    = w_P [U(k_x)]
    \label{eq:dP=wP},
\end{align}
where $w_P$ is the projected winding number defined by
\begin{align}
    \label{eq:wP}
    w_P [U(k_x)] & = \int_0^{2\pi} \frac{d k_x}{2\pi} \tr_{y,\alpha} \left[ \hat{P}_{\rm edge} U^\dag(k_x) i\partial_{k_x} U(k_x) \right],
\end{align}
with the projection operator 
$
\hat{P}_{\rm edge}=\ket{y=1}\bra{y=1}
$
at the edge.
Similarly, if the system has disorders, 
we have a generalization of Eq.~(\ref{eq:dP=wPhi}) with the projected winding number for Eq.~(\ref{eq:w1mag}).

\begin{figure}[t]
\centering
\includegraphics[width=85mm]{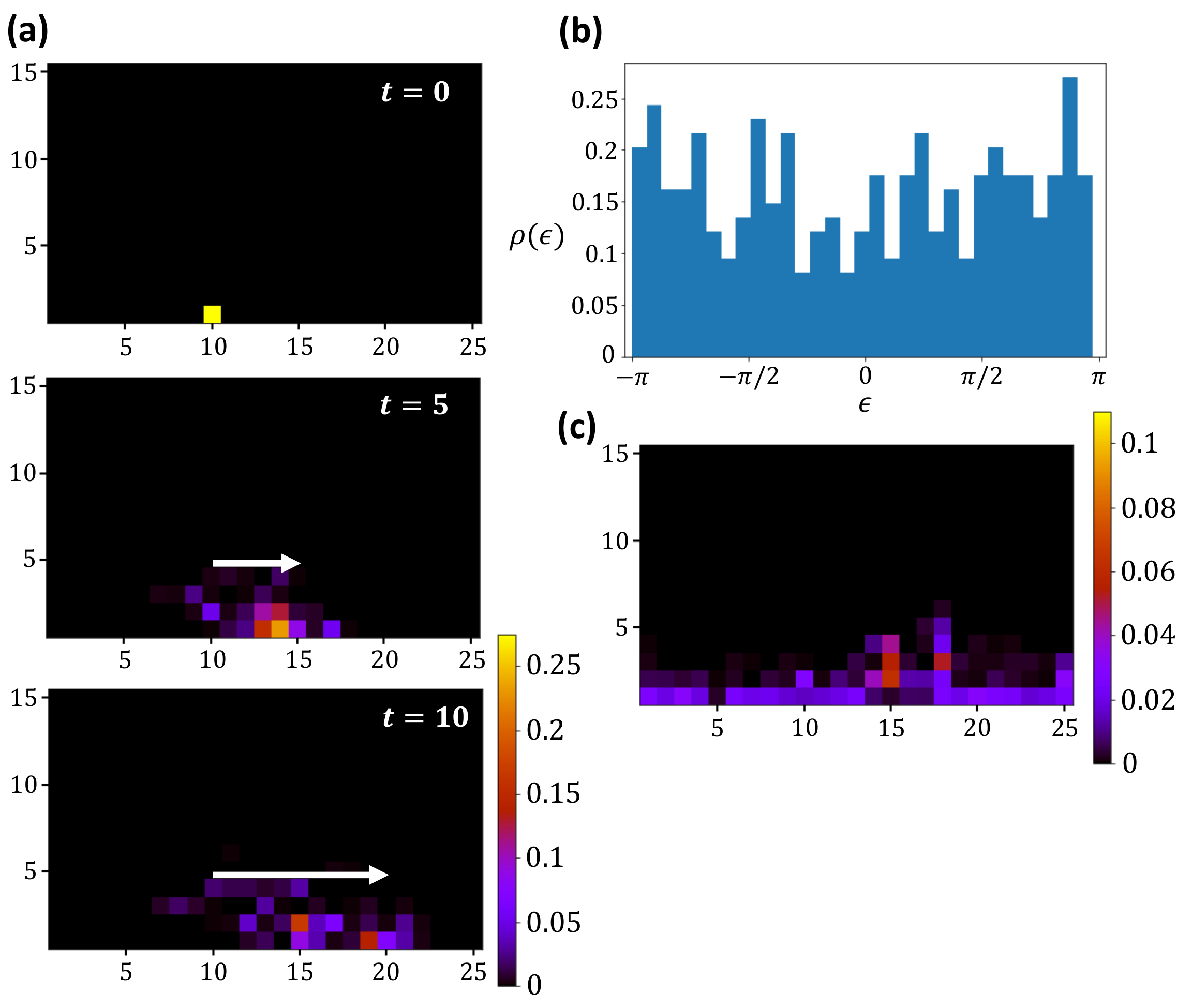} 
\caption{{\it In case of no band gap:} (a) Dynamics of a wave packet, (b) DOS and (c) a delocalized eigenstate of the Anderson model with the extrinsic edge mode in Eq.~(\ref{eq:Anderson_edge}). The parameters are $J=1$, $W=10$, $L_x=25$, $L_y=15$ and $T=1$. The color scale is different between (a) and (c). The initial state of the wave packet is $\ket{x=10,y=1}$. 
Due to $W \gg J$, all bulk eigenstates are localized. Whereas no band gap exists due to the strong random potential, the extended edge state survives and enables a directed wave packet motion.
}
	\label{fig:Anderson_model_edge_gapclosing}
\end{figure}

We remark that the projected winding numbers are not quantized in general because edge modes at $y=1$ can diffuse into the bulk. 
However, if bulk states are gapped or localized, edge modes rarely diffuse into the bulk, so the quantization of the projected winding numbers is almost recovered.
Under such situations, an extrinsic chiral mode 
induces a directed movement of wave packets at the edge
since it has a non-trivial winding number.
For instance, our model in 2D has a well-localized bulk state as shown in Fig.~\ref{fig:Anderson_model} (c), 
and thus the above mechanism explains the wave packet dynamics in Fig.~\ref{fig:Anderson_model_A} (a). 
Here note that the argument here does not necessarily require a bulk gap.
In Fig. \ref{fig:Anderson_model_edge_gapclosing}, 
we show the wave packet dynamics in  the model of Eq.~(\ref{eq:Anderson_edge}) with $W=10,J=1$, where the bulk band covers the whole energy range from $-\pi$ to $\pi$.
Even in this case, we observe that wave packets propagate to the $+x$-direction and there exists delocalized eigenstates along the edge.

\begin{figure}[t]
\centering
\includegraphics[width=85mm]{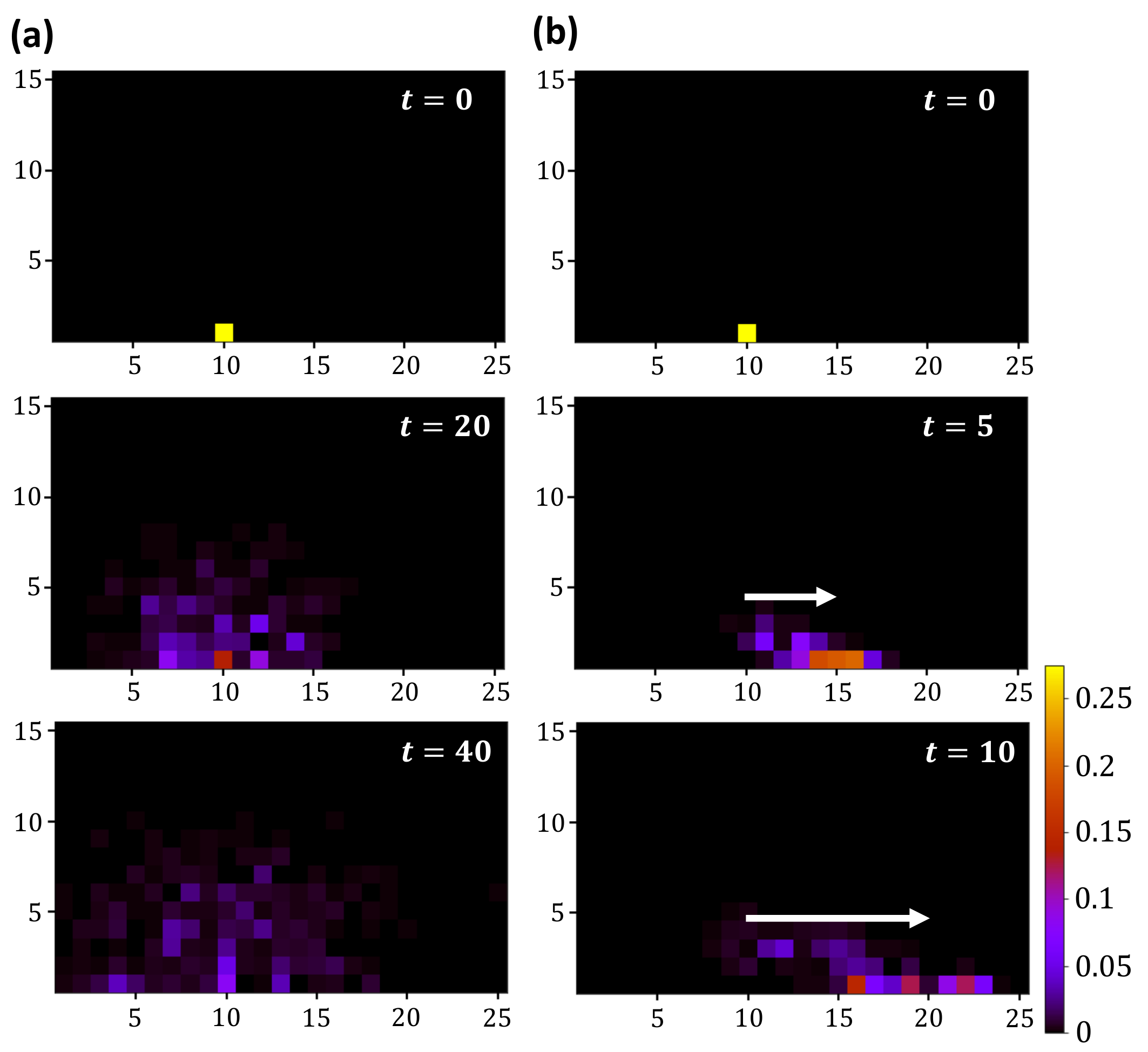}
\caption{{\it In case of noisy environment (spatial and temporal disorder):} Dynamics of a wave packet (a) for the Anderson model in Eq.~(\ref{eq:time-dep_Anderson}), and (b) that with the extrinsic edge mode in Eq.~(\ref{eq:time-dep_Anderson_edge}), where the random potentials change at each time step. The parameters are $J=0.2$, $W=1$, $L_x=25$, $L_y=15$ and $T=1$. (a) For the Anderson model without the extrinsic edge mode, 
a wave packet starting at the edge diffuses into the bulk.
(b) For the Anderson model with the extrinsic edge mode, 
the wave packet propagates to the $+x$-direction during the time scale shorter than diffusion.
}
	\label{fig:Anderson_model_time-dep_w=1}
\end{figure}

Finally, we consider a noisy environment where the random potential in Eq.~(\ref{eq:Anderson}) fluctuates in each time step \cite{Konno04,Joye10,Joye11,Evensky90,Obuse11}.
Our numerical simulations show that the directed wave packet dynamics due to the extrinsic chiral edge mode survives during the time scale shorter than diffusion. 
In Fig.~\ref{fig:Anderson_model_time-dep_w=1}, 
we compare the $t$ steps dynamics without the edge mode,
\begin{align}
    U_{\rm w/o }= \prod_{s=1}^t  e^{-iH_{A}(s) T},
    \label{eq:time-dep_Anderson}
\end{align}
to that with the edge mode,
\begin{align}
    U_{\rm w/e}= \prod_{s=1}^t A \cdot e^{-iH_{A}(s) T}.
        \label{eq:time-dep_Anderson_edge}
\end{align}
Here $H_{A}(s)$ at time step $s$ has the same form as in Eq.~(\ref{eq:Anderson}) but the random potential $V_{x,y}$ changes at each step $s$.
As seen in Fig.~\ref{fig:Anderson_model_time-dep_w=1} (a), 
the time dependent randomness leads to diffusion. 
The details of the diffusive behavior is given in the Appendix \ref{sec:diffusion}.
In the case with the extrinsic chiral edge mode
[Fig.~\ref{fig:Anderson_model_time-dep_w=1} (b)], on the other hand, we find again a directed wave packet displacement due to the nontrivial winding number $w_1[U_{\rm w/e}(\Phi)]$.
Such robustness in a noisy environment would enable extrinsic modes to realize fault-tolerance quantum devices.

\subsection{class AIII the split step quantum walk in 1D: cancelation of the edge mode}

Boundary modes of chiral-symmetric quantum walks in 1D have been experimentally observed as a localization of dynamics \cite{Kitagawa12,Barkhofen17}.
The extrinsic topological phase, however, can cancel the boundary states, leading to a dynamics with delocalized behaviors.

\begin{figure}[t]
\centering
\includegraphics[width=86mm]{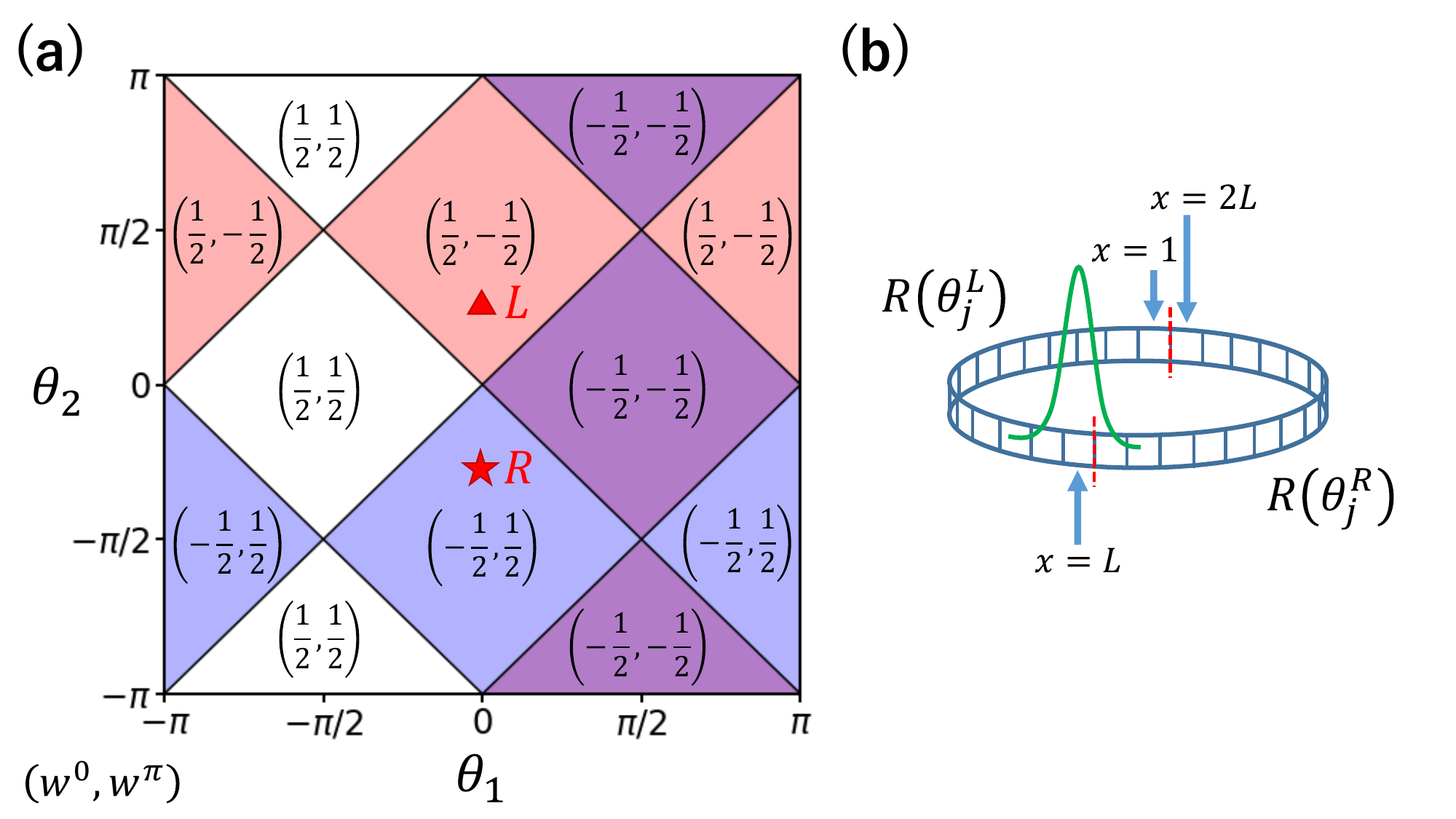}
\caption{(a) Topological phase diagram of the split-step quantum walk. $(w^0,w^\pi)$ represent the bulk topological invariants given by Eq.~(\ref{eq:w0wpi}). The red triangle and star represent the parameters for the left chain, $(\theta_1^L,\theta_2^L)=(0,\pi/4)$, and the right one, $(\theta_1^R,\theta_2^R)=(0,-\pi/4)$, used in the numerical simulations in Figs.~\ref{fig:split_step} and \ref{fig:split_step_edge}.
(b) The setup of the split-step quantum walk. The left and right regions of the quantum walk are joined at two edges at $x=1$ and $L$.
}
	\label{fig:phase_diagram}
\end{figure}

Let us study the split step quantum walk model in 1D \cite{Kitagawa10,Kitagawa12,Obuse15}:
\begin{align}
    &U_{\text{QW}} = U_2 U_1,
    \nonumber \\
    &U_1 = R_2^{1/2} S_- R_1^{1/2},\
    U_2 = R_1^{1/2} S_+ R_2^{1/2}.
    \label{eq:split_step}
\end{align}
Here, $S_+$ and $S_-$ are the shift operators, and $R_j=R(\theta_j)$ is the spin rotation coin operator, which are defined as follows:
\begin{align}
    &S_+(k) = \pmat{ e^{-ik} & 0 \\ 0 & 1},\
    S_-(k) = \pmat{ 1 & 0 \\ 0 & e^{ik}},
    \\
    &R(\theta)  =
    \pmat{\cos \theta & -\sin \theta \\ \sin \theta & \cos \theta}.
\end{align}
This model has the decomposed CS in Eq.~(\ref{eq:CSasboth}) with $\Gamma=\sigma_x$.
After performing a unitary transformation of the basis so that $\Gamma$ becomes $\Gamma = \sigma_z$, 
we calculate the bulk topological numbers $(w^0,w^\pi)$ defined by the right hand side of Eq.~(\ref{eq:BECasboth}), 
\begin{align}
w^0=\frac{w_1[b]-w_1[c]}{2},
\quad
w^\pi=\frac{w_1[a]-w_1[d]}{2},
\label{eq:w0wpi}
\end{align}
for the split step quantum walk in Eq.~(\ref{eq:split_step}). 
The obtained topological numbers are summarized in the phase diagram in Fig.~\ref{fig:phase_diagram} (a).

\begin{figure}[t]
\centering
\includegraphics[width=86mm]{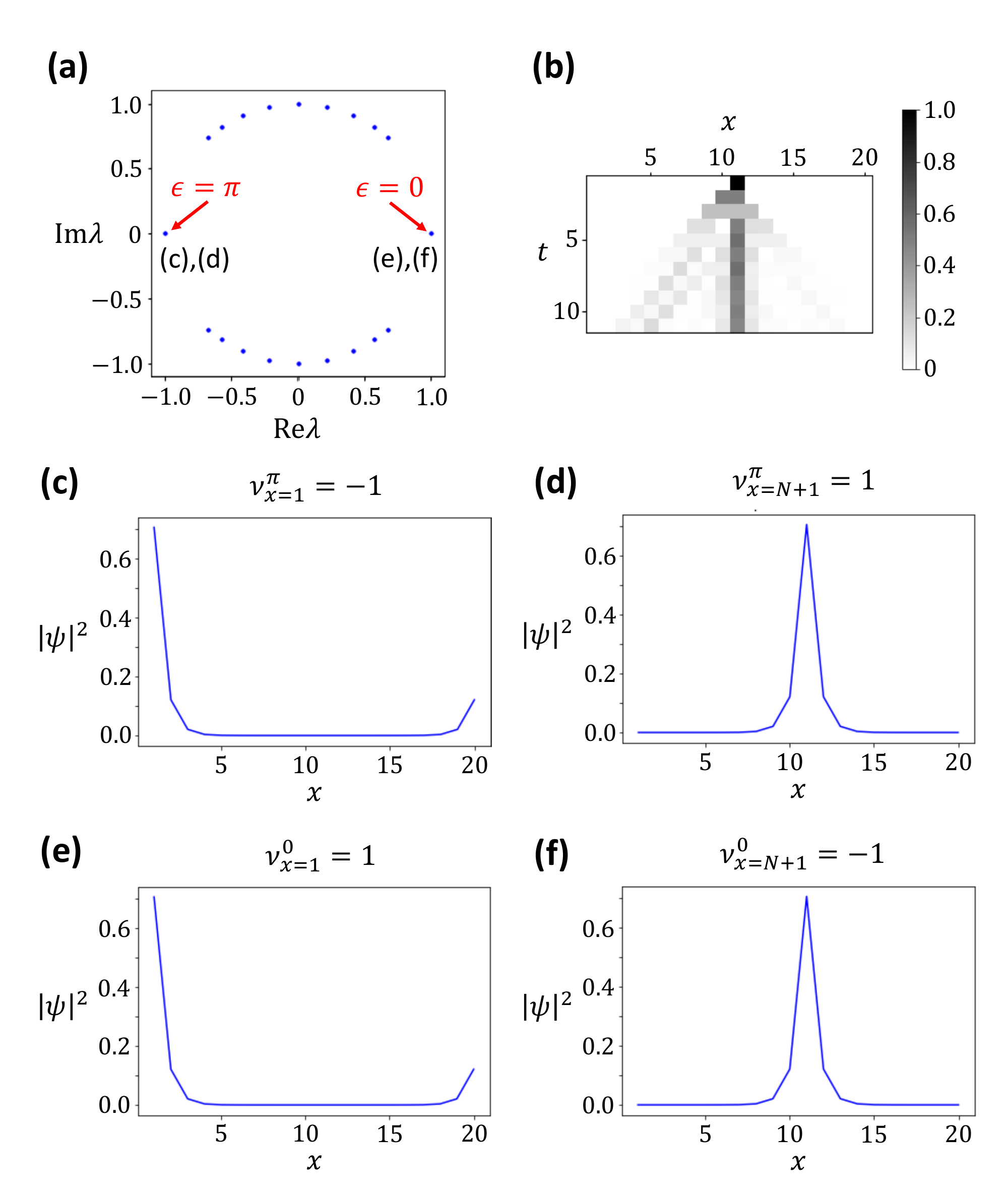}
\caption{(a) Energy spectrum, (b) dynamics, and (c-f) eigenstates of the split step walk in Eq.~(\ref{eq:split_step}). The total system size is $2N=20$. The parameters in the left half chain are $(\theta_1^L,\theta_2^L)=(0,\pi/4)$, while those in the right half chain are $(\theta_1^R,\theta_2^R)=(0,-\pi/4)$. The initial state is $\ket{x=11,\downarrow}$.  At $x=N+1$, we have a single $\epsilon=0$ mode and an $\epsilon=\pi$ mode.
The topological number of the $\epsilon=0$ mode 
is $-1$, while that of the $\epsilon=\pi$ mode
is $+1$.
}
	\label{fig:split_step}
\end{figure}

To examine boundary modes of the system, we consider the loop configuration shown in Fig.~\ref{fig:phase_diagram} (b).
The loop consists of left and right chains with the same length $N$,  
where $U_{\rm QW}$ in the left (right) chain has the coin operators with the parameter $(\theta_1^L,\theta_2^L)=(0,\pi/4)$ ($(\theta_1^R,\theta_2^R)=(0,-\pi/4)$).  
Boundary modes appear at the interfaces between the left and right chains. 

Figure \ref{fig:split_step} (a) shows
eigenvalues $\lambda$ of the eigenequation $U_{\text{QW}}\ket{\psi}=\lambda \ket{\psi}$ on the loop configuration,  
where $\lambda= 1$ ($\lambda=-1$) indicates the $\epsilon=0$ ($\epsilon=\pi$) boundary modes.  
Both of the interfaces at $x=1,N+1$ host a single $\epsilon=0$ mode $\ket{\psi^{0}_{x= 1,N+1}}$ and a $\epsilon=\pi$ one 
$\ket{\psi^{\pi}_{x= 1,N+1}}$
with the topological number $\nu_{x=1,N+1}^{0,\pi}=\langle \psi^{0,\pi}_{x=1,N+1}|\Gamma|\psi_{x= 1, N+1}^{0,\pi}\rangle$
shown in Figs.~\ref{fig:split_step} (c)-(f). 
We note that the boundary gapless modes satisfy the bulk-boundary correspondence:
\begin{align}
    \nu_{x=1}^{0,\pi}=-\nu^{0,\pi}_{x=N+1}
    =w^{0,\pi}_L-w^{0,\pi}_R,
\end{align}
where $w^{0,\pi}_{R, L}$ is the bulk topological number in
the right and left chains.
Here the original bulk-boundary correspondence in Eq.~(\ref{eq:BECasboth}) is slightly modified because we have considered interfaces between two topologically non-trivial chains.
Due to the existence of the boundary modes, 
a wave packet initially localized at one of the interfaces remains localized after time evolution [Fig.~\ref{fig:split_step}(b)]. 

As discussed in Sec.~\ref{sec:CSQW}, if one relaxes the decomposed CS in Eq.~(\ref{eq:CSasboth}) by the original CS  in Eq.~(\ref{eq:CSQW}), 
we can change the number of boundary gapless modes by using the extrinsic topological phase. 
To see this, we introduce the following unitary operator $A$ 
\begin{align}
    A =\sum_{x\neq N}^{2N} \ket{x}\bra{x}\otimes \sigma_0 
    + \ket{x=N}\bra{x=N}\otimes U_{\text{BDQW}},
\end{align}
where the boundary unitary operator $U_{\text{BDQW}}$ is taken to be nontrivial
\begin{align}
    U_{\text{BDQW}}=\sigma_x.
\end{align}
We insert $A$ between $U_1$ and $U_2$ in Eq.~(\ref{eq:split_step}),
\begin{align}
    U_{\text{QW}} = U_2 A U_1.
\label{eq:UAU}
\end{align}
Since $A$ satisfies $\Gamma A^{\dagger} \Gamma^{-1}=A$, $U_{\rm QW}$ in the above has the original CS.
We can show that $U_{\text{BDQW}}$ has nontrivial extrinsic boundary states. From the Hermitian matrix $U_{\text{BDQW}}\Gamma= \sigma_0$, we have a nontrivial topological invariant in Eq.~(\ref{eq:nFAIII0d}),
\begin{align}
    n= \frac{1}{2} [N_+(U_{\text{BDQW}}\Gamma) - N_-(U_{\text{BDQW}}\Gamma)]=1.
\end{align}
Therefore, from the extended Nielsen-Ninomiya theorem in Eq.~(\ref{eq:NN}), we obtain extrinsic boundary modes at $x=N$ with the energies $\epsilon=0,\pi$ and the topological numbers $\nu^0=-\nu^\pi =1$.
As shown below, these extrinsic boundary modes may cancel the original edge modes at the interface at $x=N+1$.

\begin{figure}[t]
\centering
\includegraphics[width=86mm]{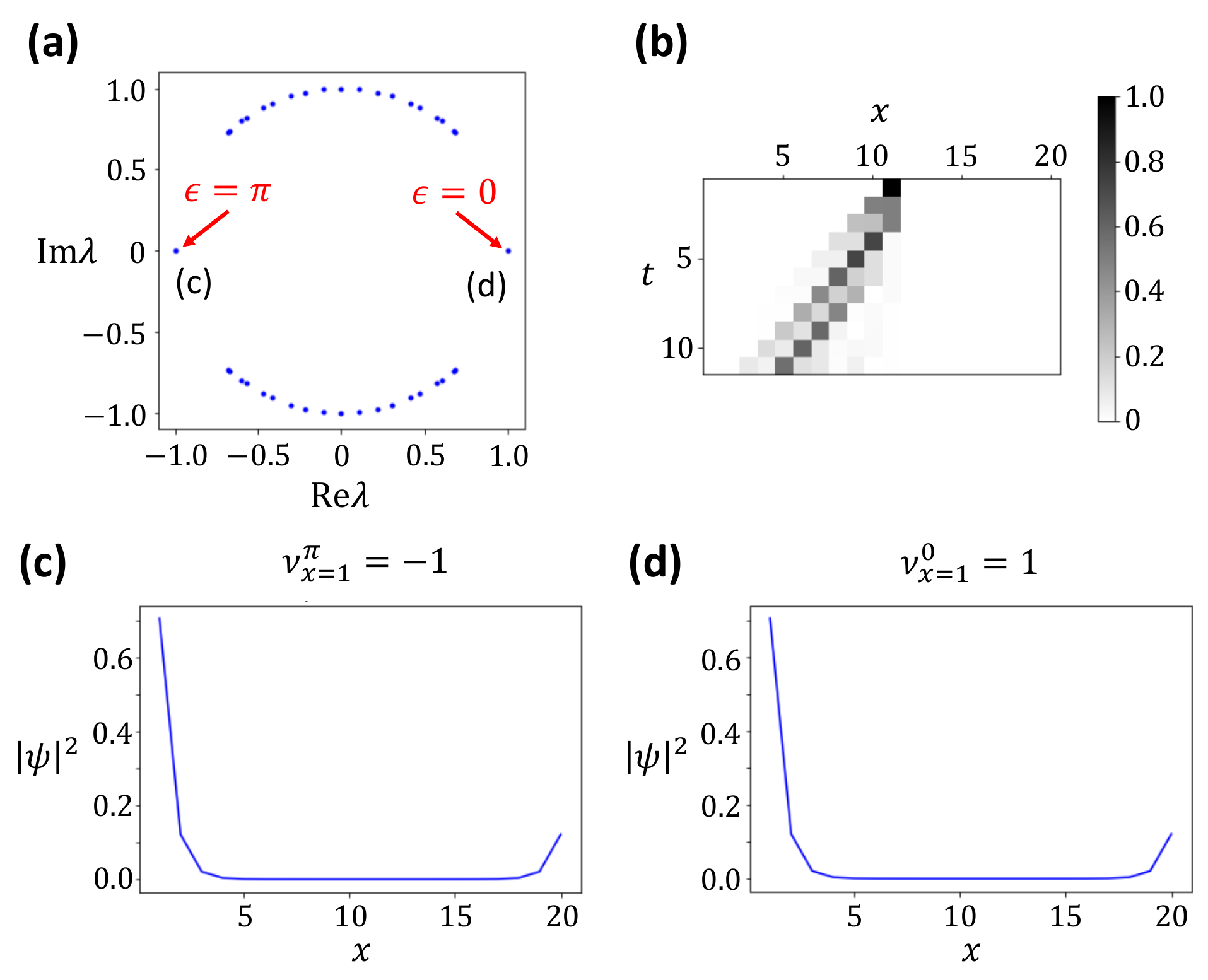}
\caption{(a) Energy spectrum, (b) dynamics, and (c,d) eigenstates of the decorated split step walk in Eq.~(\ref{eq:UAU}). The total system size is $2N=20$. The parameters in the left half chain is $(\theta_1^L,\theta_2^L)=(0,\pi/4)$, while those in the right half chain is $(\theta_1^R,\theta_2^R)=(0,-\pi/4)$. The initial state is $\ket{x=11,\downarrow}$. 
No localized state at boundary $x=N+1$ is found both in the spectrum and the dynamics.
}
	\label{fig:split_step_edge}
\end{figure}

The eigenenergies of $U_{\rm QW}$ in Eq.~(\ref{eq:UAU})
is shown in Fig.~\ref{fig:split_step_edge} (a).
While the spectrum has $\epsilon=0$ and $\epsilon=\pi$ modes, they are localized near the interface at $x=1$,
as shown in Fig.~\ref{fig:split_step_edge} (c) and (d).
No gapless edge mode exists near the interface at $x=N+1$.
As a result, in contrast to the previous case, a wave packet initially localized at $x= N+1$ spreads after the time evolution [Fig.~\ref{fig:split_step_edge} (b)].

\subsection{class A in 2D: cancelation of the chiral edge mode}
Floquet topological phases may host chiral edge modes without the  Chern number \cite{Kitagawa10_Floquet,Rudner13,Asboth15}. 
The Floquet anomalous edge states originate from the bulk topological invariant of the time-evolution operator $U(\bm{k},t)$ in Eq.~(\ref{eq:time_evolution}), 
and have been
experimentally realized in photonic systems \cite{Maczewsky17,Mukherjee17,Chen18}. 
In this section, we demonstrate that the Floquet anomalous edge states can be eliminated by
using the extrinsic topology of quantum walks in 2D.

We consider the model in 2D with Floquet anomalous edge states in Ref.~\cite{Rudner13}:
\begin{align}
H(t) = H_j, \ t\in [(j-1)T/5,jT/5],
\end{align}
with
\begin{align}
    & H_{j=1,2,3,4} = J e^{i\bm{b}_j\cdot \bm{k}} \sigma_+  + J e^{-i\bm{b}_j\cdot \bm{k}} \sigma_- + \delta_{AB} \sigma_z,
    \nonumber \\
    & H_5 = \delta_{AB} \sigma_z.
\end{align}
Here $\bm{b}_1 = -\bm{b}_3 =(a,0)$ and $\bm{b}_2 = -\bm{b}_4 =(0,a)$, $\sigma_{\pm}=(\sigma_x\pm i \sigma_y)/2$, and $\delta_{AB}$ is a real parameter. 
The one-cycle time evolution of the model is given by
\begin{align}
U_{R} = \prod_{j=1}^5 e^{-iH_j T/5}.
\label{eq:RudnerUnitary}
\end{align}

\begin{figure}[t]
\centering
\includegraphics[width=85mm]{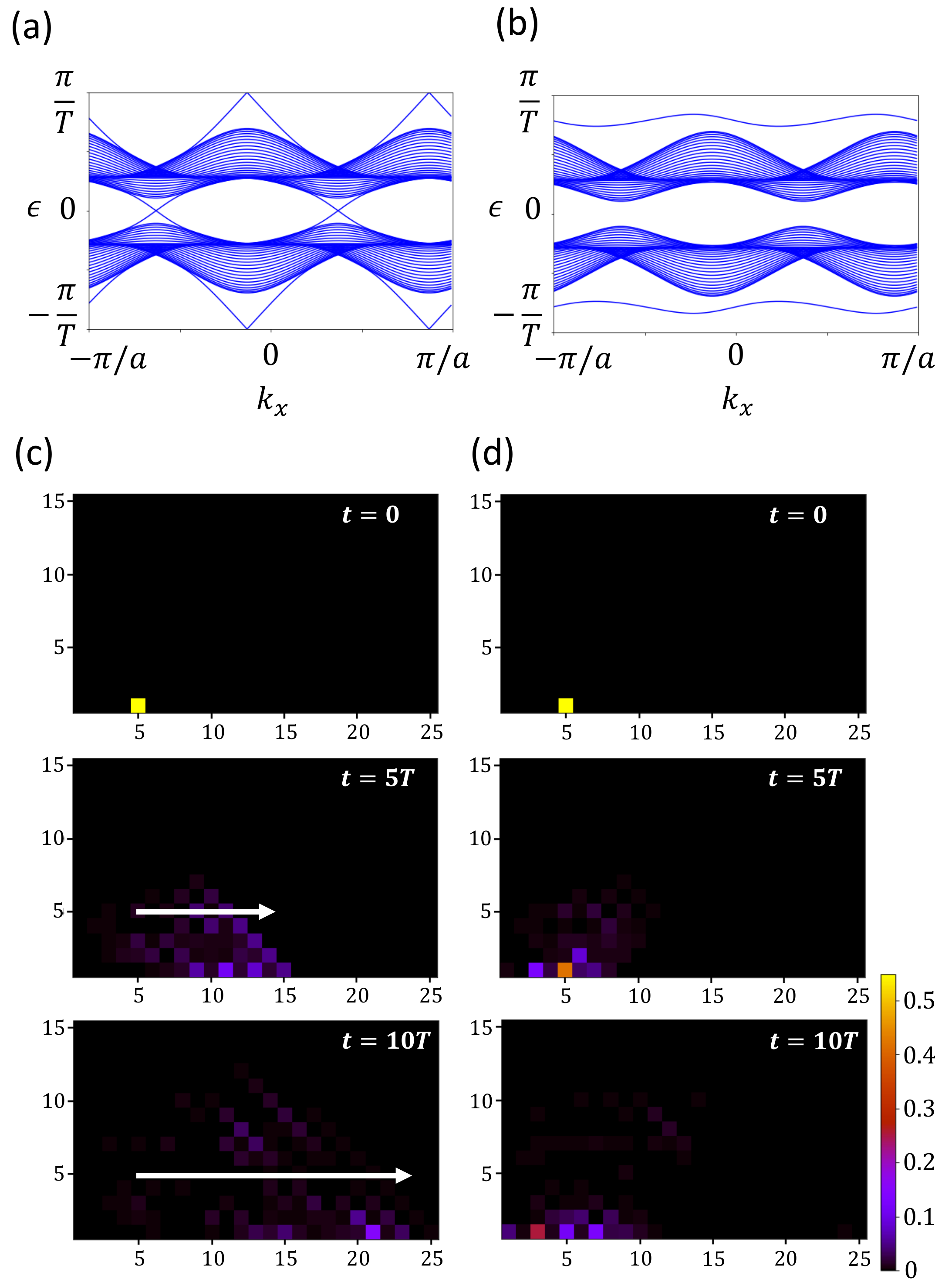} 
\caption{Energy spectrum and dynamics (a,c) without and (b,d) with the unitary operator $A$ in Eq.~(\ref{eq:A_Rudner}).
We use (a,c) the model with $U_R$ in Eq.~(\ref{eq:RudnerUnitary})  and (b,d) that with $U'_R$ in Eq.~(\ref{eq:RudnerUnitary_A}).
The parameters are $J/T=2.2\pi$, $\delta=1.3\pi$. For (a) and (b), the system size is $L_y=30$. For (c) and (d), the system size is $L_x=25$ and $L_y=15$.
The initial state is $\ket{x=5,y=1,\downarrow}$. 
We take the periodic (open) boundary condition in the $x$ ($y$) direction.
(a) The model $U_R$ in Eq.~(\ref{eq:RudnerUnitary}) has Floquet anomalous chiral edge modes.
(c) Due to the existence of chiral edge modes, a wave packet on the edge at $y=1$ propagates to the $+x$-direction.
(b) The decorated model $U'_R$ in Eq.~(\ref{eq:RudnerUnitary_A}) exhibits no chiral edge mode.
(d) Due to the disappearance of chiral edge modes, only a diffusive behaviour into the bulk is observed.
}
	\label{fig:Rudner}
\end{figure}

\begin{figure}[t]
\centering
\includegraphics[width=70mm]{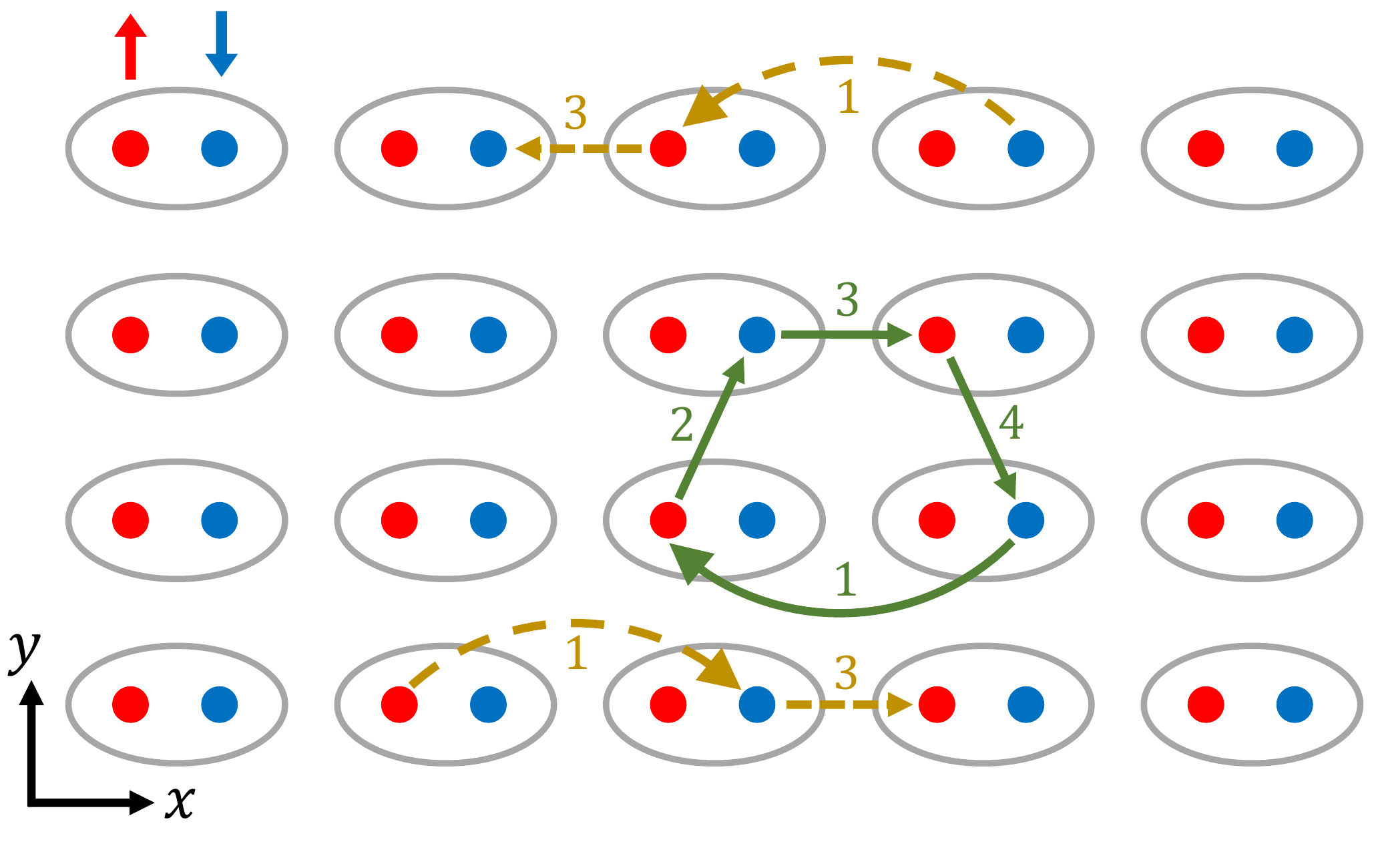} 
\caption{One-cycle dynamics of the model Eq.~(\ref{eq:RudnerUnitary}) at $JT/5=\pi/2$ and $\delta_{AB}=0$.
This model has a bipartite lattice structure with spin up (red) and spin down (blue).
After the one-cycle time evolution,
bulk states return to the same position, while a spin up (down) edge state at $y=1 \ (L_y)$ propagates in the $+(-)x$-direction.
}
	\label{fig:Rudner_simple}
\end{figure}

The energy spectrum 
and the dynamics of anomalous edge modes in the model are shown in Figs.~\ref{fig:Rudner} (a) and (c).
This model has gapless chiral edge modes both at the $\epsilon=0$ and $\epsilon=\pi$ gaps. The existence of the chiral edge modes is easily understood for 
$JT/5=\pi/2$ and $\delta_{AB}=0$.
In this case, each time-evolution unitary operator becomes
\begin{align}
&e^{-iH_{j=1,2,3,4} T/5} =- i  (e^{i\bm{b}_j\cdot \bm{k}} \sigma_+  + e^{-i\bm{b}_j\cdot \bm{k}} \sigma_- ),
\nonumber \\
&e^{-iH_5 T/5} = 1,
\end{align}
which makes the total time-evolution operator trivial $U_{R}=\hat{1}$,
but there exists a chiral mode at the boundary, as shown in Fig.~\ref{fig:Rudner_simple}.
The chiral edge mode remains even if we modify the parameters $\delta_{AB}$ and $J$ unless the energy gaps at $\epsilon=0,\pi$ are closed.

The presence of the Floquet anomalous chiral edge state is ensured by the three-dimensional winding number of the loop unitary $L({\bm k},t)$ in Eq.~(\ref{eq:decomposition}),
\begin{align}
w_3[L]
&=\frac{1}{8\pi^2} \int d k_x dk_y d t 
\nonumber\\
&\times
\tr\left(
L^{-1} \partial_{t} L 
[L^{-1} \partial_{k_x} L,
L^{-1} \partial_{k_y} L ]
\right).
\end{align}
However, as we already pointed out in Sec.~\ref{sec:Floquet_vs_qw}, $L({\bm k},t)$ is well-defined only when the microscopic Hamiltonian $H({\bm k},t)$ exists.
If one allows a general deformation of the time-evolution unitary operator, one can trivialize $L({\bm k},t)$ without closing gaps at $\epsilon=0,\pi$.
Therefore, in the framework of quantum walks, where no microscopic Hamiltonian is assumed in general,
the Floquet anomalous edge mode is not protected by the bulk topological number.

Actually, we can eliminate the anomalous edge mode in Fig.~\ref{fig:Rudner} (a)
by multiplying $U_{R}$ by $A$,
\begin{align}\label{eq:RudnerUnitary_A}
&U'_{R} = A U_{R}, \\
&A(k_x) = 
U_{\text{edge}}^{y=1}(k_x)
\otimes
\ket{y=1}\bra{y=1}
\nonumber \\
&+
U_{\text{edge}}^{y=L_y}(k_x)
\otimes
\ket{y=L_y}\bra{y=L_y}
\nonumber \\
&+
\sum_{y=2}^{L_y-1} \sigma_0 \otimes \ket{y}\bra{y},
\label{eq:A_Rudner}
\\
&U_{\text{edge}}^{y=1}(k_x)=
\pmat{
e^{2i k_x} & 0 \\
0 & 1}, \
U_{\text{edge}}^{y=L_y}(k_x)=
\pmat{
1 & 0 \\
0 & e^{-2i k_x} },
\end{align}
where we impose the open boundary conditions on $U_R$ at $y=1, L_y$. 
As shown in Figs.~\ref{fig:Rudner} (b) and (d),
no gapless chiral edge mode exists in the quasi-particle spectrum of $U_R'$, and
no directed wave packet motion is observed on the edges.

The extrinsic topology of $U_{\text{edge}}^{y=1}$ and $U_{\text{edge}}^{y=L_y}$ in $A$ explains the disappearance of the anomalous edge modes.
For these boundary unitary operators, the  winding number in Eq.~(\ref{eq:w1}) becomes nonzero:
\begin{align}
w_1\left[U_{\text{edge}}^{y=1}(k_x)\right] 
= -w_1\left[U_{\text{edge}}^{y=L_y}(k_x)\right] =2.
\end{align}
Thus, $U_{\text{edge}}^{y=1}$ and $U_{\text{edge}}^{y=L_y}$ have the extrinsic topology phase at the boundaries.
From the the extended Nielsen-Ninomiya theorem in Eq.~(\ref{eq:NN}), $U_{\text{edge}}^{y=1}$ and $U_{\text{edge}}^{y=L_y}$ provide an anomalous edge state on each boundary that has the chirality opposite to that in the original model.
As a result, $U'_R$ does not have no net topological number for the anomalous edge modes, and thus no stable anomalous edge mode remains.

We emphasize again that the boundary unitary operator $A$ affects nontrivially only on the edges, and it controls the presence and absence of anomalous gapless edge states, which implies the extrinsic nature of the edge states.

\section{Conclusion}
\label{sec:conclusion}
In this work, we argue the classification of topological phases in quantum walks.
Due to the discrete nature of quantum walk dynamics, the bulk topological invariant is insufficient to determine the boundary states.
The numbers of boundary states depend on both the bulk topology and the boundary topology.
While the conventional topological insulators and superconductors in equilibrium systems also may have similar extrinsic nature in higher-order topological phases, quantum walks may support it even in the first-order topological phases.

The extrinsic boundary states in quantum walks resemble to anomalous boundary states in Floquet systems, but
their topological origin is different.
For Floquet systems, the anomalous boundary states originate from the non-trivial topology of the bulk continuous time-evolution operator, but for quantum walks, the continuous time-evolution operator is not given.

In the previous work \cite{Asboth13}, it was shown that the bulk-boundary correspondence holds for class AIII systems in 1D with a decomposed realization of CS.
We explain how their definition assures the bulk-boundary correspondence, and discuss a similar bulk-boundary correspondence in other dimensions and symmetry classes.
We find that class CII quantum walks in 1D obey the bulk-boundary correspondence under decomposed TRS and PHS.

We also examine physical implementations of extrinsic topological phases in quantum walks.
We numerically show that the extrinsic boundary states induce the charge pump which is robust against disorders along the edge of 2D systems. We also present general arguments for the robustness of the charge pump.
Moreover, we show that using the extrinsic topology, one can eliminate the pre-existing anomalous boundary states in the class AIII 1D split step walk and a class A 2D model, respectively.
We can change the types and the numbers of gapless boundary states without changing the bulk, which implies the breakdown of the bulk-boundary correspondence in quantum walks.

In this work, we have focused on quantum walks, but most of the arguments hold for any other systems described by unitary operators such as cellular automatons. 
Applying the present arguments to such systems could be interesting.
Another promising direction is higher-order boundary states in quantum walks. While we have studied only the first-order boundary states in this work, higher-order boundary states 
may have richer extrinsic topological behaviors.
Quantum walks also support unique symmetries that have no counterpart in static systems such as time-glide symmetry \cite{Mochizuki20}.
Such symmetry could produce extrinsic topological phases unique to dynamical systems.

\smallskip
This work was supported by JST CREST Grant No. JPMJCR19T2, 
JST ERATO-FS Grant No. JPMJER2105, and
KAKENHI Grants No. JP21J11810, No. JP20H00131, No. JP20H01845, No. JP18H01140, No. JP20H01828, and No. JP21H01005 from the JSPS.


\appendix

\section{Topological classification of $U_{\text{BDQW}}(\bm{k}_\parallel)$}
\label{sec:classifyU}
As explained in the main text, we can classify $U_{\text{BDQW}}(\bm{k}_\parallel)$ by using ${\cal H}_U({\bm k}_\parallel)$ with symmetry in Eqs.~(\ref{eq:properCS_HU})-(\ref{eq:CS_HU}).
In this section, we perform the topological classification of ${\cal H}_U(\bm{k}_\parallel)$ by the extension of Clifford algebras \cite{Morimoto13,Kitaev09}.

We first review the Clifford algebra.
The Clifford algebra is a ring in mathematics, which has addition and multiplication.
The complex Clifford algebra $Cl_n$ has $n$ generators $\{e_1,\ldots,e_n\}$ satisfying
\begin{align}
    \{e_i,e_j\}=\delta_{i,j},
\end{align}
and the linear combination of the products $e_1^{p_1}e_2^{p_2}\cdots e_n^{p_n}$ form a $2^n$-dimensional complex vector space.
The real Clifford algebra $Cl_{p,q}$ has $p+q$ generators $\{e_1,\ldots,e_p;e_{p+1},\ldots,e_{p+q}\}$ satisfying
\begin{align}
    \{e_i&,e_j\} =0 \ \text{for} \ i\neq j,
    \\
    e_i^2 &=\begin{cases}
    -1 & \text{for}\ 1\leq i \leq p, \\
    1 & \text{for}\  p+1 \leq i \leq p+q,
    \end{cases}
\end{align}
and their products form a $2^{p+q}$-dimensional real vector space.
For instance, generators of $Cl_2$ are given by the Pauli matrices:
\begin{align}
    \{e_1,e_2\}=\{\sigma_x,\sigma_y\},
\end{align}
of which products provide the basis of the $2^2$-dimensional complex vector space,
\begin{align}
    1, \sigma_x,\sigma_y, i\sigma_z.
\end{align}
The space coincides with that of $2\times 2$ matrices $\mathbb{C}(2)$, and thus we obtain the isomorphism $Cl_2 \simeq \mathbb{C}(2)$.

We can obtain $Cl_{n+1}$ by adding a generator $e_0$ to a given representation of $Cl_n$. 
The problem to identify all possible representations of $e_0$ is called the extension problem of $Cl_n\to Cl_{n+1}$, and the space of the representations is called the classifying space $C_n$.
In Table \ref{tb:Cn}, we summarize the classifying space $C_n$ and the number of connected parts of $C_n$, $\pi_0(C_n)$.  $C_n$ has the Bott periodicity $C_{n+2}=C_n$.

We also have a similar extension problem for
real Clifford algebras.
For a given representation of $Cl_{p,q}$, we add a generator $e_0$ that satisfies $e_0^2=-1$ ($e_0^2=1$), then obtain the real Clifford algebra $Cl_{p+1,q}$ ($Cl_{p,q+1}$). The extension problem 
$Cl_{p,q}\to Cl_{p+1,q}$ ($Cl_{p,q}\to Cl_{p,q+1}$) defines
the classifying space $R_{p+2-q}$ ($R_{q-p}$).
Table \ref{tb:Rq} summarizes the classifying space $R_q$ and the number of connected parts of $R_q$, $\pi_0(R_q)$. $R_q$ has the Bott periodicity $R_{q+8}=R_{q}$.

\begin{table}[t]
\centering
\caption{Classifying space $C_n$.
}
\begin{tabular}{ccc}
  \hline\hline
  ~$n\mod 2$~~ & ~Classifying space $C_n$~ & ~~$\pi_0(C_n)$~ \\
  \hline
  0 & $[U(k+m)/(U(k)\times U(m))] \times \mathbb{Z}$ & $\mathbb{Z}$ \\
  1 & $U(k)$ & 0 \\
  \hline\hline
\end{tabular}
\label{tb:Cn}
\end{table}

\begin{table}[t]
\centering
\caption{Classifying space $R_q$.
}
\begin{tabular}{ccc}
  \hline\hline
  ~$q\mod 2$~~ & ~Classifying space $R_q$~ & ~~$\pi_0(R_q)$~ \\
  \hline
  0 & $[O(k+m)/(O(k)\times O(m))] \times \mathbb{Z}$ & $\mathbb{Z}$ \\
  1 & $O(k)$ & $\mathbb{Z}_2$ \\
  2 & $O(2k)/U(k)$ & $\mathbb{Z}_2$ \\
  3 & $U(2k)/Sp(k)$ & 0 \\
  4 & $[Sp(k+m)/(Sp(k)\times Sp(m))] \times \mathbb{Z}$ & $\mathbb{Z}$ \\
  5 & $Sp(k)$ & 0 \\
  6 & $Sp(k)/U(k)$ & 0 \\
  7 & $U(k)/O(k)$ & 0 \\
  \hline\hline
\end{tabular}
\label{tb:Rq}
\end{table}

\begin{table*}[t]
\centering
\caption{Clifford algebra extensions and classifying spaces for ${\cal H}_U(\bm{k}_{\parallel})$.
}
\begin{tabular}{ccccccc}
  \hline\hline
  ~AZ class~~~ & ~~$T$~~ & ~~$C$~~ & ~~$\Gamma$~~~ & ~~~~~~~~~~~~~~~~Generator~~~~~~~~~~~~~~~~ & ~~~~~~~~~~~Extension~~~~~~~~~ & ~Classifying space~ \\
  \hline
  A & 0 & 0 & 0 & $\{\gamma_1,\ldots,\gamma_{d-1},\gamma_0,\Sigma_z\}$ & $Cl_{d}\to Cl_{d+1}$ & $C_{d}$ \\
  AIII & 0 & 0 & 1 & $\{\gamma_1,\ldots,\gamma_{d-1},\gamma_0,\Sigma_z,J\Sigma_z\tilde{\Gamma}\}$ & $Cl_{d+1}\to Cl_{d+2}$ & $C_{d+1}$ \\
  \hline
  AI & $+1$ & 0 & 0 & $\{J\gamma_0;\gamma_1,\ldots,\gamma_{d-1},\Sigma_z,\tilde{T},J\tilde{T}\}$ & $Cl_{0,d+2}\to Cl_{1,d+2}$ & $R_{-d}$ \\
  BDI & $+1$ & $+1$ & 1 & $\{J\gamma_0,\tilde{\Gamma}\Sigma_z;\gamma_1,\ldots,\gamma_{d-1},\Sigma_z,\tilde{T},J\tilde{T}\}$ & $Cl_{1,d+2}\to Cl_{2,d+1}$ & $R_{1-d}$ \\
  D & 0 & $+1$ & 0 & $\{J\gamma_0,J\Sigma_z;\gamma_1,\ldots,\gamma_{d-1},\tilde{C},J\tilde{C}\}$ & $Cl_{1,d+1}\to Cl_{2,d+1}$ & $R_{2-d}$ \\
  DIII & $-1$ & $+1$ & 1 & $\{J\gamma_0,\tilde{T},J\tilde{T};\gamma_1,\ldots,\gamma_{d-1},\Sigma_z,\tilde{\Gamma}\Sigma_z\}$ & $Cl_{2,d+1}\to Cl_{3,d+1}$ & $R_{3-d}$ \\
  AII & $-1$ & 0 & 0 & $\{J\gamma_0,\tilde{T},J\tilde{T};\gamma_1,\ldots,\gamma_{d-1},\Sigma_z\}$ & $Cl_{2,d}\to Cl_{3,d}$ & $R_{4-d}$ \\
  CII & $-1$ & $-1$ & 1 & $\{J\gamma_0,\tilde{T},J\tilde{T},\tilde{\Gamma}\Sigma_z;\gamma_1,\ldots,\gamma_{d-1},\Sigma_z\}$ & $Cl_{3,d}\to Cl_{4,d}$ & $R_{5-d}$ \\
  C & 0 & $-1$ & 0 & $\{J\gamma_0,J\Sigma_z,\tilde{C},J\tilde{C};\gamma_1,\ldots,\gamma_{d-1}\}$ & $Cl_{3,d-1}\to Cl_{4,d-1}$ & $R_{6-d}$ \\
  CI & $+1$ & $-1$ & 1 & $\{J\gamma_0;\gamma_1,\ldots,\gamma_{d-1},\Sigma_z,\tilde{T},J\tilde{T},\tilde{\Gamma}\Sigma_z\}$ & $Cl_{0,d+3}\to Cl_{1,d+3}$ & $R_{7-d}$ \\
  \hline\hline
\end{tabular}
\label{tb:Cl_alg_HU}
\end{table*}

\begin{table*}[t]
\centering
\caption{Clifford algebra extensions and classifying spaces for ${\cal H}_L(\bm{k},t)$.
}
\begin{tabular}{ccccccc}
  \hline\hline
  ~AZ class~~~ & ~~$T$~~ & ~~$C$~~ & ~~$\Gamma$~~~ & ~~~~~~~~~~~~~~~~Generator~~~~~~~~~~~~~~~~ & ~~~~~~~~~~~Extension~~~~~~~~~ & ~Classifying space~ \\
  \hline
  A & 0 & 0 & 0 & $\{\gamma_1,\ldots,\gamma_d,\gamma_t, \gamma_0,\Sigma_z\}$ & $Cl_{d+2}\to Cl_{d+3}$ & $C_{d}$ \\
  AIII & 0 & 0 & 1 & $\{\gamma_1,\ldots,\gamma_d,\gamma_t,\gamma_0,\Sigma_z,J\gamma_t\tilde{\Gamma}\}$ & $Cl_{d+3}\to Cl_{d+4}$ & $C_{d+1}$ \\
  \hline
  AI & $+1$ & 0 & 0 & $\{J\gamma_0,J\Sigma_z;\gamma_1,\ldots,\gamma_d,\gamma_t,\tilde{T},J\tilde{T}\}$ & $Cl_{1,d+3}\to Cl_{2,d+3}$ & $R_{-d}$ \\
  BDI & $+1$ & $+1$ & 1 & $\{J\gamma_0,J\Sigma_z,\gamma_t\tilde{\Gamma};\gamma_1,\ldots,\gamma_d,\gamma_t,\tilde{T},J\tilde{T}\}$ & $Cl_{2,d+3}\to Cl_{3,d+3}$ & $R_{1-d}$ \\
  D & 0 & $+1$ & 0 & $\{J\gamma_0,J\gamma_t,J\Sigma_z;\gamma_1,\ldots,\gamma_d,\tilde{C},J\tilde{C}\}$ & $Cl_{2,d+2}\to Cl_{3,d+2}$ & $R_{2-d}$ \\
  DIII & $-1$ & $+1$ & 1 & $\{J\gamma_0,J\Sigma_z,\tilde{T},J\tilde{T};\gamma_1,\ldots,\gamma_d,\gamma_t,\gamma_t\tilde{\Gamma}\}$ & $Cl_{2,d+2}\to Cl_{3,d+2}$ & $R_{2-d}$ \\
  AII & $-1$ & 0 & 0 & $\{J\gamma_0,J\Sigma_z,\tilde{T},J\tilde{T};\gamma_1,\ldots,\gamma_d,\gamma_t\}$ & $Cl_{3,d+1}\to Cl_{4,d+1}$ & $R_{4-d}$ \\
  CII & $-1$ & $-1$ & 1 & $\{J\gamma_0,J\Sigma_z,\gamma_t\tilde{\Gamma},\tilde{T},J\tilde{T};\gamma_1,\ldots,\gamma_d,\gamma_t\}$ & $Cl_{4,d+1}\to Cl_{5,d+1}$ & $R_{5-d}$ \\
  C & 0 & $-1$ & 0 & $\{J\gamma_0,J\gamma_t,J\Sigma_z,\tilde{C},J\tilde{C};\gamma_1,\ldots,\gamma_d\}$ & $Cl_{4,d}\to Cl_{5,d}$ & $R_{6-d}$ \\
  CI & $+1$ & $-1$ & 1 & $\{J\gamma_0,J\Sigma_z;\gamma_1,\ldots,\gamma_d,\gamma_t,\gamma_t\tilde{\Gamma},\tilde{T},J\tilde{T}\}$ & $Cl_{1,d+4}\to Cl_{2,d+4}$ & $R_{7-d}$ \\
  \hline\hline
\end{tabular}
\label{tb:Cl_alg_HL}
\end{table*}

We perform the topological classification of ${\cal H}_U(\bm{k}_\parallel)$ as the extension problem of Clifford algebras.
Since it is enough to classify the system near topological phase transitions,
we consider ${\cal H}_U({\bm k}_\parallel)$
in the form of the Dirac Hamiltonian
\begin{align}
    {\cal H}_U(\bm{k}_\parallel)=\sum_{j=1}^{d-1} k_j \gamma_j + \gamma_0,
\end{align}
where $\gamma_j$ and $\gamma_0$ are the gamma matrices that satisfy the relation $\{\gamma_\mu,\gamma_\nu\}=\delta_{\mu,\nu}$ ($\mu,\nu=0,1,\ldots,d-1$).
Then, the gamma matrices $\gamma_\mu$, symmetry operations for ${\cal H}_U({\bm k}_\parallel)$ in Eqs.~(\ref{eq:properCS_HU})-(\ref{eq:CS_HU}), and the imaginary unit $J:=i$ if ${\cal H}_U({\bm k}_\parallel)$ has anti-unitary symmetry, form the Clifford algebra in Table \ref{tb:Cl_alg_HU}.
Possible mass terms $\gamma_0$ in the Dirac Hamiltonian provide possible topological phases, and thus the classification reduces to the extension problem of the Clifford algebra without $\gamma_0$ to that with $\gamma_0$. 
For each symmetry class, we summarize the extension and the classifying space in Table \ref{tb:Cl_alg_HU}.
Using Tables~\ref{tb:Cn} and \ref{tb:Rq}, we can specify the connected component $\pi_0$ of the classification space, 
which gives the topological numbers in Table~\ref{tb:extrinsic} in the main text.

\section{Topological classification of $L(\bm{k},t)$}
\label{sec:classifyL}

In a manner similar to Appendix \ref{sec:classifyU}, we can perform the topological classification of $L({\bm k}, t)$ by using ${\cal H }_L(\bm{k},t)$ under symmetry in Eqs.~(\ref{eq:properCS_HL})-(\ref{eq:CS_HL}).
In Table~\ref{tb:Cl_alg_HL},
we summarize the Clifford algebra, the extension by adding the mass term $\gamma_0$, and the classifying space for the "Hamiltonian"
\begin{align}
    {\cal H}_L(\bm{k},t)=\sum_{j=1}^d k_j\gamma_j + t\gamma_t + \gamma_0,
\end{align}
for each symmetry class with symmetry in Eqs.~(\ref{eq:properCS_HL})-(\ref{eq:CS_HL}). 
Note that the obtained classifying spaces are the same as those for ${\cal H}_U({\bm k}_\parallel)$ in Table \ref{tb:Cl_alg_HU}.
Therefore, the topological classification of $L(\bm{k},t)$ coincides with that of $U({\bm k}_\parallel)$.

\section{Extrinsic boundary states of quantum walks in two and three dimensions}
\label{sec:example_higher}
In this appendix, we present examples of extrinsic boundary unitary operators in 2D and 3D.
(See also Sec.~\ref{sec:simple}, where we have given an example of boundary unitary operators in 2D class A quantum walks.
)

\subsection{class AII in 2D}

An extrinsic boundary unitary operator $U_{\text{BDQW}}(k)$ of a 2D class AII quantum walk satisfies 
\begin{align}
T U_{\text{BDQW}}(k) T^{-1}=U_{\text{BDQW}}^\dag (-k),
\end{align}
where $k$ is the momentum along the boundary of the 2D quantum walk, $T$ is the time-reversal anti-unitary operator with $T^2=-1$. 
Decomposing $T$ into the unitary part ${\cal T}$ and the complex conjugation operator $K$,
\begin{align}
    T=\mathcal{T}K,
\end{align}
we find that $U_{\text{BDQW}}(k_\text{TRIM})\mathcal{T}$ at time reversal invariant momenta $k_\text{TRIM}=0,\pi$ is antisymmetric. 
Thus, we can introduce the Pfaffian ${\rm Pf}(U_{\text{BDQW}}(k_\text{TRIM})\mathcal{T})$ and define the following $\mathbb{Z}_2$ topological invariant $n$ for the boundary unitary operator,
\begin{eqnarray}
(-1)^{n}=
{\rm sgn}\left\{\frac{{\rm Pf}[U_{\text{BDQW}}(\pi)\mathcal{T}]}{{\rm Pf}[U_{\text{BDQW}}(0)\mathcal{T}]}\right. \hspace{50pt} 
\nonumber \\
\left.\times\exp\left[-\frac{1}{2}\int_{k=0}^{k=\pi}
d\log\det[U_{\text{BDQW}}(k)\mathcal{T}] \right]
\right\}.
\label{eq:nF_1dAII}
\end{eqnarray} 
When this number is non-trivial, the boundary operator hosts a Kramers pair of gapless modes in accordance with the extended Nielsen-Ninommiya theorem in Eq.~(\ref{eq:NN}). 
The presence or absence of the Kramers pair at $\epsilon=0,\pi$
defines the $\mathbb{Z}_2$ invariant $\nu^{0,\pi}$.

We can obtain a non-trivial example of the boundary operator in this class by using a $2\times 2$ unitary matrix.
For $T=i\sigma_2 K$, a simplest nontrivial model is given by
\begin{align}
    U_{\text{BDQW}}(k)=\cos k \sigma_0 +i \sin k \sigma_2.
\end{align}
Because we have
\begin{align}
    &U_{\text{BDQW}}(0)\mathcal{T}=i\sigma_2, \quad
    U_{\text{BDQW}}(\pi)\mathcal{T}=-i\sigma_2, 
    \nonumber \\
    &\det [U(k)\mathcal{T}]=-1,
\end{align}
this model gives $n=1$ (mod. $2$) in Eq.~(\ref{eq:nF_1dAII}). 
On the other hand, 
as the eigenvalues of this model are
\begin{align}
    \lambda_\pm (k)= e^{\pm ik},
\end{align}
the quasi-energies of the boundary operator are $\epsilon_\pm = \pm k$.
Thus, we have a Kramers pair of gapless modes both at $\epsilon=0, \pi$. 
As a result, this model obeys the extended Nielsen-Ninomiya theorem in Eq.~(\ref{eq:NN}) as
\begin{align}
    \sum \nu^0 = \sum \nu^\pi = n = 1.
\end{align}

\subsection{class AIII in 3D}

A boundary unitary operator $U_{\text{BDQW}}(\bm{k})$ of a 3D class AIII quantum walk has chiral symmetry
\begin{align}
\Gamma U_{\text{BDQW}}(\bm{k}) \Gamma^{-1}=U_{\text{BDQW}}^\dag (\bm{k}),
\end{align}
where ${\bm k}$ is the momentum of a 2D boundary of the quantum walk, $\Gamma$ is a unitary operator with $\Gamma^2=1$.
From this, $U_{\text{BDQW}}(\bm{k})\Gamma$ 
is a Hermitian operator, which has a gap at zero energy because of $\det [U_{\text{BDQW}}(\bm{k})\Gamma]\neq 0$.
Therefore, we can define the Chern number of $U_{\text{BDQW}}(\bm{k})\Gamma$, which gives the topological number $n$ in Eq.~(\ref{eq:NN}) for the boundary operator.
When $n$ is nonzero, the boundary unitary operator supports gapless Dirac points at $\epsilon=0, \pi$.
The topological charge $\nu^{0,\pi}$ for the Dirac points is
\begin{align}
\nu^\epsilon=\int_{S^1} \frac{d \bm{k}}{4\pi i}
\cdot \tr \left[\Gamma (H_{\text{F}}(\bm{k})-\epsilon)^{-1} \nabla (H_{\text{F}}(\bm{k})-\epsilon) \right],
\end{align}
where $S^1$ is a circle surrounding the Dirac points at $\epsilon=0,\pi$.

To obtain a non-trivial example of the boundary operator in this class, we need at least a $2\times 2$ unitary matrix.
For $\Gamma=\sigma_3$, we obtain the following nontrivial model \cite{BS20},
\begin{align}
U_{\text{BDQW}}(\bm{k}) =e^{i\theta \sigma_x/2}U_{y}^{-}(k_y/2) U_{x}^{-}(k_x) U_{y}^{+}(k_y/2)
\nonumber \\
\times U_{y}^{-}(k_y/2) U_{x}^{+}(k_x) U_{y}^{+}(k_y/2)
e^{i\theta \sigma_x/2},
\label{eq:AIII2D}
\end{align}
where
$U_{j}^{ \pm}(k_j) =P_{j}^{ \pm} e^{\mp i k_{j}}+P_{j}^{\mp}$ with $P_{j}^{ \pm}=\left(\sigma_{0} \pm \sigma_{j}\right) / 2$, 
and $\theta$ is a real parameter.
\begin{figure}[t]
\centering
\includegraphics[width=55mm]{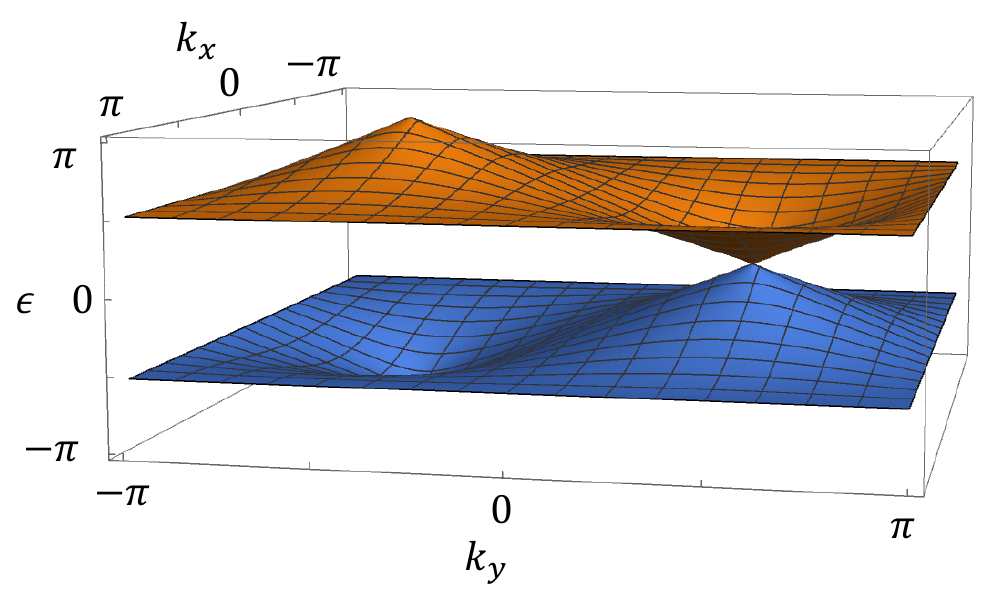} 
\caption{
The energy spectrum of Eq.~(\ref{eq:AIII2D}) at $\theta=3/4$.
There are Dirac points at $\epsilon=0$ and $\epsilon=\pi$.
}
	\label{fig:AIII_2D}
\end{figure}
We show the quasi-energy spetrum of $U_{\text{BDQW}}(\bm{k})$ in Fig.~\ref{fig:AIII_2D}.
A Dirac point exists both at $\epsilon=0$ and $\epsilon=\pi$.
We can confirm the relation in Eq.~(\ref{eq:NN}) as follows.
Near the Dirac point at $\epsilon=0$, the effective Hamiltonian has the form
\begin{align}
H_{\text{BDQW}}({\bm k})\approx (k_x-\theta) \sigma_x + k_y \left(\cos^2 \frac{\theta}{2}\right) \sigma_y,
\end{align}
which gives the topological charge $\nu^0=1$.
Near the Dirac point at $\epsilon=\pi$, the effective Hamiltonian has the form
\begin{align}
H_{\text{BDQW}}({\bm k})\approx (k_x-\theta) \sigma_x + k_y \left(\cos^2 \frac{\theta}{2}\right) \sigma_y,
\end{align}
which give the topological charge $\nu^\pi=-1$.
On the other hand, the Hermitian operator $U_{\rm BDQW}\Gamma$ takes the form 
\begin{align}
U_{\text{BDQW}}({\bm k})\Gamma
&=d_x({\bm k})\sigma_x+ (d_y({\bm k}) \cos \theta-d_z({\bm k})\sin \theta)\sigma_y
\nonumber \\
&+(d_z({\bm k})\cos\theta +d_y({\bm k})\sin\theta)\sigma_z
\end{align}
where
\begin{align}
\left\{
\begin{array}{l}
d_x({\bm k})=\cos^2 \left( k_x/2 \right)\sin k_y, \\
d_y({\bm k})=\sin k_x \cos^2 \left(k_y/2\right), \\
d_z({\bm k})=\cos k_x \cos^2 \left(k_y/2\right) -\sin^2 \left(k_y/2\right).
\end{array}
\right.
\end{align}
The Chern number of $U_{\text{BDQW}}({\bm k})\Gamma$ is equal to 1
because the unit vector $\bm{d}=(d_x,d_y,d_z)$ wraps a unit sphere one when ${\bm k}$ covers the whole 2D Brillouin zone, indicating $n=1$.
As a result, this model obeys the extended Nielsen-Ninomiya theorem in Eq.~(\ref{eq:NN}) as
\begin{align}
    \sum \nu^0 = -\sum \nu^\pi = n = 1.
\end{align}

\vspace{10pt} 

\section{The winding number $w_1[a]$, $w_1[b]$, $w_1[c]$, $w_1[d]$ for class CII in 1D}
\label{sec:CII_2Zwinding}

In this section, we show that the winding number $w_1[a]$ in Sec.~\ref{sec:CSQW} 
takes an even integer for class CII in 1D with decomposed TRS and PHS in Eqs.~(\ref{eq:TRS_CII}) and (\ref{eq:PHS_CII}).
In a similar manner, one can show that $w_1[b]$ $w_1[c]$, and $w_1[d]$ are also even integers.

For this purpose, we take the basis where CS and PHS are given by

\begin{align}
    \Gamma=\pmat{\hat{1} & 0 \\ 0 & -\hat{1}}, 
    \quad
    C = \pmat{\sigma_2 & 0 \\ 0 & -\sigma_2}K.
\end{align}
Then, PHS in Eq.~(\ref{eq:PHS_CII}) leads to
\begin{align}
    \pmat{ \sigma_2 a^*(k) \sigma_2 & -\sigma_2 b^*(k) \sigma_2 \\
    -\sigma_2 c^*(k) \sigma_2 &
    \sigma_2 d^*(k) \sigma_2}
    =
    \pmat{a(-k) & b(-k) \\
    c(-k) & d(-k)}.
\label{eq:PHS_abcd}
\end{align}
When the system has a gap at $\epsilon=0$, it holds that $\det [a(k)] \neq 0$ \cite{Mochizuki20}, and thus one can define the winding number $w_1[a]$.
To show that $w_1[a]$ is an even integer, 
we first deform $a(k)$ into a unitary matrix. During the deformation, we can retain the condition $\det [a(k)] \neq 0$, and so $w_1[a]$ does not change.
After the unitarization, $a(k)$ is diagonalizable.
Because of $\sigma_2 a^* (k) \sigma_2 = a(-k)$ in Eq.~(\ref{eq:PHS_abcd}), any eigenstate of $a(k)$ has  a Kramers partner, and thus $a(k)$ is written as
\begin{align}
    a(k) = &\sum_{n}\big[\lambda_n(k) \ket{\psi_n(k)}\bra{\psi_n(k)} 
    \nonumber \\
    &+ \lambda_n^*(-k) \sigma_2 \ket{\psi_n^*(-k)}\bra{\psi_n^*(-k)}\sigma_2\big],
\end{align}
where $|\psi_n(k)\rangle$ is an eigenstate of $a(k)$ with the eigenvalue $\lambda_n(k)$.
Then, we have
\begin{align}
\tr[a(k)^\dag \partial_k a(k)]
&=\partial_k \log \det [a(k)] 
\nonumber \\
&= \partial_k \log
\left[\prod_n \lambda_n(k) \lambda_n^*(-k)\right]
\nonumber \\
&= \partial_k \left[\sum_n \log
 \lambda_n(k) + \sum_n \log \lambda_n^*(-k)\right],
\end{align}
which leads to
\begin{align}
w_1[a]&=\int_0^{2\pi i} \frac{dk}{2\pi} \tr[a(k)^\dag \partial_k a(k)]
\nonumber \\
&= \int_0^{2\pi i} \frac{dk}{2\pi}  \partial_k \left[\sum_n \log
 \lambda_n(k) + \sum_n \log \lambda_n^*(-k)\right]
\nonumber \\
&=2 \int_0^{2\pi i} \frac{dk}{2\pi} \partial_k \sum_n \log \lambda_n(k).
\end{align}
Because the integral
\begin{align}
\int_0^{2\pi i} \frac{dk}{2\pi} \partial_k \sum_n \log \lambda_n(k)
\end{align}
is an integer due to the periodicity of $a(k)$ in  $k$, 
$w_1[a]$ is an even integer.

\section{Proof of Eq.~(\ref{eq:dP=wPhi})}
\label{sec:dP=wPhi}
In this section, we prove the formula in Eq.~(\ref{eq:dP=wPhi}) in the main text.
For a rigorous argument, 
instead of the original quantum walk operator $\hat{U}$ having $L$ sites in the $x$-direction, 
we consider $N$ copies of $\hat{U}$ arranged in a line in the $x$-direction, which is denoted by $\hat{U}_N$.
Whereas $\hat{U}$ does not have translation symmetry in the presence of onsite disorders, $\hat{U}_N$ has translation invariance with respect to $L$-sites translations.
We also assume the periodic boundary condition of $\hat{U}_N$ in the $x$-direction.
Below we show that the position displacement after one-cycle evolution by $\hat{U}_N$ is equal to the winding number of $\hat{U}(\Phi)$:
\begin{align}\label{eq:dP=wPhiS}
   & \sum_{x=1}^L \sum_\alpha \bra{x,\alpha} \hat{U}_N^\dag  \frac{\hat{x}}{L} \hat{U}_N \ket{x,\alpha}
    - \sum_{x=1}^L \sum_\alpha \bra{x,\alpha} \frac{\hat{x}}{L} \ket{x,\alpha}
    \nonumber \\
    &=
    \int_0^{2\pi} \frac{d\Phi}{2\pi} \tr \left[ \hat{U}^\dag(\Phi) i \partial_\Phi \hat{U} (\Phi) \right].
\end{align}
The right hand side of the above equation defines $P(T)-P$, {\it i.e.} an averaged version of $P_x(T)-P_x$, 
and thus Eq.~(\ref{eq:dP=wPhiS}) is the exact description of Eq.~(\ref{eq:dP=wPhi}) in the main text.

To show the above equation, we introduce $\hat{U}_N(\Phi)$ by
\begin{align}
    \hat{U}_N (\Phi) = e^{-i(\Phi/L)\hat{x}} \hat{U}_N e^{i(\Phi/L)\hat{x}},
\end{align}
where $\Phi$ takes the discrete values of $\Phi=2\pi p/N$ with integers $p$.
For these values of $\Phi$, 
$\hat{U}_N(\Phi)$ keeps the same periodic boundary condition as $\hat{U}_N$, and equals to the flux inserted $\hat{U}_N$ by the uniform gauge potential $A_x=\Phi/L$. 
Below, we assume the large $N$ limit, and
treat $\Phi$ as a continuous real number.

The flux inserted operator obeys the Heisenberg equation
\begin{align}
    i \partial_\Phi \hat{U}_N (\Phi) = \left[ \frac{\hat{x}}{L} , \hat{U}_N \right].
\end{align}
Therefore, we obtain
\begin{align}
    \hat{U}_N^\dag(\Phi) i \partial_\Phi \hat{U}_N (\Phi) = \hat{U}_N^\dag (\Phi) \frac{\hat{x}}{L} \hat{U}_N (\Phi) - \frac{\hat{x}}{L},
\end{align}
which leads to
\begin{align}
    \sum_{x=1}^L \sum_\alpha \bra{x,\alpha} \hat{U}_N^\dag(\Phi) i \partial_\Phi \hat{U}_N (\Phi) \ket{x,\alpha}
    \nonumber \\
    = \sum_{x=1}^L \sum_\alpha \bra{x,\alpha} \hat{U}_N^\dag (\Phi) \frac{\hat{x}}{L} \hat{U}_N (\Phi) \ket{x,\alpha} 
    \nonumber \\
    - \sum_{x=1}^L \sum_\alpha \bra{x,\alpha}  \frac{\hat{x}}{L} \ket{x,\alpha}.
    \label{eq:magnetic_route1}
\end{align}
From translation symmetry of $\hat{U}_N(\Phi)$ under $L$ sites translations ${\cal T}_L$,
\begin{align}
\mathcal{T}_L^\dag \hat{U}_N(\Phi) \mathcal{T}_L = \hat{U}_N (\Phi),     
\end{align}
the left hand side of Eq.~(\ref{eq:magnetic_route1}) becomes
\begin{align}
    &\sum_{x=1}^L \sum_\alpha \bra{x,\alpha} \hat{U}_N^\dag(\Phi) i \partial_\Phi \hat{U}_N (\Phi) \ket{x,\alpha}
    \nonumber \\
    =& \frac{1}{N} \sum_{x=1}^L \sum_{n=1}^N \sum_\alpha \bra{x,\alpha} \left(\mathcal{T}_L^\dag\right)^n \hat{U}_N^\dag(\Phi) i \partial_\Phi \hat{U}_N (\Phi) \left( \mathcal{T}_L \right)^n \ket{x,\alpha}
    \nonumber \\
    =& \frac{1}{N} \sum_{x=1}^{NL} \sum_\alpha \bra{x,\alpha} \hat{U}_N^\dag(\Phi) i \partial_\Phi \hat{U}_N (\Phi) \ket{x,\alpha}
    \nonumber \\
    =& \frac{1}{N} \tr_N \left[ \hat{U}_N^\dag(\Phi) i \partial_\Phi \hat{U}_N (\Phi) \right],
\end{align}
where ${\rm tr}_N$ is the trace over the Hilbert space for $\hat{U}_N$.
On the other hand, the first term of the right hand side in Eq.~(\ref{eq:magnetic_route1}) is recast into
\begin{align}
    & \sum_{x=1}^L \sum_\alpha \bra{x,\alpha} \hat{U}_N^\dag (\Phi) \frac{\hat{x}}{L} \hat{U}_N (\Phi) \ket{x,\alpha}
    \nonumber \\
    =& \sum_{x=1}^L \sum_\alpha \bra{x,\alpha} e^{-i(\Phi/L)\hat{x}} \hat{U}_N^\dag e^{i(\Phi/L)\hat{x}} \frac{\hat{x}}{L} e^{-i(\Phi/L)\hat{x}} 
    \nonumber \\
    & \quad \quad \quad \times \hat{U}_N (\Phi) e^{i(\Phi/L)\hat{x}} \ket{x,\alpha}
    \nonumber \\
    =& \sum_{x=1}^L \sum_\alpha \bra{x,\alpha} \hat{U}_N^\dag  \frac{\hat{x}}{L} \hat{U}_N \ket{x,\alpha}.
\end{align}
Therefore, Eq.~(\ref{eq:magnetic_route1}) leads to
\begin{align}
    & \frac{1}{N} \int_0^{2\pi} \frac{d\Phi}{2\pi} \tr_N \left[ \hat{U}_N^\dag(\Phi) i \partial_\Phi \hat{U}_N (\Phi) \right]
    \nonumber \\
    =& \sum_{x=1}^L \sum_\alpha \bra{x,\alpha} \hat{U}_N^\dag  \frac{\hat{x}}{L} \hat{U}_N \ket{x,\alpha}
    \nonumber \\
    &- \sum_{x=1}^L \sum_\alpha \bra{x,\alpha} \frac{\hat{x}}{L} \ket{x,\alpha}.
\end{align}
Finally, it holds that \footnote{Note that Eq.~(\ref{eq:trNtr}) can be easily shown for $U(k)=e^{\pm ik}$. In general, any model can be continuously deformed into a direct product of $U(k)=e^{\pm ik}$ up to addiction or subtraction of trivial models, and thus Eq.~(\ref{eq:trNtr}) holds. 
}
\begin{align}
    \frac{1}{N} \int_0^{2\pi} \frac{d\Phi}{2\pi} \tr_N \left[ \hat{U}_N^\dag(\Phi) i \partial_\Phi \hat{U}_N (\Phi) \right]
    \nonumber \\
    = 
    \int_0^{2\pi} \frac{d\Phi}{2\pi} \tr \left[ \hat{U}^\dag(\Phi) i \partial_\Phi \hat{U} (\Phi) \right],
\label{eq:trNtr}
\end{align}
and thus we have Eq.~(\ref{eq:dP=wPhiS}).

\section{Proof of Eq.~(\ref{eq:dP=wP})}
\label{sec:dP=wP}

The polarization at $t=T$ is rewritten as
\begin{align}
    &P_x (T)|_{y=1} 
    = \sum_{\alpha} \bra{x,y=1,\alpha} \hat{U}^\dag \hat{x} \hat{U} \ket{x,y=1,\alpha}
    \nonumber \\
    &= \sum_{\alpha} \left[ \frac{1}{\sqrt{L_x}} \sum_{k_x} e^{ik_x x} \bra{k_x} \right]\bra{y=1,\alpha} \hat{U}^\dag \left[ \sum_{x'} x' \ket{x'}\bra{x'} \right]
    \nonumber \\
    & \quad \times \hat{U}  \left[ \frac{1}{\sqrt{L_x}} \sum_{k'_x} e^{-ik'_x x} \ket{k'_x} \right] \ket{y=1,\alpha}
    \nonumber \\
    &= \frac{1}{L_x^2} \sum_{\alpha,k_x,k'_x,x'} e^{ik_x x} \bra{y=1,\alpha} U^\dag(k_x) e^{-ik_x x'} [-i\partial_{k'_x} e^{ik'_x x'}]
    \nonumber \\
    & \quad \times U(k'_x) e^{-ik'_x x} \ket{y=1,\alpha}
    \nonumber \\
    &= \frac{1}{L_x} \sum_{\alpha,k_x} e^{ik_x x} \bra{y=1,\alpha} U^\dag(k_x) i\partial_{k_x}[ U(k_x) e^{-ik_x x}] \ket{y=1,\alpha}
    \nonumber \\
    &= \frac{1}{L_x} \sum_{\alpha,k_x} \bra{y=1,\alpha} U^\dag(k_x) [i\partial_{k_x} U(k_x)] \ket{y=1,\alpha}
    + \sum_{\alpha} x
    \nonumber \\
    &= \int_0^{2\pi} \frac{d k_x}{2\pi} \tr_{y,\alpha} \left[ \hat{P}_{\rm edge} U^\dag(k_x) i\partial_{k_x} U(k_x) \right] + P_x|_{y=1},
\end{align}
with $\hat{P}_{\rm edge}=|y=1\rangle\langle y=1|$, 
which means Eq.~(\ref{eq:dP=wP}).

To check Eq.~(\ref{eq:dP=wP}), we consider the  model with 2 sites in the $y$-direction,
\begin{align}
    &U(k_x)
    \nonumber\\
    &= e^{-ik_x} \cos \theta \ket{y=1}\bra{y=1}
    - e^{-ik_x} \sin \theta \ket{y=1}\bra{y=2}
    \nonumber\\
    &
    + \sin \theta \ket{y=2}\bra{y=1}
    + \cos \theta \ket{y=2}\bra{y=2}.
    \label{eq:2site}
\end{align}
The model provides one-site displacement in the $x$-direction when a particle is located at $y=1$ after the one-cycle time evolution.
If the particle starts at $y=1$, only the first term in the right hand side of Eq.~(\ref{eq:2site}) gives the one-site displacement. 
Thus, the particle is expected to move in the $x$-direction at rate $\cos^2\theta$.

Equation (\ref{eq:dP=wP}) reproduces this result.  
Actually, from direct calculations
\begin{align}
    \hat{P}_{\rm edge} U^{\dagger}(k_x) 
    &= e^{ik_x} \cos \theta \ket{y=1}\bra{y=1} 
    \nonumber \\
    &+ \sin \theta \ket{y=1}\bra{y=2},
    \nonumber \\
    i \partial_{k_x} U(k_x)
    &=  e^{-ik_x} \cos \theta \ket{y=1}\bra{y=1}
    \nonumber\\
    &-e^{-ik_x}\sin\theta\ket{y=1}\bra{y=2},
\end{align}
we have
\begin{align}
    w_P[U(k_x)] &= \int_0^{2\pi} \frac{\mathrm{d}k_x}{2\pi} \tr_{y,\alpha} \left[ \hat{P}_{\rm edge} U^\dag(k_x) i\partial_{k_x} U(k_x) \right]
    \nonumber \\
    &=
    \int_0^{2\pi} \frac{\mathrm{d}k_x}{2\pi} \tr_{y,\alpha} \left[  \cos^2 \theta \ket{y=1}\bra{y=1} \right] 
    \nonumber\\
    &= \cos^2 \theta.
\end{align}

\section{Effects of random phases of the boundary unitary operator in the Anderson model}
\label{sec:random_phase}
\begin{figure}[thbp]
\centering
\includegraphics[width=85mm]{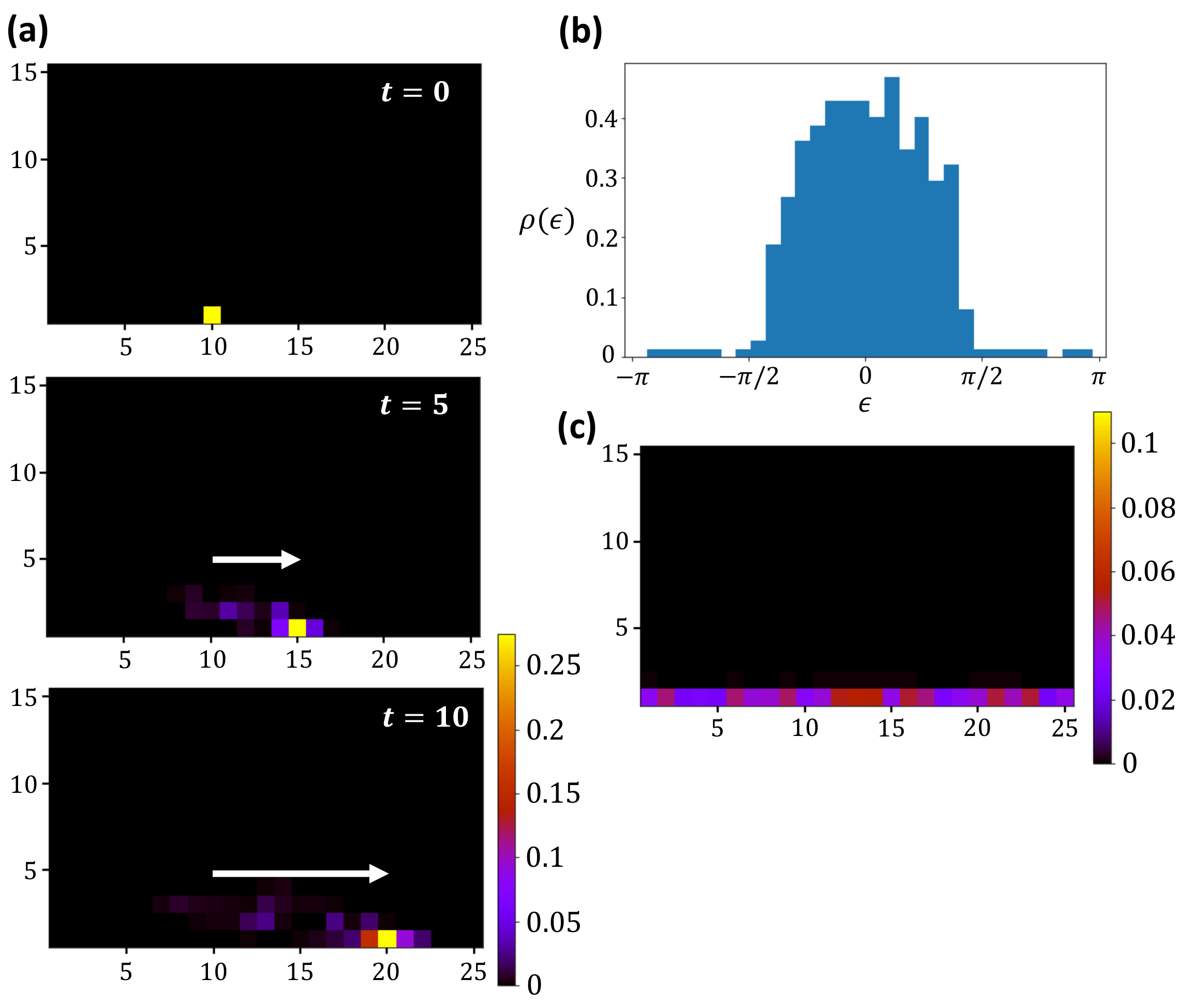} 
\caption{(a) Dynamics of a wave packet, (b) DOS and (c) a delocalized eigenstate of the Anderson model with  the disordered extrinsic edge mode
in Eq.~(\ref{eq:Anderson_edge_phase}). The parameters are $J=0.2$, $W=1$, $L_x=25$, $L_y=15$ and $T=1$. The color scale is different between (a) and (c). The initial state of the wave packet is $\ket{x=10,y=1}$. 
We take the periodic (open) boundary condition in the $x$ ($y$) direction.
Even in the presence of random phases on the boundary unitary operator, the anomalous edge state survives, and enables a directed wave packet motion.
}
	\label{fig:Anderson_A_phase}
\end{figure}

In this section, we numerically check that the extrinsic edge mode in Eq.~(\ref{eq:Anderson_edge}) is robust against random phases on the edge.
For this purpose, we consider $U_A''$ below, instead of $U_A'$,
\begin{align}
    U_{A}'' &= A' \cdot e^{-iH_{A}T},
    \nonumber \\
    A' &= \sum_{x=1}^{L_x} e^{-i\phi_x}\ket{x+1}\bra{x}\otimes \ket{y=1}\bra{y=1}
    \nonumber \\
    & + \sum_{x=1}^{L_x} \sum_{y=2}^{L_y} \ket{x}\bra{x}\otimes \ket{y}\bra{y},
    \label{eq:Anderson_edge_phase}
\end{align}
where $\phi_x$ is uniformly distributed in a range $[0,2\pi]$.
The resultant dynamics of a wave packet starting at an edge of the system, the DOS histogram, and a typical eigenstate profile are shown in Fig. \ref{fig:Anderson_A_phase}.  
We can see the directed wave packet motion along $y=1$ occurs even in this case.
Actually, this behavior is assured by the formulae Eqs.~(\ref{eq:dP=wPhi}) and (\ref{eq:dP=wP}).

\section{Diffusive behavior of the time-dependent Anderson model}
\label{sec:diffusion}

Here we show diffusive behaviors of our time-dependent Anderson models in Eqs.~(\ref{eq:time-dep_Anderson}) and (\ref{eq:time-dep_Anderson_edge}). 
For this purpose, we numerically examine $\rho(y,t)$ and $\langle y^2 \rangle$ defined by
\begin{align}
&\rho(y,t)=\int \rho(x,y,t) dx,
\nonumber\\
&\langle y^2\rangle= \int y^2 \rho(x,y,t) dx dy,
\end{align}

\noindent
where $\rho(x,y, t)=|\psi(x,y,t)|^2$ is the density distribution.
If $\rho(x,y)$ obeys the diffusion equation
\begin{align}
    \frac{\partial \rho}{\partial t} = \frac{D}{2}\left(
    \frac{\partial^2 \rho}{\partial x^2}
    +\frac{\partial^2 \rho}{\partial y^2}
    \right),
\end{align}

\begin{figure}[H]
\centering
\includegraphics[width=86mm]{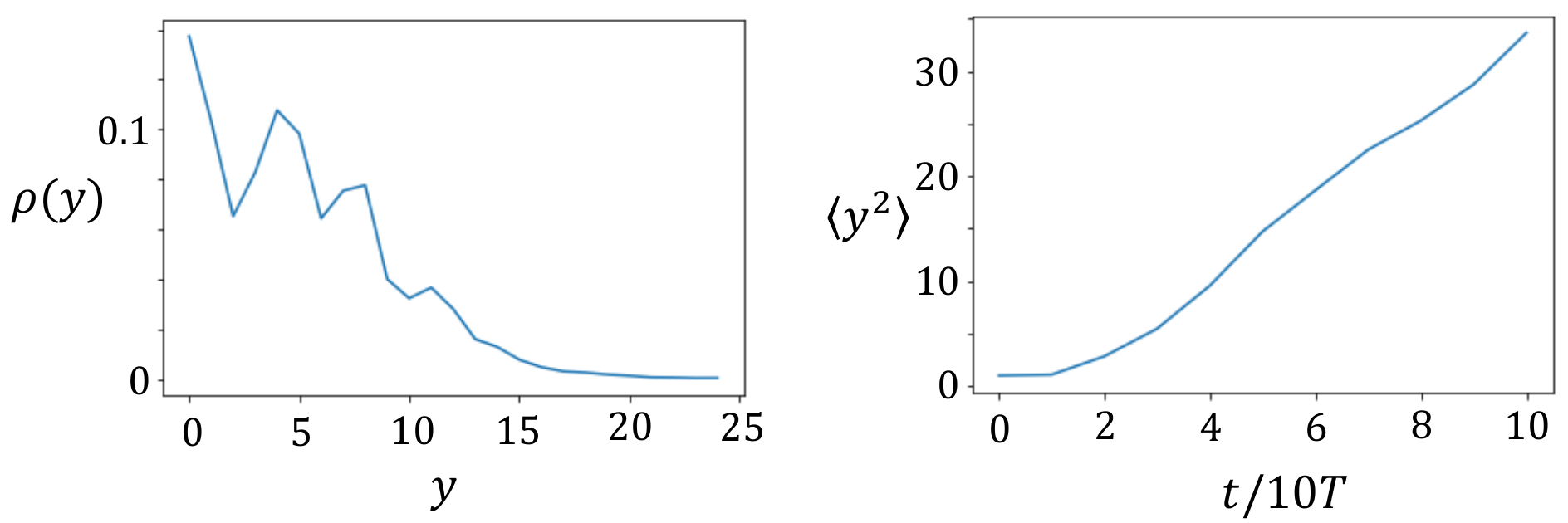}
\caption{Density distribution $\rho(y,t)$ at $t=100T$ and the mean squared distance $\braket{y^2}$ of the time-dependent Anderson model in Eq.~(\ref{eq:time-dep_Anderson}). Parameters are $L_x =40$, $L_y=25$, $J = 0.2$, $W = 1$, and $T = 1$. The initial state of the wave packet is $\ket{x=20, y=1}$. The density distribution $\rho(y,t)$ exhibits a Gaussian tail, and the mean squared distance $\braket{y^2}$ shows a nearly linear behavior in time.} 
	\label{fig:time-dep_Anderson}
\end{figure}
\begin{figure}[H]
\centering
\includegraphics[width=86mm]{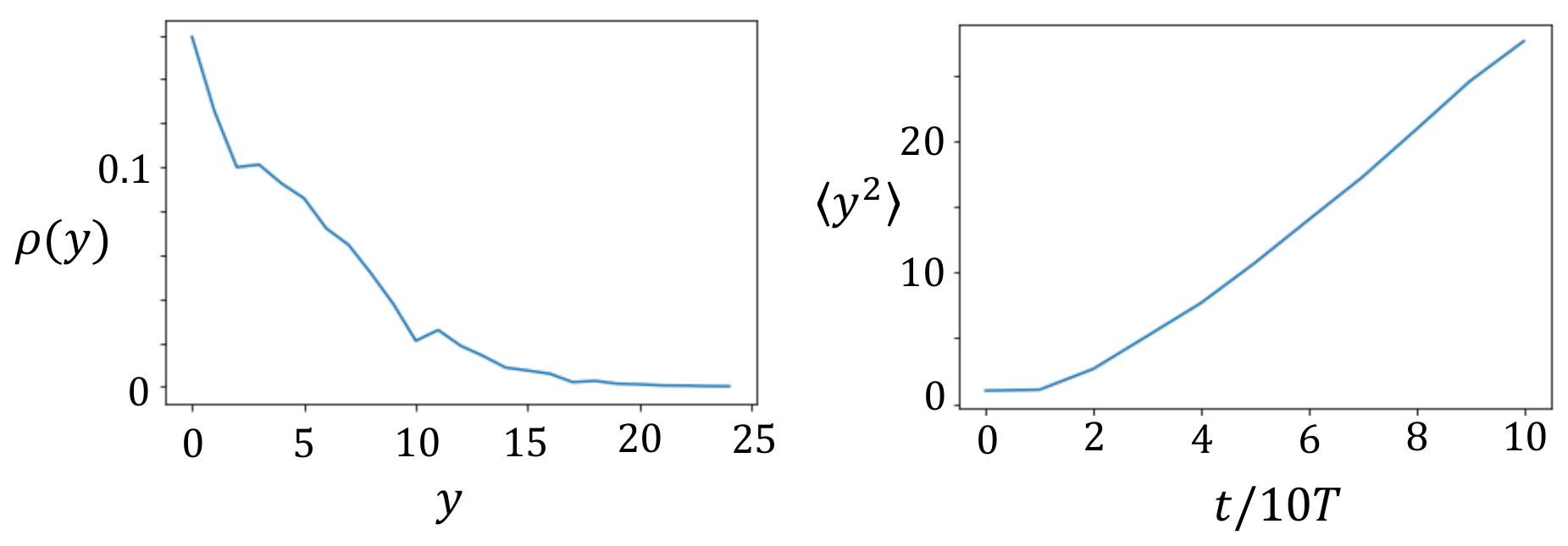}
\caption{Density distribution $\rho(y,t)$ at $t=100T$ and the mean squared distance $\braket{y^2}$ of the time-dependent Anderson model with the extrinsic edge mode in Eq.~(\ref{eq:time-dep_Anderson_edge}). Parameters are $L_x =40$, $L_y =25$, $J = 0.2$, $W = 1$, and $T = 1$. The initial state of the wave packet is $\ket{x=20, y=1}$.
Even in this case, the density distribution $\rho(y,t)$ exhibits a Gaussian tail, and the mean squared distance $\braket{y^2}$ shows a nearly linear behavior in time.}
	\label{fig:time-dep_Anderson_edge}
\end{figure}


\noindent
with a diffusion constant $D$, then the typical solution takes the form of the Gaussian distribution:
\begin{align}
    \rho(x,y,t) = \frac{1}{2\pi \sigma_t^2} 
    \exp\left[ - \frac{x^2+y^2}{2\sigma_t^2}
    \right]
    , \ \sigma_t = \sqrt{Dt},
\end{align}
which leads to
\begin{align}
    \rho (y,t) = 
    \frac{1}{\sqrt{2\pi \sigma_t^2}} 
    \exp\left[ - \frac{y^2}{2\sigma_t^2}
    \right],
    \quad 
\braket{y^2}=Dt.
\label{eq:rhoy2}
\end{align}
We show our numerical results of $\rho(y,t=100T)$ and $\braket{y^2}$ for the time-dependent Anderson models in Eqs.~(\ref{eq:time-dep_Anderson}) and (\ref{eq:time-dep_Anderson_edge}) in Figs.~\ref{fig:time-dep_Anderson} and \ref{fig:time-dep_Anderson_edge}, respectively.
These numerical results display Gaussian tails for $\rho(y,t=100T)$
and nearly linear behaviors in time for $\langle y^2\rangle$, which are consistent with the diffusive behaviors in Eq.~(\ref{eq:rhoy2}).
Similar results have been reported for models in Refs.~\cite{Konno04,Joye10,Joye11,Evensky90}.

\bibliographystyle{apsrev4-1}
\bibliography{QwalkBEC}

\end{document}